\documentclass[10pt,twocolums]{iopart}
\usepackage{epsf}

\begin{document}

\topical[Vortex matter in S/F hybrids]{Nucleation of
superconductivity and vortex matter in superconductor --
ferromagnet hybrids}

\author{A Yu Aladyshkin$^{1,2}$, A V Silhanek$^{1}$, W Gillijns$^{1}$,  and V V Moshchalkov$^{1}$, }
\address{$^{(1)}$ INPAC -- Institute for Nanoscale Physics and Chemistry, Nanoscale
Superconductivity and Magnetism  and Pulsed Fields Group,
K.U.Leuven, Celestijnenlaan 200D, B--3001 Leuven, Belgium \\
$^{(2)}$ Institute for Physics of Microstructures, Russian Academy
of Sciences, 603950, Nizhny Novgorod, GSP-105, Russia}

\ead{aladyshkin@ipm.sci-nnov.ru,
alejandro.silhanek@fys.kuleuven.be}

\begin{abstract}
The theoretical and experimental results concerning the
thermodynamical and low-frequency transport properties of hybrid
structures, consisting of spatially-separated conventional
low-temperature superconductor (S) and ferromagnet (F), is
reviewed. Since the superconducting and ferromagnetic parts are
assumed to be electrically insulated, no proximity effect is
present and thus the interaction between both subsystems is
through their respective magnetic stray fields. Depending on the
temperature range and the value of the external field $H_{ext}$,
different behavior of such S/F hybrids is anticipated. Rather
close to the superconducting phase transition line, when the
superconducting state is only weakly developed, the magnetization
of the ferromagnet is solely determined by the magnetic history of
the system and it is not influenced by the field generated by the
supercurrents. In contrast to that, the nonuniform magnetic field
pattern, induced by the ferromagnet, strongly affect the
nucleation of superconductivity leading to an exotic dependence of
the critical temperature $T_{c}$ on $H_{ext}$. Deeper in the
superconducting state the effect of the screening currents cannot
be neglected anymore. In this region of the phase diagram various
aspects of the interaction between vortices and magnetic
inhomogeneities are discussed. In the last section we briefly
summarize the physics of S/F hybrids when the magnetization of the
ferromagnet is no longer fixed but can change under the influence
of the superconducting currents. As a consequence, the
superconductor and ferromagnet become truly coupled and the
equilibrium configuration of this ``soft'' S/F hybrids requires
rearrangements of both, superconducting and ferromagnetic
characteristics, as compared with ``hard'' S/F structures.
\newline
\newline Some figures in this paper are in color only in the
electronic version.
\end{abstract}

\maketitle

\tableofcontents

\newpage
\section*{List of main notations}

\vspace*{0.1cm} \noindent {\it Acronyms}

\begin{tabbing}
  $\alpha$, $\beta$ \= c;  \kill
    GL                  \> $\quad$ Ginzburg-Landau, \\
    DWS                 \> $\quad$ domain-wall superconductivity, \\
    F                   \> $\quad$ ferromagnet or ferromagnetic, \\
    OP                  \> $\quad$ order parameter, \\
    RDS                 \> $\quad$ reverse-domain superconductivity, \\
    S                   \> $\quad$ superconductor or superconducting \\
    1D                  \> $\quad$ one-dimensional \\
    2D                  \> $\quad$ two-dimensional \\
\end{tabbing}

\vspace*{0.1cm} \noindent {\it Latin letters}

\begin{tabbing}
  $\alpha$, $\beta$ \= c;  \kill
  ${\bf A}$             \> $\quad$ vector potential, corresponding to the total magnetic field: ${\bf B}={\rm rot}\,{\bf A}$, \\
  ${\bf a}$             \> $\quad$ vector potential, describing to the nonuniform component of the magnetic field, ${\bf b}={\rm rot}\,{\bf a}$, \\
  ${\bf B}$             \> $\quad$ total magnetic field: ${\bf B}={\bf H}_{ext}+{\bf b}$, \\
  ${\bf b}$             \> $\quad$ nonuniform component of the magnetic field induced by ferromagnet, \\
  $c$                   \> $\quad$ speed of light, \\
  $D_s$                 \> $\quad$ thickness of the superconducting film, \\
  $D_f$                 \> $\quad$ thickness of the ferromagnetic film (or single crystal), \\
  $f$                   \> $\quad$ absolute value of the normalized OP wave function: $f=\sqrt{({\rm Re}\,\psi)^2+({\rm Im}\,\psi)^2}$, \\
  ${\bf j}_{ext}$       \> $\quad$ the density of the external current: ${\rm rot\,}{\bf H}_{ext}=(4\pi/c)\,{\bf j}_{ext}$, \\
  ${\bf j}_s$           \> $\quad$ the density of superconducting currents, \\
  $G_{sf}$              \> $\quad$ free (Gibbs) energy of the S/F hybrid, \\
  $G_{m}$               \> $\quad$ term in the free energy functional accounting for the spatial variation of the magnetization,\\
  ${\bf H}_{ext}$       \> $\quad$ external magnetic field, \\
  ${\bf H}_{ex}$        \> $\quad$ exchange field, \\
  $H_{c1}$              \> $\quad$ lower critical field: $H_{c1}=\Phi_0\,\ln(\lambda/\xi)/(4\pi\lambda^2)$, \\
  $H_{c2}$              \> $\quad$ upper critical field: $H_{c2}=\Phi_0/(2\pi\xi^2)=H^{(0)}_{c2}\,(1-T/T_{c0})$, \\
  $H^{(0)}_{c2}$        \> $\quad$ upper critical field at $T=0$: $H^{(0)}_{c2}=\Phi_0/(2\pi\xi_0^2)$, \\
  $h$                   \> $\quad$ separation between superconducting and ferromagnetic films, \\
  $L$                   \> $\quad$ angular momentum of Cooper pairs (vorticity): $\psi=f(r)\,e^{iL\varphi}$, \\
  $\ell_H$              \> $\quad$ magnetic length: $\ell_H=\sqrt{\Phi_0/(2\pi|H_{ext}|)}$, \\
  $\ell_b^*$            \> $\quad$ effective magnetic length determined by a local magnetic field $b_z^*$: $\ell_b^*=\sqrt{\Phi_0/(2\pi|b_z^*|)}$, \\
  $\ell_{\psi}$         \> $\quad$ typical width of the localized OP wave function, \\
  ${\bf M}$             \> $\quad$ magnetization of the ferromagnet, \\
  $M_s$                 \> $\quad$ magnetization of the ferromagnet in saturation, \\
  ${\bf m}_0$           \> $\quad$ dipolar moment of a point-like magnetic particle, \\
  $R_s$                 \> $\quad$ radius of the superconducting disk, \\
  $R_f$                 \> $\quad$ radius of the ferromagnetic disk-shaped dots, \\
  ${\bf R}_d$           \> $\quad$ position of a point-magnetic dipole: ${\bf R}_d=\{X_d,Y_d,Z_d\}$, \\
  $T_{c0}$              \> $\quad$ superconducting critical temperature at $B=0$, \\
  $w$                   \> $\quad$ period of the one-dimensional domain structures in ferromagnet \\
\end{tabbing}

\vspace*{0.1cm} \noindent {\it Greek letters}

\begin{tabbing}
  $\alpha$, $\beta$ \= c;  \kill
  $\alpha$, $\beta$     \> $\quad$ constants of the standard expansion of the density of the free energy with respect to $|\Psi|^2$, \\
  $\epsilon^{(0)}_v$    \> $\quad$ self-energy of the vortex line per unit length: $\epsilon^{(0)}_v=(\Phi_0/4\pi\lambda)^2\,\ln \lambda/\xi$, \\
  $\Theta$              \> $\quad$ the OP phase: $\Theta=\arctan\, ({\rm Im}\,\psi/{\rm\, Re}\,\psi)$, \\
  $\lambda$             \> $\quad$ temperature-dependent magnetic field (London) penetration length: $\lambda=\lambda_0/\sqrt{1-T/T_{c0}}$, \\
  $\lambda_0$           \> $\quad$ magnetic field penetration length at $T=0$, \\
  $\xi$                 \> $\quad$ temperature-dependent superconducting coherence length: $\xi=\xi_0/\sqrt{1-T/T_{c0}}$, \\
  $\xi_0$               \> $\quad$ Ginzburg-Landau coherence length at $T=0$, \\
  $\pi$                 \> $\quad$ 3.141592653..., \\
  $\rho$                \> $\quad$ electrical resistivity, \\
  $\Phi_0$              \> $\quad$ magnetic flux quantum: $\Phi_0=\pi\hbar c/e\simeq2.07$ Oe$\cdot$cm$^2$, \\
  $\Psi$                \> $\quad$ superconducting order parameter (OP) wave function, \\
  $\Psi_0$              \> $\quad$ OP saturated value, $\Psi_0=\sqrt{-\alpha/\beta}$, \\
  $\psi$                \> $\quad$ normalized OP wave function, $\psi=\Psi/\Psi_0$ \\
\end{tabbing}

\vspace*{0.1cm} \noindent {\it Coordinate systems}

Throughout this paper we use both cartesian reference system
$(x,y,z)$ and cylindrical reference system $(r,\varphi,z)$, where
$z-$axis is always taken perpendicular to the superconducting
film/disk.

\section{Introduction}
\label{Introduction}

According to the classical Bardeen--Cooper--Schrieffer theory of
superconductivity, the ground state of the superconducting
condensate consists of electron pairs with opposite spins (the
so-called spin-singlet state) bounded via phonon interactions
\cite{Bardeen-PR-57a,Bardeen-PR-57b}. As early as 1956, Ginzburg
\cite{Ginzburg-JETP-56} pointed out that this fragile state of
matter could be destroyed by the formation of a homogeneous
ferromagnetic ordering of spins if its corresponding magnetic
field exceeds the thermodynamical critical field of the
superconductor. Later on, Matthias {\it et al.}
\cite{Matthias-PRL-58a,Matthias-PRL-58b,Matthias-PRL-60}
demonstrated that besides the orbital effect (i.e. a pure
electromagnetic interaction between the ferromagnetic and
superconducting subsystems), there is also a strong suppression of
superconductivity arising from the exchange interaction which
tends to align the spins of the electrons in detriment of
Cooper-pair formation. Anderson and Suhl \cite{Anderson-PR-59}
predicted that a compromise between these antagonistic states can
be achieved if the ferromagnetic phase is allowed to break into
domains of size much smaller than the superconducting coherence
length $\xi$ in such a way that from the superconductivity point
of view, the net magnetic moment averages to zero. Alternatively,
Larkin and Ovchinnikov \cite{Larkin-JETP-65} and Fulde and Ferrel
\cite{Fulde-PR-64}, theoretically predicted that superconductivity
can survive in a uniform ferromagnetic state if the
superconducting order parameter is spatially modulated.

In general terms, the effective polarization of the conduction
electrons, either due to the external field $H_{ext}$ (orbital
effect) or the exchange field $H_{ex}$ (paramagnetic effect),
leads to a modification (suppression and modulation) of the
superconducting order parameter. Typically, in ferromagnetic
metals the exchange field is considerably higher than the internal
magnetic field and it dominates the properties of the system.
However, in some cases, where both fields can have opposite
directions, an effective compensation of the conduction electrons
polarization can occur and consequently superconductivity can be
recovered at high fields \mbox{$H_{ext}\simeq-H_{ex}$} (Jaccarino
and Peter \cite{Jaccarino-PRL-62}). Bulaevskii {\it et al.}
\cite{Bulaevskii-AdvPhys-84} gave an excellent overview of both
experimental and theoretical aspects of coexistence of
superconductivity and ferromagnetism where both orbital and
exchange effects are taken into account.

    \begin{figure}[b!]
    \begin{center}
    \epsfxsize=80mm \epsfbox{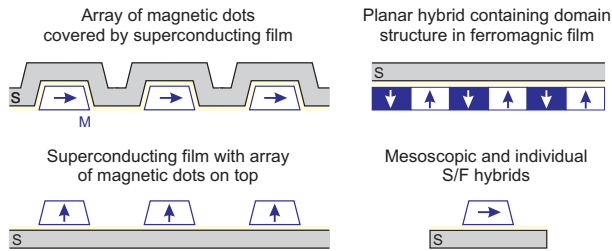}
    \end{center}
    \caption{(color online) Typical examples of considered S/F hybrid systems with dominant orbital interaction.}
    \label{Fig:Geometry}
    \end{figure}

The progressive development of material deposition techniques and
the advent of refined lithographic methods have made it possible
to fabricate superconductor-ferromagnet structures (S/F) at
nanometer scales. Unlike the investigations dealing with the
coexistence of superconductivity and ferromagnetism in
ferromagnetic superconductors (for review see Flouquet and Buzdin
\cite{Flouquet-PW-02}), the ferromagnetic and supercondicting
subsystems in the artificial heterostructures can be physically
separated. As a consequence, the strong exchange interaction is
limited to a certain distance around the S/F interface whereas the
weaker electromagnetic interaction can persist to longer distances
into each subsystem. In recent reviews, Izyumov {\it et al.}
\cite{Izyumov-UFN-02}, Buzdin \cite{Buzdin-RMP-05} and Bergeret
{\it et al.} \cite{Bergeret-RMP-05} discussed in detail the role
of proximity effects in S/F heterostructures dominated by exchange
interactions\footnote[1] {In particular, trilayered S/F/S
    structures with transparent S/F interfaces allow to realize
    Josephson junctions with an arbitrary phase difference between the
    superconducting electrodes, which depend on the thickness of the
    ferromagnetic layer (see, e.g., papers of Proki\'{c} {\it et al.}
    \cite{Prokic-PRB-99},  Ryazanov {\it et al.}
    \cite{Ryazanov-PRL-01}, Kontos {\it et al.} \cite{Kontos-PRL-02},
    Buzdin and Baladie \cite{Buzdin-PRB-03a}, Oboznov {\it et al.}
    \cite{Oboznov-PRL-06} and references therein). The antipode F/S/F
    heterostructures attract a considerable attention in connection
    with the investigation of unusual properties of such layered hybrid structures governed by the
    mutual orientation of the vectors of the magnetization in the ``top'' and ``bottom'' ferromagnetic layers
    (see, e.g., papers of Deutscher and Meunier \cite{Deutscher-PRL-69}, Ledvij {\it et al.} \cite{Ledvij-PRB-91},
    Buzdin {\it et al.} \cite{Buzdin-EPL-99}, Tagirov
    \cite{Tagirov-PRL-99},
    Baladi\'{e} {\it et al.} \cite{Baladie-PRB-01},
    Gu {\it et al.} \cite{Gu-PRL-02,Gu-JAP-03}, Pe\~{n}a {\it et al.} \cite{Pena-PRL-05},
    Moraru {\it et al.} \cite{Moraru-PRL-06}, Rusanov {\it et al.}
    \cite{Rusanov-PRB-06}, Steiner and Ziemann \cite{Steiner-PRB-06},
    Singh {\it et al.} \cite{Singh-PRB-07}).}.
In order to unveil the effect of electromagnetic coupling it is
imperative to suppress proximity effects by introducing an
insulating buffer material between the S and F films. In an
earlier report, Lyuksyutov and Pokrovsky
\cite{Lyuksyutov-AdvPhys-05} addressed the physical implications
of both electromagnetic coupling and exchange interaction in the
S/F systems deep into the superconducting state.

\begin{figure}[t!]
    \begin{center}
    \epsfxsize=80mm \epsfbox{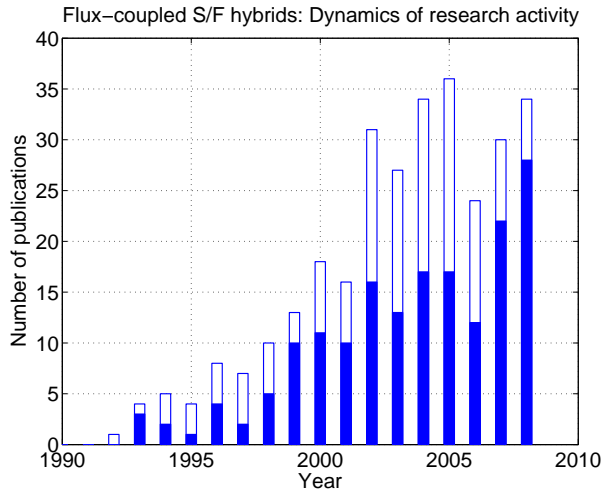}
    \end{center}
    \caption{(color online) The histogram shows an increase of the number of publications
    dealing with the investigations of the S/F hybrids where the conventional low$-T_c$ superconductors interact
    with magnetic textures mainly via stray magnetic fields: blue bars correspond to the
    experimental papers, while white bars refer to pure theoretical contributions.}
    \label{Fig:Publications}
    \end{figure}

In the present review we are aiming to discuss the thermodynamic
and low-frequency transport phenomena in the S/F hybrid structures
dominated by electromagnetic interactions. We focus only on the
S/F hybrids consisting of conventional low$-T_c$ superconductors
without weak links\footnote[2]{Ferromagnetic dots are shown to
    induce an additional phase difference in
    Josephson junctions, leading to a significant modification of the
    dependence of the Josephson critical current $I_c$ on the external
    magnetic field $H_{ext}$ (so-called Fraunhofer diffraction
    pattern, see, e.g., textbook of Barone and Paterno \cite{Barone-Paterno}), which becomes sensitive to the
    magnetization of ferromagnetic particle (Aladyshkin {\it et al.} \cite{Aladyshkin-JMMM-03}, Vdovichev {\it
    et al.} \cite{Vdovichev-JETPLett-04}, Fraerman {\it et al.}
    \cite{Fraerman-PRB-06}, Held {\it et al.} \cite{Held-APL-06},
    Samokhvalov \cite{Samokhvalov-JETP-07}). }. The S/F heterostructures with pure
electromagnetic coupling can be described phenomenologically using
Ginzburg-Landau and London formalisms rather than sophisticated
microscopical models. Some typical examples of such structures
found in the literature are shown schematically in
Fig.~\ref{Fig:Geometry}. As an illustration of the continuous
growth of interest in S/F heterostructures with suppressed
proximity effect we refer to Fig. \ref{Fig:Publications}, which
shows the number of the publications during the last two decades.

The review is organized as follows. Section \ref{Part-I} is
devoted to the nucleation of the superconducting order parameter
under inhomogeneous magnetic fields, induced by single domain
walls and periodic domain structures in plain ferromagnetic films
or by magnetic dots. A similar problem for individual symmetric
microstructures was reviewed by Chibotaru {\it et al.}
\cite{Chibotaru-JMP-05}. Section \ref{LondonSection} is devoted to
the static and dynamic properties of S/F systems at low
temperatures when the superconducting OP becomes fully developed
and the screening effects cannot be disregarded any longer. The
vortex pinning properties of the S/F hybrids have been recently
analyzed by V\'{e}lez {\it et al.} \cite{Velez-JMMM-08} and the
fabrication of ordered magnetic nanostructures has been earlier
considered by Mart\'{\i}n {\it et al.} \cite{Martin-JMMM-03}. In
the last section \ref{Soft-SF-systems} we briefly introduce the
problem of ``soft'' magnets in combination with superconducting
materials, where now the superconducting currents and the magnetic
stray field emanating from the ferromagnetic material mutually
influence each other. In the conclusion, we formulate a number of
relevant issues that, to our understanding, remain unsettled and
deserve further investigations. The appendix summarizes the
experimental and theoretical research activities on the considered
S/F heterostructures, where we present a classification based on
the choice of materials for the experimental research and on the
used model for the theoretical treatment.

Importantly, we would like to note already in the Introduction
that the literature and references used by the authors in this
review by no mean can be considered as a complete set. Due to
dynamic and rather complex character of the subject and also to
the limited space in this review, inevitably quite a lot of
important and interesting contributions could have been missed
and, therefore, in a way, the used references reflect the
``working list'' of publications the authors of this review
are dealing with.

\section{Nucleation of superconductivity in S/F hybrids (high-temperature limit)}
\label{Part-I}

\subsection{Ginzburg-Landau description of a magnetically coupled S/F hybrid system}

{\it Derivation of the Ginzburg-Landau  equations}

\noindent In order to describe hybrid structures, consisting of a
type-II superconductor and a ferromagnet, for the case that no
diffusion of Cooper pairs from superconductor to ferromagnet takes
place, the phenomenological Ginzburg-Landau (GL) theory can be
used. As a starting point we consider the properties of S/F
hybrids for external magnetic fields ${\bf H}_{ext}$ below the
coercive field of the ferromagnet which is assumed to be
relatively large. In this case the magnetization of the
ferromagnet ${\bf M}$ is determined by the magnetic history only
and it does neither depend on ${\bf H}_{ext}$ nor on the
distribution of the screening currents inside the superconductor.
Such a ``hard-magnet approximation'' is frequently used for a
theoretical treatment and it appears to be approximately valid for
most part of the experimental studies presented in this section.
The review of the properties of hybrid S/F systems consisting of
superconductors and soft magnets will be presented later on in
section \ref{Soft-SF-systems}.

Following Landau's idea of phase transitions of the second kind,
the equilibrium properties of a system close to the phase
transition line can be obtained by minimization of the free energy
functional (see, e.g, textbooks of Abrikosov
\cite{Abrikosov-book}, Schmidt \cite{Schmidt-book}, Tinkham
\cite{Tinkham-book}):
    \begin{eqnarray}
    \label{GibbsEnergyExpansion1} \nonumber
    G_{sf}=G_{s0}+G_m+\int\limits_{V}
    \left\{\alpha\,|\Psi|^2 + \frac{\beta}{2}\,|\Psi|^4
    +\frac{1}{4m}\left|-i\hbar\nabla\Psi-\frac{2e}{c}{\bf
    A}\Psi\right|^2 \right. \\ \left. + \frac{{\bf B}^2}{8\pi}- {\bf
    B}\cdot{\bf M} - \frac{{\bf
    B}\cdot{\bf H}_{ext}}{4\pi}  \right\} \, dV,
    \end{eqnarray}
where the integration should be performed over the entire
space\footnote[1]{
    Hereafter we used the Gauss (centimeter-gram-second) system of units, therefore
    all vectors ${\bf B}$, ${\bf M}$, and ${\bf H}_{ext}$  have the same dimensionality:
    [$B$]=Gauss (G), [$M$]=Oersted (Oe), [$H_{ext}$]=Oersted (Oe).}.
Here $G_{s0}$ is a field-- and temperature--independent part of
the free energy, $\alpha=\alpha_0\,(T_{c0}-T)$,  $\alpha_0$ and
$\beta$ are positive temperature--independent constants, $\Psi$ is
an effective wave function of the Cooper pairs, ${\bf B}({\bf
r})={\rm rot}\,{\bf A}({\bf r})$ is the magnetic field and the
corresponding vector potential, $T_{c0}$ is the critical
temperature at $B=0$, $e$ and $m$ are charge and mass of carriers
(e.g., electrons), and $c$ is the speed of light. The term
$G_{m}$, which will be explicitly introduced in the last section
\ref{Soft-SF-systems}, accounts for the self-energy of the
ferromagnet which depends on the particular distribution of
magnetization. This term seems to be constant for hard
ferromagnets with a fixed distribution of magnetization, therefore
it does not influence the OP pattern and the superconducting
current distribution in hard S/F hybrids. Introducing a
dimensionless wave function $\psi=\Psi/\Psi_0$, normalized by the
OP value $\Psi_0=\sqrt{a_0(T_{c0}-T)/b}\,$ in saturation, one can
rewrite Eq. (\ref{GibbsEnergyExpansion1}) in the following form
    \begin{eqnarray}
    \label{GibbsEnergyExpansion2} \nonumber G_{sf}=G_{s0}+G_m+\int\limits_{V}\left\{\frac{\Phi^2_0}{32\pi^3
    \lambda^2}\left(-\frac{1}{\xi^2}|\psi|^2  +
    \frac{1}{2\xi^2}|\psi|^4
    +\left|\nabla \psi + i \frac{2\pi}{\Phi_0}{\bf
    A}\psi\right|^2  \right) \right. \\ \left. + \frac{{\bf B}^2}{8\pi} - {\bf
    B}\cdot{\bf M} - \frac{{\bf B}\cdot{\bf H}_{ext}}{4\pi} \right\} \, dV,
    \end{eqnarray}
expressed via the temperature-dependent coherence length
$\xi^2=\hbar/[4m \,a_0 (T_{c0}-T)]$, the London penetration depth
$\lambda^2=mc^2b/[8\pi e^2 \,a_0 (T_{c0}-T)]$, and the magnetic
flux quantum $\Phi_0=\pi\hbar c/|e|$.

Although the Ginzburg-Landau model was proven to be consistent
only at temperatures close to the superconducting critical
temperature (Gorkov \cite{Gorkov-JETP-59}), the applicability of
this model seems to be much broader, at least from a qualitative
point of view. After minimization of the free energy functional
Eq. (\ref{GibbsEnergyExpansion2}) with respect to the order
parameter (OP) wave function $\psi$ and ${\bf A}$ respectively,
one can derive the two coupled Ginzburg-Landau equations
\cite{Abrikosov-book,Schmidt-book,Tinkham-book}:
    \begin{eqnarray}
    \label{FullGLEquation-1}
    -\xi^2\left(\nabla+ i\frac{2\pi}{\Phi_0}\,{\bf A}\right)^2\psi
    -
    \psi + |\psi|^2\psi = 0, \\  \label{FullGLEquation-2} {\rm rot~}{\rm rot~}{\bf A} =
    \frac{4\pi}{c}\,{\bf j}_s + 4\pi\,{\rm rot\,}{\bf M} + \frac{4\pi}{c}\,{\bf j}_{\,ext} ,
    \end{eqnarray}
where
    \begin{eqnarray}
    \label{FullGLEquation-3}
    \nonumber {\bf j}_s= \frac{c}{4\pi}\, \frac{|\psi|^2}{\lambda^2}\,\left(\frac{\Phi_0}{2\pi}\nabla\Theta - {\bf
    A}\right)
    \end{eqnarray}
represents the density of superconducting currents, while ${\bf
j}_{\,ext}=(c/4\pi)\,{\rm rot}\,{\bf H}_{ext}$ is the density of
the currents corresponding to external sources, and $\Theta$ is
the OP phase, $\psi({\bf r})=f({\bf r})\,e^{i\Theta({\bf r})}$.

\vspace*{0.3cm}

\noindent {\it Linearized GL equation}

\noindent It is quite natural to expect that at the initial stage
of the formation of superconductivity [i.e. close to the phase
transition line $T_c(H_{ext})$, which separates the normal and
superconducting state in the $T-H_{ext}$ plane], the density of
the superconducting condensate will be much smaller than the fully
developed OP value: \mbox{$|\psi|^2\ll 1$}. This allows one to
neglect: (i) the nonlinear term $|\psi|^2 \psi$ in
Eq.~(\ref{FullGLEquation-1}) and, (ii) the corrections to the
vector potential ${\bf A}$ caused by the screening currents in
Eq.~(\ref{FullGLEquation-2}), since the supercurrents ${\bf j}_s$
are also proportional to $|\psi|^2$. Thus, the nucleation of
superconductivity can be analyzed in the framework of the
linearized GL equation
\cite{Abrikosov-book,Schmidt-book,Tinkham-book}:
    \begin{eqnarray}
    \label{LinearizedGLEquation-1} -\left(\nabla +
    i\frac{2\pi}{\Phi_0}\,{\bf A}\right)^2 \psi = \frac{1}{\xi^2}\, \psi,
    \end{eqnarray}
in a given magnetic field described by the vector potential distribution
    \begin{eqnarray}
    \label{LinearizedGLEquation-2}
    {\bf A} = \frac{1}{c}\int\limits \frac{{\bf j}_{\,ext}({\bf r}^{\prime})}{|{\bf r}-{\bf
    r}^{\prime}|}\, d^3{\bf r}^{\prime} + \int \frac{{\rm rot}{\bf\,M}({\bf r}^{\prime})}{|{\bf r}-{\bf
    r}^{\prime}|}\, d^3{\bf r}^{\prime},
    \end{eqnarray}

\noindent The solution of Eq.~(\ref{LinearizedGLEquation-1})
consists of a set of eigenvalues $(1/\xi^2)_n$, corresponding to
the appearance of certain OP pattern $\psi_{n}$, for every value
of the applied magnetic field $H_{ext}$. The critical temperature
of the superconducting transition $T_c$ is determined by the
lowest eigenvalue of the problem: $T_c
=\,T_{c0}\left\{1-\xi_0^{2}\,(1/\xi^2)_{min}\right\}$.

\vspace*{0.3cm}

\noindent {\it The phase boundary for plain superconducting films}

\noindent First, we would like to present the well-known solution
of the linearized GL equation Eq. (\ref{LinearizedGLEquation-1})
corresponding to the OP nucleation in a plain superconducting
film, infinite in the lateral direction and placed in a transverse
\emph{uniform} magnetic field ${\bf H}_{ext}=H_{ext}\,{\bf z}_0$
\cite{Abrikosov-book,Schmidt-book,Tinkham-book}. Taking the gauge
$A_y=xH_{ext}$, one can see that
Eq.~(\ref{LinearizedGLEquation-1}) depends explicitly on the
$x-$coordinate only, therefore its general solution can be written
in the form: $\psi=f(x)\,e^{iky+iqz}$, where the wave vectors $k$
and $q$ should adjust themselves to provide the maximization of
the $T_c$ value. Using this representation in
Eq.~(\ref{LinearizedGLEquation-1}), it is easy to see that the
spectrum of eigenvalues $(1/\xi^{2})_n$ is similar to the energy
spectrum of the harmonic oscillator but shifted:
$(1/\xi^{2})_n=2\pi(2n+1) H_{ext}/\Phi_0+q^{2}$ and the
$(1/\xi^{2})$ minimum (the maximum of $T_c$) corresponds to $n=0$
and $q=0$ for any $H_{ext}$ value, $(1/\xi^{2})_{min}=2\pi
H_{ext}/\Phi_0$. The critical temperature of the superconducting
transition\footnote[1]{It is well known that
    superconductivity
    nucleates in the form of a Gaussian-like OP wave function
    $\psi(x,y)=e^{-(x-x_0)^2/2\ell_H^2}\,e^{iky}$, localized in the
    lateral direction at distances of the order of the so-called
    magnetic length $\ell_H^2=\Phi_0/(2\pi |H_{ext}|)$ and uniform
    over the film thickness. The oscillatory factor $e^{iky}$ describes the
    displacement of the OP maximum positioned at $x_0=k\Phi_0/(2\pi
    H_{ext})$ without a change of the $(1/\xi^{2})_{min}$ value.
    It is interesting to note that the
    confinement of the OP wave function is determined by the magnetic length
    $\ell_H$, i.e. the OP width is a function of the external field $H_{ext}$.
    On the other hand, the temperature-dependent coherence length
    $\xi$ is a natural length scale describing the spatial OP
    variations. The equality $\ell^2_H=\xi^2$ defines the same phase
    boundary in the $T-H_{ext}$ plane as that given by Eq. (\ref{Hc2}).}
    as a function of a
    uniform transverse magnetic field is given by $T_c
    =\,T_{c0}\left[1-2\pi\xi_0^{2}\,H_{ext}/\Phi_0\right]$, or
    \begin{eqnarray}
    \label{Hc2}
    1-\frac{T_c}{T_{c0}}= \frac{|H_{ext}|}{H_{c2}^{(0)}},
    \end{eqnarray}
where $H_{c2}^{(0)}=\Phi_0/(2\pi\xi_0^2)$ is the upper critical
field at $T=0$. The inversely proportional dependence of the shift
of the critical temperature $1-T_c/T_{c0}$ on the square of the OP
width $\ell_H^2$ can be interpreted in terms of the quantum-size
effect for Cooper pairs in a uniform magnetic field. It should be
mentioned that the effect of sample's topology on the eigenenergy
spectrum $(1/\xi^{2})_n$ becomes extremely important for
mesoscopic superconducting systems, whose lateral dimensions are
comparable with the coherence length $\xi$. Indeed, this
additional confinement of the OP wave function significantly
modifies the OP nucleation in mesoscopic superconductors and the
corresponding phase boundaries $T_c(H_{ext})$ differ considerably
from that typical for bulk samples and films infinite in the
lateral directions (see Chibotaru {\it et al.}
\cite{Chibotaru-JMP-05}, Moshchalkov {\it et al.}
\cite{Moshchalkov-Nature-95,Moshchalkov-monograph-00}, Berger and
Rubinstein \cite{Berger-book}).

\subsection{Magnetic confinement of the OP wave function in an inhomogeneous magnetic field: general considerations}
\label{GeneralConsideration}

The main focus in this section is to describe the nucleation of
the superconducting order parameter in a static \emph{non-uniform}
magnetic field ${\bf H}_{ext}+{\bf b}({\bf r})$ based on a simple
approach\footnote[1]{We introduce the
    following notations: ${\bf b}={\rm rot}\,{\bf a}$ characterizes
    the non-uniform component of the magnetic field only,
    while ${\bf B}={\bf H}_{ext}+{\bf b}={\rm rot}\,{\bf A}$
    is the total magnetic field distribution, the external field ${\bf H}_{ext}$
    is assumed to be uniform.}.
This method makes it possible to see directly a correspondence
between the position of the maximum of the localized wave function
$\psi$ and the critical temperature $T_c$ in the presence of
spatially-modulated magnetic field ${\bf b}({\bf r})$, generated
by ferromagnet. For simplicity, we assume that the thin
superconducting film is infinite in the $(x,y)-$plane, i.e.
perpendicular to the direction of the external field ${\bf
H}_{ext}=H_{ext}\,{\bf z}_{0}$. This allows us to neglect the
possible appearance of superconductivity in the sample perimeter
(surface
superconductivity~\cite{Abrikosov-book,Schmidt-book,Tinkham-book})
and focus only on the effect arising from the nonuniform magnetic
field.

\vspace*{0.2cm}

\noindent {\it Importance of out-of-plane component of the field}

\noindent It should be emphasized that the formation (or
destruction) of superconductivity in thin superconducting films is
sensitive to the spatial variation of the out-of-plane component
of the total magnetic field. Indeed, the upper critical fields
$H_{c2}^{\perp}$ and $H_{c2}^{\|}$ for the out-of-plane and
in-plane orientation for a uniform applied magnetic field can be
estimated as follows
\cite{Abrikosov-book,Schmidt-book,Tinkham-book}:
    \begin{eqnarray}
    H_{c2}^{\perp} \sim \frac{\Phi_0}{\xi^2},\qquad H_{c2}^{\|}
    \sim \frac{\Phi_0}{\xi D_s},
    \end{eqnarray}
where $D_s$ is the thickness of the superconducting sample. For
rather thin superconducting films and/or close to the
superconducting critical temperature \mbox{$D_s\ll \xi=\xi_0
(1-T/T_{c0})^{-1/2}$}, therefore $H_{c2}^{\perp}\ll H_{c2}^{\|}$.
In other words, superconductivity will generally be destroyed by
the out-of-plane component of the magnetic field rather than by
the in-plane component, and thus, to a large extent, the spatial
distribution of the out-of-plane component determines the OP
nucleation in thin-film structures. Since a uniform magnetic field
is known to suppress the critical temperature, one can expect that
the highest $T_c$ value should correspond to the OP wave function
localized near regions with the lowest values of the perpendicular
magnetic field $|B_z({\bf r})|$ provided that
$B_z=H_{ext}+b_z({\bf r})$ varies slowly in space.

If the field $H_{ext}$ exceeds the amplitude of the internal field
modulation (i.e. \mbox{$H_{ext}<-{\rm max\,}b_z$} and
\mbox{$H_{ext}>-{\rm min\,}b_z$}), the total magnetic field is
non-zero in the whole sample volume, and the favorable positions
for the OP nucleation are at the locations of minima of
\mbox{$|B_z({\bf r})|=|H_{ext}+b_z({\bf r})|$}. If the
characteristic width $\ell_{\psi}$ of the OP wave function, which
will be defined later, is much less than the typical length scale
$\ell_{b}$ of the magnetic field variation, then locally the
magnetic field can be considered as uniform at distances of the
order of $\ell_{\psi}$ and it approximately equals to \mbox{${\rm
min\,}|H_{ext}+b_z({\bf r})|$}. Then, using the standard
expression for the upper critical field Eq. (\ref{Hc2}) and
substituting the effective magnetic field instead of the applied
field, one can obtain the following estimate for the phase
boundary
    \begin{eqnarray}
    \label{MagneticBias}
    1-\frac{T_c}{T_{c0}} \simeq {\rm min}\,\frac{|H_{ext}+b_z({\bf
    r})|}{H_{c2}^{(0)}},\quad |H_{ext}|\gg {\rm max~}|b_z|.
    \end{eqnarray}
According to this expression, the dependence $T_c(H_{ext})$ is
still linear asymptotically even in the presence of a nonuniform
magnetic field. However, the critical field will be shifted
upwards (for $H_{ext}>0$) and downward (for $H_{ext}<0$) by an
amount close to the amplitude of the field modulation. Generally
speaking, such a ``magnetic bias" can be asymmetric with respect
to the $H_{ext}=0$ provided that ${\rm max\,}b_z(x)\neq |{\rm
min\,}b_z(x)|$.

\begin{figure*}[t!]
    \begin{center}
    \epsfxsize=75mm \epsfbox{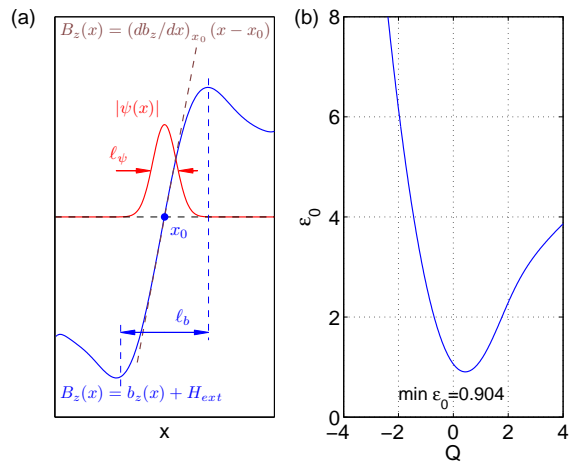}
    \end{center}
    \caption{(color online) (a) Schematic representation of the OP wave function
    $\psi(x)$ localized near the point $x_0$, where the
    $z-$component of the total magnetic field $B_z=H_{ext}+b_z$ vanishes.
    Provided that the OP width is much smaller than the typical length
    scales of the magnetic field ($\ell_{\psi}\ll \ell_b$),
    the actual field distribution $B_z(x)$ can be
    approximated by a linear dependence $B_z(x)\simeq
    \left(db_z/dx\right)_{x_0} (x-x_0)$. \newline
    (b). Energy spectrum $\varepsilon_0$ vs. $Q$ of the model problem Eq. (\ref{Bi-quadratic}),
    after Aladyshkin {\it et al.} \cite{Aladyshkin-JPCM-03}.}
    \label{Fig:LocalizedWavefunction}
    \end{figure*}

For relatively low $H_{ext}$ values, when the absolute value of
the external field is less than the amplitude of the field
modulation (\mbox{$-{\rm max\,}b_z<H_{ext}<-{\rm min\,}b_z$}), the
$z-$component of the total magnetic field $H_{ext}+b_z(x)$ becomes
zero locally somewhere inside the superconducting film. As a
result, superconductivity is expected to appear first near the
positions where $H_{ext}+b_z({\bf r}_0)=0$ [see panel (a) in Fig.
\ref{Fig:LocalizedWavefunction}]. It is natural to expect that the
details of the OP nucleation depend strongly on the exact topology
of the stray field as well as the field gradient near the lines of
zero field. As an example we will analyze the formation of
superconductivity in a thin superconducting film placed in an
inhomogeneous magnetic field modulated along certain direction.

\vspace*{0.2cm}

\noindent {\it OP nucleation in a magnetic field modulated in one
direction}

\noindent Following Aladyshkin {\it et al.}
\cite{Aladyshkin-PRB-03}, we estimated the dependence of
$T_{c}(H_{ext})$ for a thin superconducting film in the presence
of a non-uniform magnetic field modulated along the $x-$direction,
where $(x,y,z)$ is the Cartesian reference system. Let the
external field be oriented perpendicular to the plane of the
superconducting film, ${\bf H}_{ext}=H_{ext}\,{\bf z}_0$, while
the $z-$component of the total magnetic field vanishes at the
point $x_0$, i.e. $H_{ext}+b_z(x_0)=0$. The vector potential,
corresponding to the field distribution
\mbox{$B_z(x)=H_{ext}+b_z(x)$}, can be chosen in the form
$A_y(x)=x H_{ext}+a_y(x)$, where $b_z=da_y/dx$. Since there is no
explicit dependence on the $y-$coordinate in the linearized GL
equation Eq. (\ref{LinearizedGLEquation-1}), the solution uniform
over the sample thickness can be generally found as
$\psi(x,y)=f_k(x)\,e^{iky}$ and the absolute value of the OP
satisfies the following equation:
    \begin{eqnarray}
    \label{Linearized-GL-x}
    -\frac{d^2f_k}{dx^2}+\left(\frac{2\pi}{\Phi_0} x H_{ext}
    + \frac{2\pi}{\Phi_0} a_y(x) -k\right)^2 f_k=\frac{1}{\xi^2}f_k.
    \end{eqnarray}
Now the parameter $k$ cannot be excluded by a shift of the origin
of the reference system, therefore one should determine the
particular $k$ value in order to minimize
$(1/\xi_0^2)(1-T_c/T_{c0})$ and thus, to maximize the $T_c$ value.

For an unidirectional modulation of the field, the curves of zero
field, where we expect the preferable OP nucleation, are straight
lines parallel to the $y-$axis, and their positions depend on the
external field, $x_0=x_{0}(H_{ext})$. Expanding the vector
potential inside the superconducting film in a power series around
the point $x_0$, one can get
    \begin{eqnarray}
    \nonumber
    A_y(x)\simeq x_0 H_{ext}+a_y(x_0)+\frac{1}{2}\, b_z^{\prime}(x_0)(x-x_0)^2 + ...
    \end{eqnarray}
This local approximation is valid as long as
\mbox{$\left|b_z^{\prime\prime}(x_0)\ell_{\psi}/b_z^{\prime}(x_0)\right|\ll
1.$} Introducing a new coordinate $\tau=(x-x_0)/\ell_{\psi}$ and
the following auxiliary parameters $\ell_{\psi}$ and $Q_k$
    \begin{eqnarray}
    \nonumber \ell_{\psi} = \sqrt[3]{\frac{\Phi_0}{\pi|b_z'(x_0)|}}, \qquad
    Q_k = -\sqrt[3]
    {\frac{\Phi_0}{\pi
    b_z'(x_0)}}\left(\frac{2\pi}{\Phi_0}x_0H_{ext} + \frac{2\pi}{\Phi_0}a_y(x_0) - k\right),
    \end{eqnarray}
we can reduce Eq.~(\ref{Linearized-GL-x}) to the bi-quadratic dimensionless equation
    \begin{eqnarray}
    \label{Bi-quadratic}
    -\frac{d^2 f}{d\tau^2} + (\tau^2-Q_k)^2 f = \varepsilon f \ , \quad
    \mbox{where}\quad
    \varepsilon = \frac{\ell_{\psi}^2}{\xi_0^2}\left(1-\frac{T}{T_{c0}}\right).
    \end{eqnarray}
Thus, the problem of the calculation of the highest $T_c$ value in
the presence of an arbitrary slowly-varying magnetic field as a
function of both the external field and the parameters of the
``internal'' fields is reduced to the determination of the lowest
eigenvalue $\varepsilon_0=\varepsilon_0(Q)$ of the model equation
Eq. (\ref{Bi-quadratic}). As was shown in
Ref.~\cite{Aladyshkin-JPCM-03}, the function $\epsilon_0(Q)$ is
characterized by the following asymptotical behavior:
\mbox{$\varepsilon_0(Q)\simeq Q^2+\sqrt{-Q}$} for \mbox{$Q\ll -1$}
and \mbox{$\epsilon_0(Q)\simeq 2\sqrt{Q}$} for \mbox{$Q\gg 1$} and
it has the minimum value $\varepsilon_{min}=0.904$ [the panel (b)
in Fig. \ref{Fig:LocalizedWavefunction}].

Extracting $T_c$ from
$\varepsilon_0=(1-T_c/T_{c0})\cdot\ell_{\psi}^2/\xi_0^2$, the
approximate expression of the phase boundary takes the following
form:
    \begin{eqnarray}
    \label{Tc-approx-x}
    1-\frac{T_c}{T_{c0}}\simeq  \frac{\xi_0^2}{\ell_{\psi}^2}\,
    \min_{k}
    \varepsilon_0(Q_k) \simeq
    {\xi_0^2}\left(\pi\frac{|b_z'(x_0)|}{\Phi_0}\right)^{2/3}, \quad -{\rm max~}\,b_z<H_{ext}<-{\rm min~}\,b_z.
    \end{eqnarray}
If there are several points $x_{0,i}$ where the external field
compensates the field generated by the ferromagnetic structure,
then the right-hand part of Eq. (\ref{Tc-approx-x}) should be
minimized with respect to $x_{0,i}$. The application of
Eq.(\ref{Tc-approx-x}) for the model cases $b_z=4M_s
\arctan(D_f/x)$ (single domain wall) and $b_z=B_0 \sin(2\pi x/w)$
(periodic domain structure) were given in Ref.
\cite{Aladyshkin-PRB-03}.

\vspace*{0.2cm}

\noindent {\it OP nucleation in axially-symmetrical magnetic
field}

\noindent Similar to the discussion above, one can expect that in
the presence of an axially-symmetrical magnetic field
superconductivity will nucleate in the form of ring-shaped
channels of radius $r=r_0$, where $H_{ext}+b_z(r_0)=0$. The
independence of the linearized GL equation Eq.
(\ref{LinearizedGLEquation-1}) on the angular $\varphi-$coordinate
results in a conservation of the angular momentum (vorticity) $L$
of the superconducting wave function. Thus, a nonuniform magnetic
field makes it possible to have an appearance of giant
(multiquanta) vortex states, which are energetically unfavorable
in plain (non-perforated) large-area superconducting films, but
have been observed in mesoscopic superconductors (Moshchalkov {\it
et al.} \cite{Moshchalkov-Nature-95,Moshchalkov-monograph-00},
Berger and Rubinstein \cite{Berger-book}) and nanostructured films
with antidot lattices (Baert {\it et al.} \cite{Baert-PRL-95},
Moshchalkov {\it et al.} \cite{Moshchalkov-PRB-98}). Expanding the
vector potential in the vicinity of $r_0$ and repeating similar
transformations as above, one can get the following approximate
expression\footnote[1]{Note that the case of a point magnetic
    dipole approximation fails for $L=0$ since the maximum of the OP wave
    function is located at $r=0$ and this theory cannot correctly describe neither the OP
    nucleation nor the phase boundary $T_c(H_{ext})$ at negative $H_{ext}$ values
    close to the compensation field $B_0=-\max b_z(r)$.} for the
phase transition line (Aladyshkin {\it et al.}
\cite{Aladyshkin-JPCM-03,Aladyshkin-PRB-07})
    \begin{eqnarray}
    \label{Eq:Perturb-theory-14}
    \nonumber
    1-\frac{T_c}{T_{c0}} \simeq \frac{\xi_0^2}{\ell_{\psi}^2}
    \,
    \left( \min_L \varepsilon_0(Q_L)-\frac{\ell_{\psi}^2}{4r_0^2}\right), \quad -{\rm max~}\,b_z<H_{ext}<-{\rm
    min~}\,b_z,
    \end{eqnarray}
where the parameters $\ell_{\psi} = \sqrt[3]{\Phi_0/(\pi
\left|db_z/dr\right|_{r_0}})$ and $Q_L=-\big[2\pi r_0
A_{\varphi}(r_0)/\Phi_0-L\big]\,\ell_{\psi}/r_0$ depend on the
external field. Thus, this model predicts field-induced
transitions between giant vortex states with different
vorticities. Since the vorticity $L$ is a discrete parameter, the
changes of the favorable $L$ value while sweeping the external
field leads to abrupt changes in $dT_c/dH_{ext}$. Similar periodic
oscillations of $T_c$ were originally observed by Little and Parks
\cite{Little-PRL-62,Parks-PR-64} for a superconducting cylinder in
a parallel magnetic field and later for any mesoscopic
superconductor in a perpendicular magnetic field (for a review see
Chibotaru {\it et al.} \cite{Chibotaru-JMP-05}).

\vspace*{0.2cm}

\noindent {\it Effect of nonuniform magnetic field on
two-dimensional electron gas}

\noindent It is interesting to note that there is a formal
similarity between the linearized GL equation
(\ref{LinearizedGLEquation-1}) and the stationary
\mbox{Schr\"{o}dinger} equation (see, e.g., Ref.
\cite{Landau-Lifshitz-III}) for a charged free spinless particle
in a magnetic field
    \begin{eqnarray}
    \label{StationarySchrodingerEquation}
    -\frac{\hbar^2}{2m}\left(\nabla -
    \frac{i e}{\hbar c}{\bf A}\right)^2 \psi = E \psi,
    \end{eqnarray}
where $\psi$ is the single particle wave function, $e$ is the
charge and $m$ is the mass of this particle. Based on this
analogy, one can map the results, obtained for a normal electronic
gas in the ground state, on the properties of the superconducting
condensate near the ``superconductor--normal metal'' transition on
the $T-H_{ext}$ diagram.

In particular, the effect of a unidirectional magnetic field
modulation on the energy spectrum of a two-dimensional electronic
gas (2DEG) was analyzed by \mbox{M\"{u}ller} \cite{Muller-PRL-92},
Xue and Xiao \cite{Xue-PRB-92}, Peeters and Vasilopoulos
\cite{Peeters-PRB-93a}, Peeters and Matulis
\cite{Peeters-PRB-93b}, Wu and Ulloa \cite{Wu-PRB-93}, Matulis
{\it et al.} \cite{Matulis-PRL-94}, Ibrahim and Peeters
\cite{Ibrahim-PRB-95}, Peeters {\it et al.}
\cite{Peeters-PhysB-96}, Gumbs and Zhang \cite{Gumbs-SSC-00},
Reijniers and Peeters \cite{Reijniers-JPCM-00,Reijniers-PRB-01a},
Nogaret {\it et al.} \cite{Nogaret-PRB-97,Nogaret-PRL-00}.
Remarkably, in a magnetic field varying linearly along certain
direction, quasi-classical electronic trajectories propagating
perpendicularly to the field gradient $\nabla B_z$ are confined to
a narrow one-dimensional channel localized around the region where
$B_z=0$ \cite{Muller-PRL-92}. A more detailed numerical treatment
revealed a lifting of the well-known degeneracy of the Landau
states on the centers of the Larmor's orbit, inherent to electrons
in a uniform magnetic field. Periodic magnetic field patterns with
zero and nonzero average value was shown to transform the standard
Landau spectrum $E=\hbar\omega_c(n+1/2)$ into a periodic $E(k_y)$
dependence, describing the broadening of the discrete Landau
levels into minibands as in the case of one-dimensional potential
(here $\omega_c=|e|H_{ext}/(mc)$ is the cyclotron frequency). An
oscillatory change of the width of energy bands as $H$ is swept
was shown to give rise to oscillations in the magnetoresistance of
2DEG at low $H_{ext}$ values, which reflect the commensurability
between the diameter of cyclotron orbit at the Fermi level and the
period of the magnetic field modulation \cite{Xue-PRB-92}. The
mentioned oscillatory magnetoresistance due to commensurability
effects was later on corroborated experimentally by Carmona {\it
et al.} \cite{Carmona-PRL-95}. The influence of two-dimensional
magnetic modulations on the single-particle energy spectrum was
theoretically considered by Hofstadter \cite{Hofstadter-PRB-76}. A
modification of the scattering of two-dimension electrons due to
the presence of either ferromagnetic dots or superconducting
vortices was shown to lead to a nontrivial change of the
conductivity in various hybrid systems: 2DEG/superconductor (Geim
{\it et al.} \cite{Geim-PRL-92}, Brey and Fertig
\cite{Brey-PRB-93}, Nielsen and Hedegard \cite{Nielsen-PRL-95},
Reijniers {\it et al.} \cite{Reijniers-PRB-99}) and
2DEG/ferromagnet (Khveshchenko and Meshkov
\cite{Khveshchenko-PRB-93}, Ye {\it et al.} \cite{Ye-PRL-95},
Solimany and Kramer \cite{Solimany-SSC-95},
 Ibrahim {\it et al.}
\cite{Ibrahim-PRB-98}, Sim {\it et al.} \cite{Sim-PRB-98}, Dubonos
{\it et al.} \cite{Dubonos-PhysE-00}, Reijniers {\it et al.}
\cite{Reijniers-PhysE-00,Reijniers-PRB-01b}). The effect of
inhomogeneous magnetic field on the weak-localization corrections
to the classical conductivity of disordered 2DEG was considered by
Rammer and Shelankov \cite{Rammer-PRB-87}, Bending
\cite{Bending-PRB-94}, Bending {\it et al.}
\cite{Bending-PRL-90,Bending-PRB-90,Bending-PRB-92}, Mancoff {\it
et al.} \cite{Mancoff-PRB-95}, Shelankov \cite{Shelankov-PRB-00},
Wang \cite{Wang-PRB-02}.

\subsection{Planar S/F hybrids with ferromagnetic bubble domains: theory}
\label{Theory-planarSF-nucleation}

The aforementioned Ginzburg-Landau formalism can be applied in the
case of a non-uniform magnetic field generated by the domain
structure of a plain magnetic film. The problem of the OP
nucleation in planar S/F hybrid structures was theoretically
analyzed for hard ferromagnets characterized by an out-of-plane
magnetization ${\bf M}=M_z(x)\,{\bf z}_0$ by Aladyshkin {\it et
al.} \cite{Aladyshkin-PRB-03}, Buzdin and Melnikov
\cite{Buzdin-PRB-03}, Samokhin and Shirokoff
\cite{Samokhin-PRB-05}, Aladyshkin {\it et al.}
\cite{Aladyshkin-PRB-06}, Gillijns {\it et al.}
\cite{Gillijns-PRB-07a}. It is also worth mentioning the pioneer
paper of Pannetier {\it et al.} \cite{Pannetier-mohograph-95},
where the OP nucleation in a periodic sinusoidal magnetic field
generated by a meander-like lithographically-prepared metallic
coil was considered.


The distribution of the vector potential ${\bf a}({\bf r})$ can be
obtained, either by integration of the last term in the r.h.s. of
Eq.~(\ref{LinearizedGLEquation-2}) or by a direct consideration of
the magnetostatic problem. Provided that the width of the domain
walls $\delta$ is much smaller than other relevant length scales,
the field distributions can be calculated analytically for some
simple configurations (Aladyshkin {\it et al.}
\cite{Aladyshkin-PRB-06}, Sonin
\cite{Sonin-JTP-88,Sonin-PRB-02b}). Choosing the gauge
$A_y=x\,H_{ext}+a_y(x,z)$, one can easily see that the linearized
GL equation Eq. (\ref{LinearizedGLEquation-1}) does not depend on
the $y-$coordinate, hence we can generally find the solution in
the form $\psi(x,y,z)=f_k(x,z)\,\exp(i k y)$, where the function
$f_k(x,z)$ should be determined from the following 2D equation:
    \begin{eqnarray}
    \label{Eq:GL-x}
    - \frac{\partial^2 f_{k}}{\partial x^2} - \frac{\partial^2 f_{k}}{\partial z^2}
    + \left[\frac{2\pi}{\Phi_0}a_y(x,z)+\frac{2\pi}{\Phi_0} x H_{ext}
    -k\right]^2 f_{k} =
    \frac{1}{\xi^2}f_{k},
    \end{eqnarray}
If the superconducting film has insulating interfaces at the top
and bottom surfaces and ${\bf A}\cdot {\bf n}=0$ at the surface
$\partial V_s$ of the superconductor, one should apply the
standard boundary conditions: $\partial f_{k}/\partial
n\Big|_{\partial V_s}=0$ (here ${\bf n}$ is the normal vector).

\begin{figure*}[t!]
    \begin{center}
    \epsfxsize=80mm \epsfbox{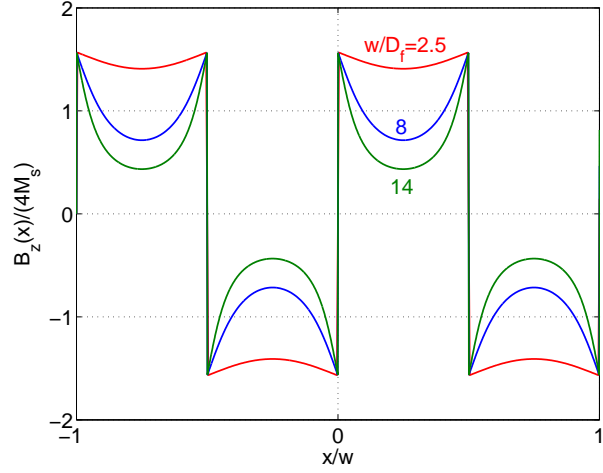}
    \end{center}
    \caption{(Color online) Transverse $z-$component of the magnetic
    field, induced by one-dimensional periodic distribution of
    magnetization with the amplitude $M_s$ and the period $w$,
    calculated at the distance $h\ll D_f$ above the ferromagnetic
    film of a thickness $D_f$, adapted from  Aladyshkin and Moshchalkov \cite{Aladyshkin-PRB-06}.}
    \label{Fig:FieldDistribution}
    \end{figure*}

The spatial distribution of the magnetic field, induced by the
periodic 1D domain structure, strongly depends on the relationship
between the width of the magnetic domains $w$ and the thickness of
the ferromagnetic film $D_f$ as well as on the distance $h$
between superconductor and ferromagnet (Fig.
\ref{Fig:FieldDistribution}). If the superconducting film
thickness $D_s$ is much smaller than the typical length scales of
the nonuniform magnetic field ($w$ and $D_f$), in a first
approximation one can neglect the OP variations in the
$z-$direction and omit the term ${\partial^2 f_{k}}/{\partial
z^2}$ in Eq.~(\ref{Eq:GL-x}). As a result, the OP nucleation in a
thin superconducting film is determined by the spatial profile of
the perpendicular magnetic field only, while the effect of the
parallel field can be ignored.

\vspace*{0.2cm}

    \begin{figure*}[b!]
    \begin{center}
    \epsfxsize=80mm \epsfbox{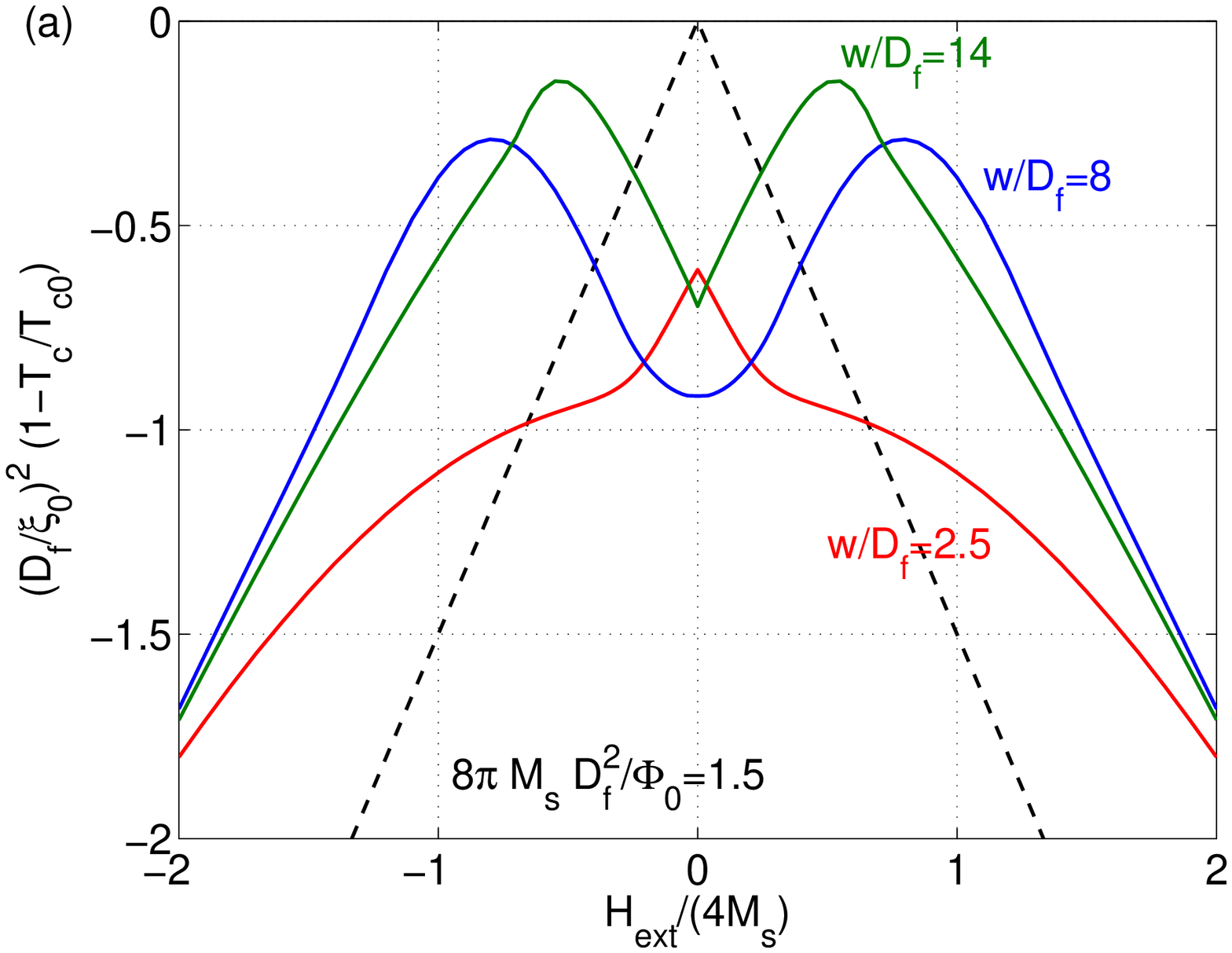}
    \epsfxsize=80mm \epsfbox{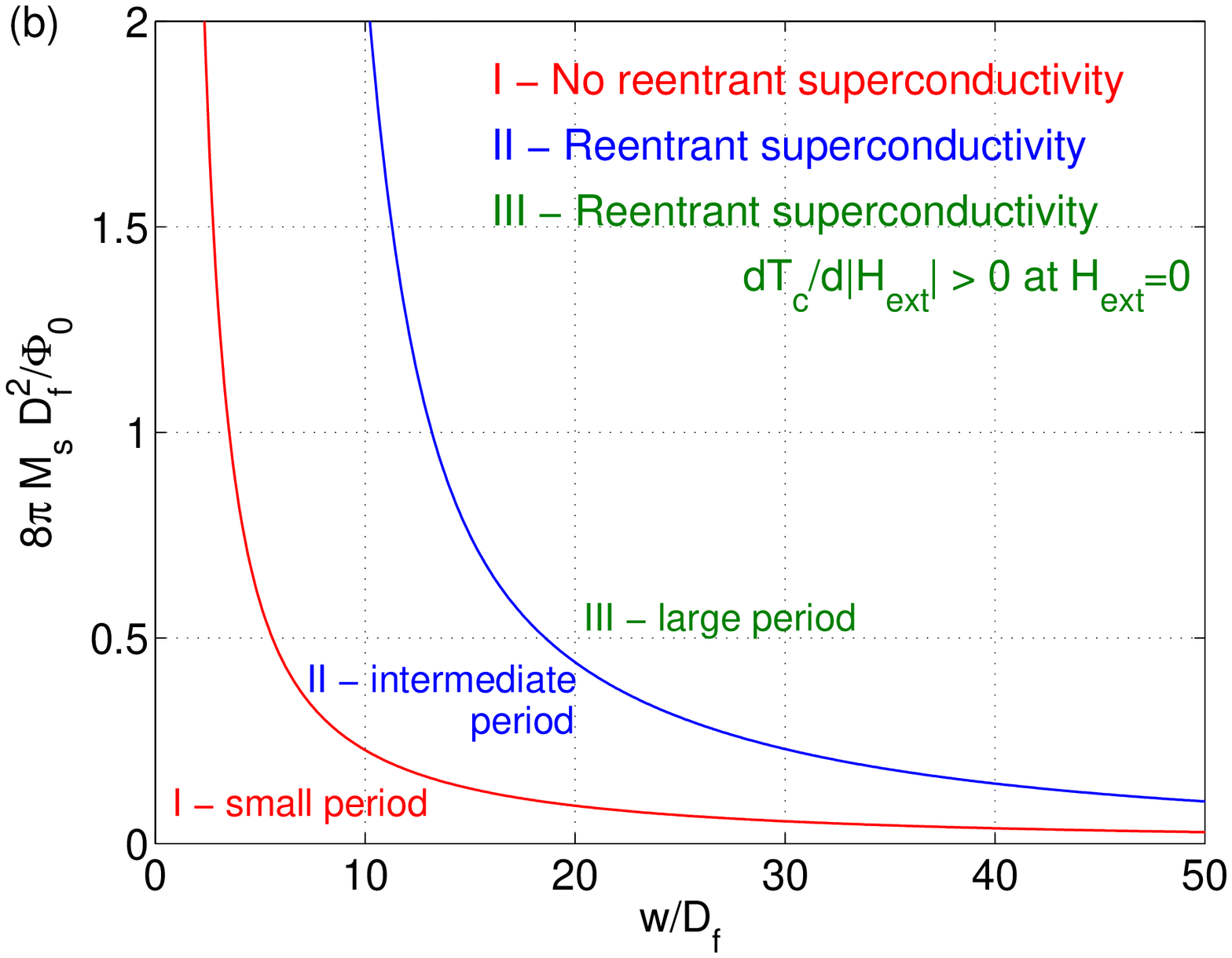}
    \end{center}
    \caption{(Color online) (a) Examples of the phase transition lines $T_c(H_{ext})$
    for a planar S/F structure containing a periodic 1D domain
    structure ($M_s$ is the saturated magnetization, $D_f$ is the ferromagnetic film thickness,
    $w$ is the period of the domain structure,
    the thickness of the superconducting film $D_s\ll (D_f,w)$ and
    the separation between superconducting and ferromagnetic films $h/D_f\ll 1$),
    adapted from  Aladyshkin and Moshchalkov \cite{Aladyshkin-PRB-06}.
    Black dashed line correspond to the $T_c(H_{ext})$ dependence in the absence of the
    non-uniform field.
    \newline
    (b) Different regimes of localized superconductivity in the presence of a 1D domain structure in the $M_s-w$ plane,
    obtained numerically for $D_s\ll (D_f,w)$ and $h/D_f\ll 1$,
    adapted from  Aladyshkin and Moshchalkov \cite{Aladyshkin-PRB-06}.
    In regions II and III the phase boundary $T_c(H_{ext})$ exhibits
    reentrant superconductivity. The slope $dT_c/d|H_{ext}|$ at
    $H_{ext}=0$ can be positive (III), zero (II) or negative (II, near the separating line I--II).
    Region I corresponds to the monotonic $T_c(H_{ext})$ dependence.}
    \label{Fig:Aladyshkin-PRB-2006}
    \end{figure*}

\noindent  {\it Criterium for the development of domain-wall
superconductivity}

\noindent In this section we will discuss the possibility of
localizing the OP wave function near a domain wall at zero
external field. Obviously, the regime of domain-wall
superconductivity (DWS) can be achieved only if the typical
$b^*_z$ value inside the magnetic domains is rather large in order
to provide an exponential decay of the OP, described by the
effective magnetic length $\ell^*_{b}=\sqrt{\Phi_0/2\pi |b^*_z|}$,
within a half-width of the domain: $\ell^*_{b}<w/2$. For thick
ferromagnetic films \mbox{($w/D_f\ll 1$)} the magnetic field
inside domains is almost uniform  and it can be estimated as
$b^*_z\simeq 2\pi M_s$, giving us the rough criterion of the
realization of the DWS regime and the critical temperature
$T_c^{(0)}$ at $H_{ext}=0$: $\pi^2 M_s w^2/\Phi_0>1$ and
$(T_{c0}-T_c^{(0)})/T_{c0}\simeq 2\pi M_s/H_{c2}^{(0)}$. Of
course, $2\pi M_s/H_{c2}^{(0)}$ should be less than unity
otherwise superconductivity will be totally suppressed.

By applying an external field on the order of the compensation
field, $H_{ext}\simeq 2\pi M_s$, one can get local compensation of
the field above the domains with opposite polarity and a doubling
of the field above the domains of the same polarity. Since
superconductivity is expected to form at regions with zero field
(which are $w$ wide), the maximal critical temperature can be
estimated as follows: $(T_{c0}-T_c^{max})/T_{c0}\simeq
\xi_0^2/w^2$ (a consequence of the quantum-size effect for Cooper
pairs in nonuniform magnetic field). Therefore, $T_c^{max}$ will
exceed $T_c^{(0)}$, pointing out to the non-monotonous
$T_c(H_{ext})$ dependence for the same $M_s$ and $w$ parameters,
which are necessary to have the DWS regime at $H_{ext}=0$. The
typical phase boundary $T_c(H_{ext})$, corresponding to the DWS
regime at $H_{ext}=0$ and shown in panel (a) in
Fig.~\ref{Fig:Aladyshkin-PRB-2006}, is characterized by the
presence of a pronounced reentrant behavior and the parabolic
dependence of $T_c$ on $H_{ext}$ at low fields (curve labelled
$w/D_f=8$). This type of phase boundary was predicted by Pannetier
{\it et al.} \cite{Pannetier-mohograph-95} for a superconducing
film in a field of parallel metallic wires carrying a dc current,
and by Buzdin and Melnikov \cite{Buzdin-PRB-03} and Aladyshkin
{\it et al.}~\cite{Aladyshkin-PRB-03,Aladyshkin-PRB-06} for planar
S/F hybrids.

\vspace*{0.2cm}

\noindent  {\it Localized superconductivity in S/F hybrids for
$w/D_f\ll 1$ and $w/D_f\gg 1$}

\noindent For S/F hybrids with smaller periods of the field
modulation (\mbox{$\pi^2 M_s w^2/\Phi_0\ll 1$}) the OP
distribution cannot follow the rapid field variations and, as a
consequence, at \mbox{$H_{ext}\simeq 0$} there is a broad OP wave
function, spreading over several domains and resulting in an
effective averaging of the nonuniform magnetic field. In this case
the critical temperature was shown to decrease monotonically with
increasing $|H_{ext}|$ [curve labelled $w/D_f=2.5$ in
Fig.~\ref{Fig:Aladyshkin-PRB-2006}(a)], similar to the case of
superconducting films in a uniform magnetic field. By applying an
external field, one can shrink the width of the OP wave function
and localize it within one half-period above the domains with
opposite magnetization. The interplay between both, the external
field and the periodic magnetic field, which determines the
resulting OP width, leads to a sign change of the second
derivative $d^2T_c/dH_{ext}^2$. At high $H_{ext}$ values the width
of the OP wave function, positioned at the center of the magnetic
domain, is determined by the local field $B_{loc}\simeq
|H_{ext}|-2\pi M_s$, therefore we come to a biased linear
dependence $1-T_c/T_{c0}\simeq \big||H_{ext}|-2\pi
M_s\big|/H_{c2}^{(0)}$. These qualitative arguments were supported
by numerical solutions of the linearized GL equation
\cite{Aladyshkin-PRB-06}.

The case $w/D_f\gg 1$ should be treated separately since the
$z-$component of the field inside the magnetic domains is very
inhomogeneous: the absolute value $|b_z(x)|$ reaches a minimum
$b^*_z=8\pi M_sD_f/w$ at the domain center, while the maximal
value is still equal to $2\pi M_s$ at the domain walls. It was
shown that the $|b_z(x)|$ minima are favorable for the OP
nucleation at $H_{ext}=0$. In this regime, the OP localization in
the center of the domains at $H_{ext}=0$ is possible as long as
$2\pi^2 M_s w D_f/\Phi_0>1$. At the same time the nucleation near
domain walls is suppressed by the mentioned field enhancement near
the domain walls. The sudden displacement of the localized OP wave
function between the centers of the domains of positive and
negative magnetization, when inverting the $H_{ext}$ polarity,
results in a new type of phase boundary $T_c(H_{ext})$ with a
singularity at $H_{ext}=0$ \cite{Aladyshkin-PRB-06}. It is
important to note that, for $w/D_f\gg 1$ and $H_{ext}=0$  the
critical temperature {\it increases} linearly with almost the same
slope $dT_c/d|H_{ext}|=T_{c0}/H_{c2}^{(0)}$  as the $T_c$ value
decreases in an applied uniform magnetic field [curve labelled
$w/D_f=14$ in Fig.~\ref{Fig:Aladyshkin-PRB-2006} (a)].

\subsection{Planar S/F hybrids with ferromagnetic bubble domains: experiments}
\label{Experiment-planarSF-nucleation}

\vspace*{0.2cm}

\noindent  {\it OP nucleation in perpendicular magnetic field}

\noindent To the best of our knowledge, the first observation of
reentrant superconductivity\footnote[1]{
    The experimental observation of the influence of a periodic magnetic field, generated by an
    array of parallel wires with current $I$ flowing alternatively in
    opposite directions, on the properties of an Al superconducting bridge was
    reported by Pannetier {\it et al.} \cite{Pannetier-mohograph-95}. Since the ${\rm max} |b_z|\propto I$,
    reentrant superconductivity can be realized for rather high $I$ values, as it was shown experimentally.
    It should be noted that already in sixties Artley {\it et al.} \cite{Artley-APL-66} experimentally studied
    the effect of the domain walls in a thin permalloy film on the
    superconducting transition of a thin indium film.}
in planar S/F hybrids was reported by Yang {\it et al.}
\cite{Yang-Nature-04} who measured the electrical resistance of a
superconducting Nb film grown on top of a ferromagnetic
BaFe$_{12}$O$_{19}$ substrate characterized by an out-of-plane
magnetization. Later on, the same system Nb/BaFe$_{12}$O$_{19}$
was examined by Yang {\it et al.} in Ref. \cite{Yang-PRB-06}. From
the parameters typical for the domain structure in
BaFe$_{12}$O$_{19}$ single crystals and Nb films (\mbox{$M_s\simeq
10^2~$Oe}, \mbox{$w\simeq 2~\mu$m}, \mbox{$D_f\simeq 90~\mu$m},
$H_{c2}^{(0)}\simeq 30~$kOe), the following estimates can be
obtained: $w/D_f\simeq 0.02$, $\pi^2M_s w^2/\Phi_0>10^2$ and $2\pi
M_s/H_{c2}^{(0)}\simeq 0.02$. Therefore, such a ferromagnet is
suitable for the realization of the DWS regime at $H_{ext}=0$. The
appearance of these localized superconducting paths guided by
domain walls was shown to result in a broadening of the
superconducting resistive transition at low magnetic fields. As
the field $H_{ext}$ is ramped up, the superconducting areas shift
away from the domain walls towards the wider regions above the
domains with an opposite polarity (so-called reversed-domain
superconductivity, RDS) where the absolute value of the total
magnetic field is minimal because of the compensation effect. As a
consequence, the superconducting critical temperature $T_c$
increased with increasing $|H_{ext}|$ up to 5 kOe. Once the
external field exceeds the saturation field $H_s$ of the
ferromagnet ($H_s\simeq 5.5$~kOe at low temperatures), the domain
structure in the ferromagnet disappears and the phase boundary
abruptly returns back to the standard linear dependence
$(1-T_c/T_{co})\simeq |H_{ext}|/H_{c2}^{(0)}$. Since the width and
the shape of the magnetic domains continuously depend on the
external field, the theory developed in section
\ref{GeneralConsideration} is not directly applicable for the
description of the experiment, although it qualitatively explains
the main features of the OP nucleation in such S/F systems.

Substituting Nb by a superconductor with a smaller $H_{c2}^{(0)}$
value (e.g., Pb with $H_{c2}^{(0)}\simeq 1.8$~kOe) allows one to
study the effect of the superconducting coherence length
$\xi_0=\sqrt{\Phi_0/2\pi H_{c2}^{(0)}}$ on the localization of the
OP. It was shown by Yang {\it et al.} \cite{Yang-APL-06}  that the
increase of the $M_s/H_{c2}^{(0)}$ ratio suppresses the critical
temperature of the formation of domain-wall superconductivity at
zero external field\footnote[2]{In section
    \ref{Theory-planarSF-nucleation}
    we argued that the critical temperature, $T_c^{(0)}$, at zero external field
    is proportional to $1-2\pi M_s/H_{c2}^{(0)}$.},
therefore superconductivity in Pb/BaFe$_{12}$O$_{19}$ hybrids
appeared only near the compensation fields above the reversed
domains.

    \begin{figure}[t!]
    \begin{center}
    \epsfysize=65mm \epsfbox{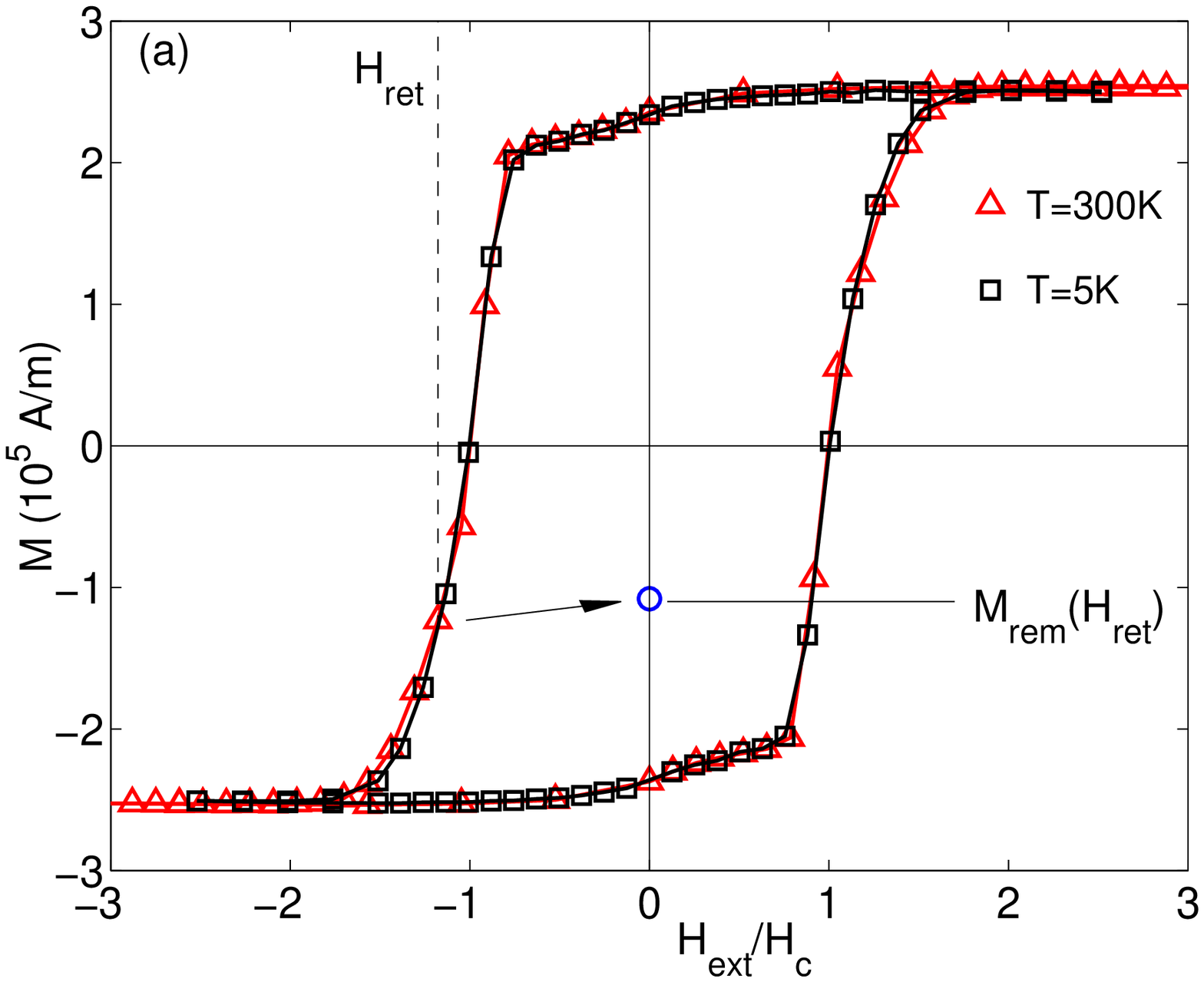}
    \epsfysize=65mm \epsfbox{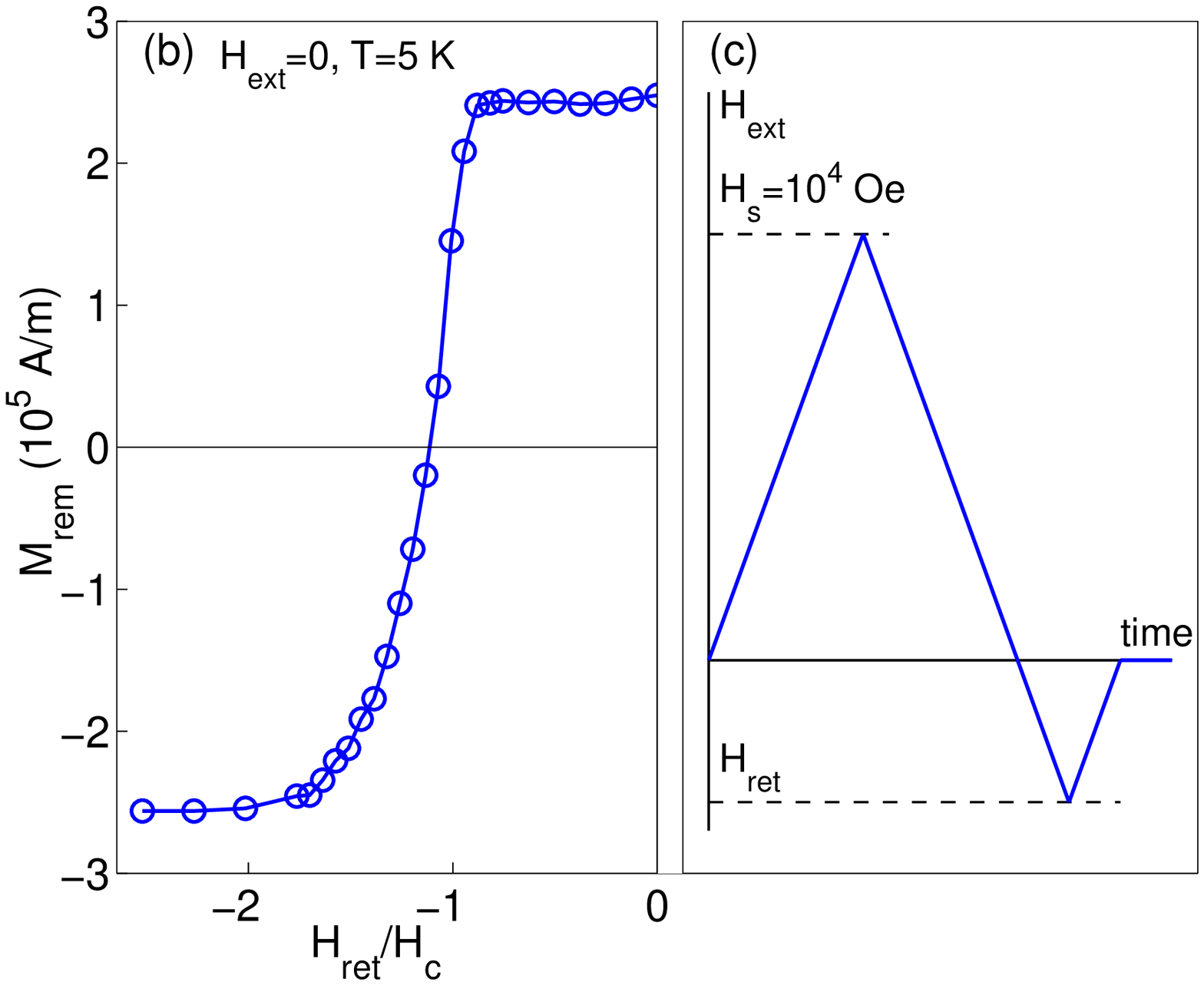}
    \end{center}
    \caption{(color online) Preparation of the magnetic state in a ferromagnetic Co/Pt film
    with a desirable remanent magnetization $M_{rem}$. \newline
    (a) Magnetization loops
    $M(H_{ext})$ at 300 K (triangles) and 5 K (squares), the magnetic
    field axis is normalized by the corresponding coercive fields $H_c$, adapted from  Gillijns {\it et al.}
    \cite{Gillijns-PRB-07a}; \newline
    (b) Remanent magnetization $M_{rem}$, measured at 5 K and $H_{ext}=0$ after
    saturation in positive fields (up to 10$^4$ Oe) and subsequent
    application of a returning field $H_{ret}$ [this procedure is
    shown schematically in panel (c)], adapted from  Gillijns {\it et al.}
    \cite{Gillijns-PRB-07a}.}
    \label{Fig:Gillijns-PRB-07-abc}
    \end{figure}

Direct visualization of localized superconductivity in
Nb/PbFe$_{12}$O$_{19}$ structures was performed by Fritzsche {\it
et al.} \cite{Fritzsche-PRL-06}. The basic idea of this technique
is the following: if the sample temperature becomes close to a
local critical temperature at a certain position $(x,y)$, then a
laser pulse, focused on that point, can induce the local
destruction of superconductivity due to heating. The observed
increase of the global resistance $R$ of the superconducting
bridge can be associated with the derivative $dR(x,y)/dT$. By
varying the temperature and scanning the laser beam over the Nb
bridge under investigation, it is possible to image the areas with
different critical temperatures. For example, it allows one to
attribute the formation of well-defined regions with rather high
local critical temperatures above magnetic domains at the
compensation field with the appearance of reversed-domain
superconductivity.

The effect of the amplitude of the field modulation on the OP
nucleation was considered by Gillijns {\it et al.}
\cite{Gillijns-PRL-05,Gillijns-PhysC-06} on thin-film trilayered
hybrid F/S/F structures. In contrast to the BaFe$_{12}$O$_{19}$
single crystal discussed above, the multilayered Co/Pd films are
characterized by a high residual out-of-plane magnetization,
$M_s\sim 10^2$ Oe, almost independent on the external field at
$|H_{ext}|<H_{coer}$, where \mbox{$H_{coer}\simeq 10^3$~Oe} is the
typical coercive field at low temperatures. The use of two
ferromagnetic films with slightly different coercive fields
allowed them to prepare different magnetic configurations and thus
to control the amplitude of the nonuniform field inside the
superconductor due to the superposition of the partial stray
fields via an appropriate demagnetizing procedure. The effective
doubling of the amplitude of the internal field for a
configuration with two demagnetized ferromagnetic films
(containing bubble domains) leads to a broadening of the
temperature interval, where the $T_c(H_{ext})$ line demonstrates
the non-monotonous behavior. In other words, the critical
temperature of domain-wall superconductivity at $H_{ext}=0$
expectedly decreases as the effective magnetization increases. In
addition, the enhancement of the internal field results in a shift
of the $T_c$ maxima to higher $H_{ext}$ values, what corresponds
to reversed-domain superconductivity.

\begin{figure}[t!]
    \begin{center}
    \epsfxsize=80mm \epsfbox{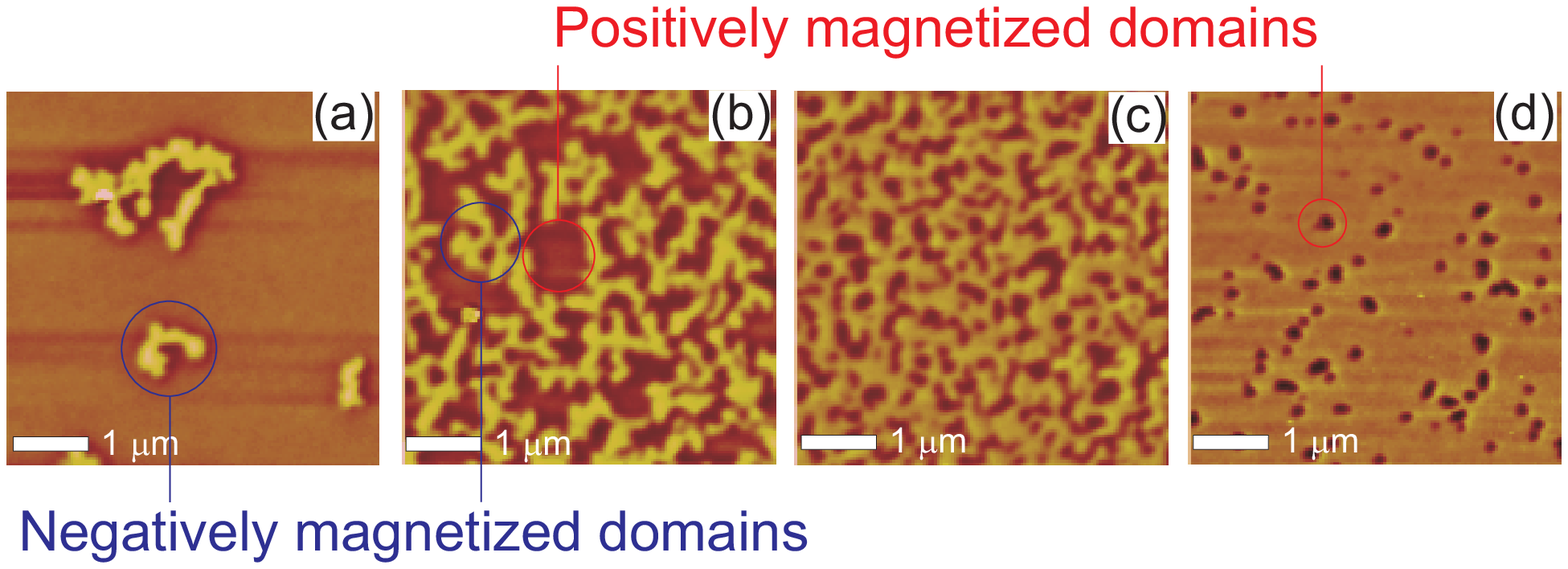}
    \epsfxsize=80mm \epsfbox{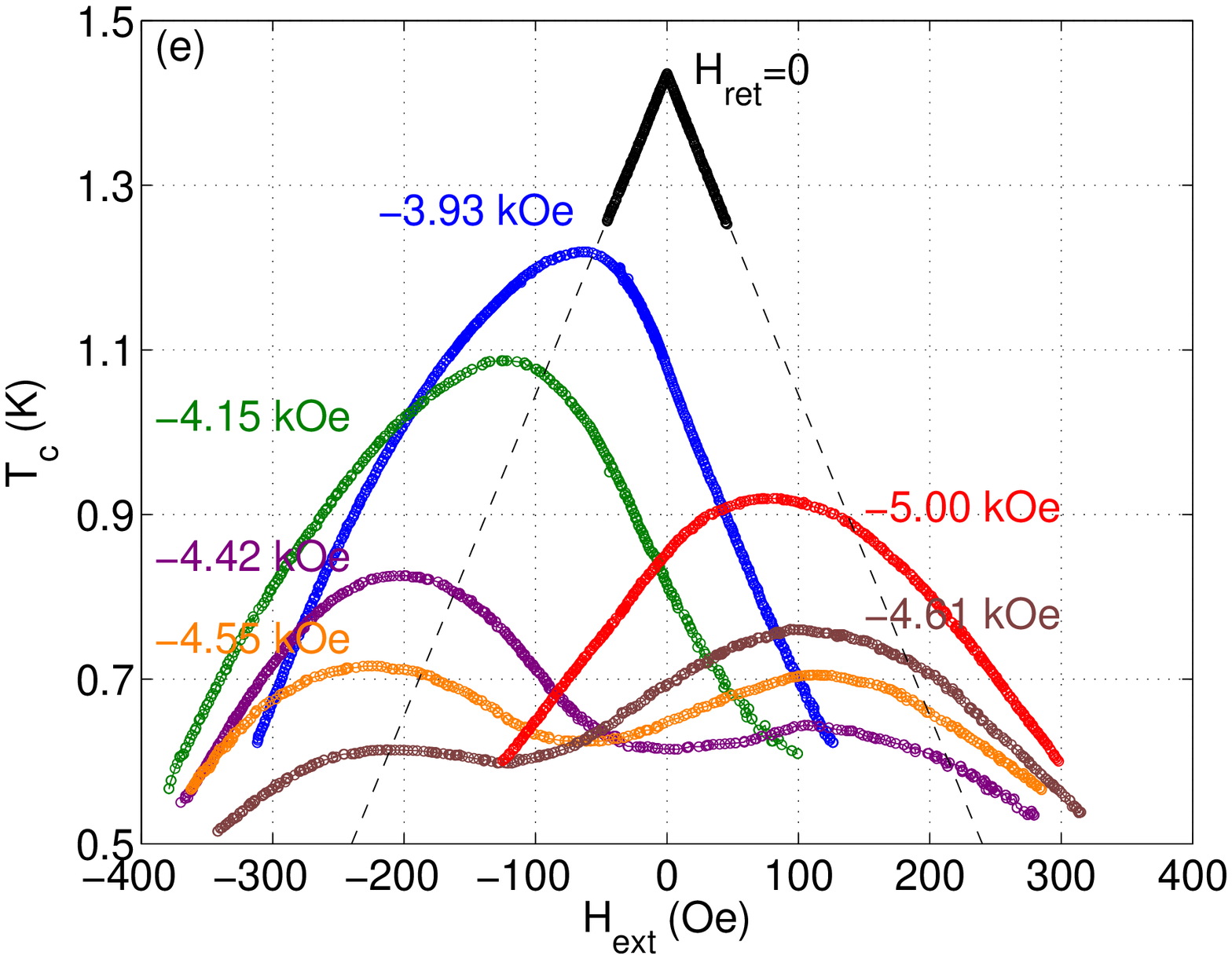} \caption{(color
    online) (a)--(d) MFM images obtained at $T=300$ K for $H_{ret}$ values equal to
    -1.75 kOe (a), -2.00 kOe (b), -2.50 kOe (c), -3.00 kOe (d), the coercive field $H^{300 K}_c=1.91$
    kOe, adapted from
    Gillijns {\it et al.} \cite{Gillijns-PRB-07a}.
    The dark (bright) color
    represents domains with positive (negative) magnetization.
    \newline (e) A set of experimental phase boundaries $T_c(H_{ext})$
    obtained for the same bilayered S/F sample (a superconducting Al
    film  on top of a Co/Pt multilayer) in various magnetic states
    measured after the procedure of an incomplete
    demagnetization: $H_{ext}=0\Rightarrow$ $H_{ext}= 10$ kOe
    $\Rightarrow$ $H_{ext}=H_{ret}\Rightarrow$ $H_{ext}=0$ for
    various returning fields $H_{ret}$ indicated on the diagram, the coercive field $H^{5 K}_c=3.97$
    kOe, adapted from
    Gillijns {\it et al.} \cite{Gillijns-PRB-07a}.}
    \label{Fig-Gillijns-PRB-2007-d}
    \end{center}
    \end{figure}

The spatial extension where the field compensation takes place, is
a crucial parameter defining the nucleation of superconductivity:
an OP trapped in a broader region results in a higher $T_{c}$
value and vice versa (Gillijns {\it et al.}
\cite{Gillijns-PRB-07a}, Aladyshkin {\it et al.}
\cite{Aladyshkin-PhysC-08}). The magnetic state of the ferromagnet
can be reversibly changed after the following procedure of an
incomplete demagnetization : $H_{ext}=0\Rightarrow$ $H_{ext}=H_s$
$\Rightarrow H_{ext}=H_{ret} \Rightarrow H_{ext}=0$, where $H_s$
is the saturation field (see Fig. \ref{Fig:Gillijns-PRB-07-abc}).
As a result, one can obtain any desirable remanent magnetization
\mbox{$-M_s<M_{rem}<M_s$} (as well as any average width of the
magnetic domains) by varying the $H_{ret}$ value. At $H_{ret}<0$
the formation of the negative domains decreases the average width
of the positive domains and it causes a drastic lowering of the
height of the $T_c$ peak, positioned at negative fields and
attributed to the appearance of superconductivity above large
positive domains [curves $H_{ret}=-3.93$ kOe and $H_{ret}=-4.15$
kOe in Fig. \ref{Fig-Gillijns-PRB-2007-d}]. This observation is a
direct consequence of the increase of the ground energy of the
``particle-in-a-box" with decreasing width of the box. When
\mbox{$M_{rem}$} is close to zero, thus indicating the presence of
an equal distribution of positive and negative domains, a nearly
symmetric phase boundary with two maxima of the same amplitude is
recovered [curve $H_{ret}=-4.55$ kOe in Fig.
\ref{Fig-Gillijns-PRB-2007-d}]. For higher $H_{ret}$ values, the
first peak, located at negative fields, disappears, whereas the
peak at positive fields shifts up in temperature and is displaced
to lower magnetic field values [curves $H_{ret}=-4.61$ kOe and
$H_{ret}=-5.00$ kOe in Fig. \ref{Fig-Gillijns-PRB-2007-d}]. This
second peak eventually evolves into a linear phase boundary when
the ferromagnetic film is fully magnetized in the negative
direction.

\vspace*{0.2cm}

\noindent {\it OP nucleation in parallel magnetic field}

\noindent We would like to mention a few related papers devoted to
the nucleation of superconductivity in various planar S/F
structures, where superconducting and ferromagnetic layers were
not electrical insulated and thus an effect of exchange
interaction can not be excluded.

An appearance of domain walls in permalloy film leading to dips in
a field dependence of the resistivity of a superconducting Nb film
was observed by Rusanov {\it et al.} \cite{Rusanov-PRL-04}. The
position of these resistivity minima were found to be dependent on
the sweep direction of the in-plane oriented external field. An
opposite effect (maxima of resistivity at temperatures below
$T_{c0}$) was observed by Bell {\it et al.} \cite{Bell-PRB-06} for
thin-film amorphous S/F structures consisting of superconducting
MoGe films and ferromagnetic GdNi layers. It was interpreted as
flow of weakly pinned vortices induced by the stray field of
magnetic domains in the ferromagnetic layers. Zhu {\it et al.}
\cite{Zhu-PRL-08} demonstrated that the domain structure in
multilayered CoPt films can be modified by applying in-plane
external field during the deposition process. This deposition
field does not change the overall perpendicular magnetic
anisotropy of the Co/Pt films but it induces a weak an in-plane
magnetic anisotropy and eventually alters the domain patterns.
Indeed, after demagnetizing with an in-plane ac magnetic field
oriented along the deposition field direction, one can prepare the
domain structure in a form of largely parallel stripe domains. In
contrast to that, the same multilayered structure fabricated at
zero field or demagnetized with out-of-plane ac field exhibits a
nearly random labyrinth-type domain pattern. Sweeping the external
field $H^{\parallel}_{ext}$, one can control the arrangement of
domain walls, and drive the S/F hybrid from normal to
superconducting state for the same temperature and magnetic field.

\subsection{S/F hybrids with 2D periodic magnetic field: theory and experiments}

In this section we continue the discussion concerning the
properties of large-area S/F hybrids in which the nonuniform
magnetic field is created by regular arrays of ferromagnetic dots.
The fabrication of such magnetic structures makes it possible to
achieve full control of the spatial characteristics of the
nonuniform magnetic field (both the topology and the period),
which can be eventually designed practically at will. One can
expect that due to the field-compensation effect the inhomogeneous
magnetic field modulated in both directions will affect the OP
nucleation in the same way as for the planar S/F structures.
However, the 2D periodicity of the magnetic field naturally leads
to the appearance of well-defined commensurability effects for
such hybrid systems. In other words, a resonant change in the
thermodynamical and transport properties of the superconducting
films appears as a function of the external magnetic field
$H_{ext}$, similar to that observed for superconductors with
periodic spatial modulation of their properties (e.g., perforated
superconducting films \cite{Baert-PRL-95,Moshchalkov-PRB-98}).
These matching phenomena take place at particular $H_{ext,n}$
values that can be used as indicators allowing us to find a
relationship between the most probable microscopic arrangement of
the vortices in the periodic potential and the global
characteristics of the considered S/F hybrids measured in the
experiments.

\vspace*{0.3cm}

\noindent  {\it Commensurate solutions of the GL equations.}

\noindent The periodic solutions of the GL equations in the
presence of a nonuniform 2D periodic magnetic field can be
constructed by considering one or more magnetic unit cells (of
total area $S$) and applying the following boundary conditions:
    \begin{eqnarray}
    \label{Gauge-Transformation}
    {\bf A}({\bf r}+{\bf b}_k)={\bf A}({\bf r})+\nabla\eta_k({\bf r}),
    \\\nonumber
    \Psi({\bf r}+{\bf b}_k)=\Psi({\bf r})\,\exp\left(2\pi {\rm i} \eta_k({\bf
    r})/\Phi_0\right),
    \end{eqnarray}
where ${\bf b}_k$, $k=\{x,y\}$ are the lattice vectors, and
$\eta_k$ is the gauge potential (Doria {\it et al.}
\cite{Doria-PRB-89}). The gauge transformation Eq.
(\ref{Gauge-Transformation}) is possible provided that the flux
induced by the external magnetic field $H_{ext}S$ through the
chosen area $S$ is equal to an integer number of flux quanta,
$n\Phi_0$, which gives us the matching fields $H^{ext}_n=\pm
n\Phi_0/S$.

    \begin{figure}[t!]
    \begin{center}
    \epsfxsize=120mm \epsfbox{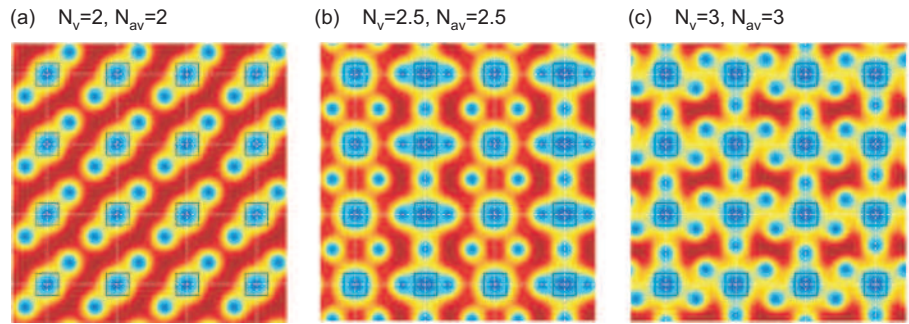} \caption{(color
    online) Contour-plots of the Cooper pair density
    $|\psi|^2$ of stable vortex phases in superconducting thin
    films in the presence of a magnetic dot array with dipolar moments oriented
    perpendicular to the film plane, adapted from
    Priour and Fertig \cite{Priour-PRL-04}. The images in (a), (b), and (c) correspond
    to magnetic dot strengths of 3.61, 4.33, and 4.95 fundamental flux units,
    respectively (the dot's dipole moment is specified by the positive flux
    passing through each unit cell). The Cooper pair density is highest
    in red and lowest in blue regions. Antivortices appear as small
    blue spots between the magnetic dots depicted by black squares.
    The number of vortices $N_v$ and antivortices $N_{av}$ per unit cell are indicated on the
    plots. The calculations were
    performed for a large 4$\times$4 supercell with periodic
    boundary conditions and for the following parameters:
    $\kappa=\lambda/\xi=4$, the period of the magnetic dot array
    is 6.25 $\xi$, the lateral size of the magnetic dot is
    2$\,\xi\times 2\,\xi$.}
    \label{Figure-Priour-PRL-04}
    \end{center}
    \end{figure}

The formation of different vortex patterns in S/F hybrids,
containing square arrays of the magnetic dots with perpendicular
magnetization, at $H_{ext}=0$ were studied by Priour and Fertig
\cite{Priour-PRL-04,Priour-PhysC-04} and Milo\v{s}evi\'{c} and
Peeters \cite{Milosevic-PRL-04,Milosevic-JLTP-05}. Since the total
flux through an underlying superconducting film, infinite in the
lateral direction, equals to zero, vortices cannot nucleate
without corresponding antivortices, keeping the total vorticity
zero. It was shown that the number of the vortex-antivortex pairs
depends on the dipolar moment of the magnetic dot. The equilibrium
vortex phase, corresponding to the minimum of the free energy
functional, can exhibit a lower symmetry than the symmetry of the
nonuniform magnetic field. In order to get such vortex states of
reduced symmetry, the GL equations with periodic boundary
conditions should be considered in a large supercell ($2 \times
2$, $4 \times 4$ etc.). In most cases vortices are confined to the
dot regions, while the antivortices, depending on the dot's
magnetization, can form a rich variety of regular lattice states
with broken orientational and mirror symmetries
(Fig.~\ref{Figure-Priour-PRL-04}). The creation of
vortex-antivortex pairs in case of ferromagnetic dots with
in-plane magnetization was considered by Milo\v{s}evi\'{c} and
Peeters \cite{Milosevic-PhysC-06}. As expected, vortex-antivortex
pairs appear under the poles of each magnet according to their
specific inhomogeneous magnetic field keeping the total flux
through the superconducting film equals to zero.

The influence of the external field on the formation of
symmetrical and asymmetrical commensurate vortex configurations in
thin superconducting films in the presence of a square array of
out-of-plane magnetized dots was investigated by Milo\v{s}evi\'{c}
and Peeters
\cite{Milosevic-PRL-04,Milosevic-JLTP-05,Milosevic-PhysC-04a,Milosevic-PRL-05a}.
The simulations were carried out only for some discrete values
$H_{ext,n}$ of the external field, corresponding to the magnetic
flux quantization per magnetic supercell of area $S$. Vortices
were shown to be attracted by the magnetic dots in the parallel
case (at $H_{ext,n}>0$ for $M_z>0$) and repelled in the
antiparallel case (at $H_{ext,n}<0$ for $M_z>0$). In the parallel
case the vortex configurations for the integer matching fields are
similar to that for the vortex pinning by regular arrays of
antidots.

The temperature dependence of the magnetization threshold for the
creation of vortex-antivortex pairs was considered in Ref.
\cite{Milosevic-PRL-05a}. It was noted that the system will not
necessarily relax to the ground state, if there are metastable
states, corresponding to local minima in the free energy. As long
as the given vortex state is still stable and it is separated from
other stable vortex configurations by a finite energy barrier
(analogous to the Bean-Livingston barrier for the vortex entry
into superconducting samples), then the vorticity remains the same
even when changing temperature. However, an increase in
temperature resulting in a decrease of the height of the energy
barriers and strengthening of the thermal fluctuations, can
eventually cause a phase transition between the vortex states with
different number of vortices. The modification of the ground state
(at $H_{ext}=0$) by the creation of extra vortex-antivortex pairs,
when changing temperature and/or increasing the $M_s$ value
manifests itself as cusps in the phase boundary separating
superconductor from the normal metal phase in the $M_s-T$ diagram,
similar to the Little-Parks oscillations in the $T_c(H_{ext})$
dependence \cite{Little-PRL-62,Parks-PR-64}.

\vspace*{0.2cm}

\noindent  {\it Oscillatory nature of the phase transition line
(in-plane dot's magnetization)}

    \begin{figure}[t!]
    \begin{center}
    \epsfxsize=80mm \epsfbox{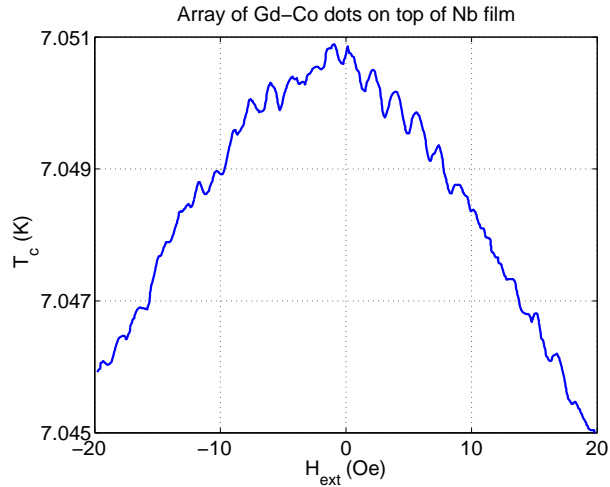} \caption{(color
    online) Phase transition line of a superconducting Nb film with
    a ferromagnetic Gd$_{33}$Co$_{67}$ particle array (the period
    4$\mu$m), adapted from Otani {\it et al.} \cite{Otani-JMMM-93}.}
    \end{center}
    \end{figure}

\noindent The influence of two-dimensional square arrays of
micron-sized, in-plane magnetized particles (SmCo, GdCo, FeNi) on
the electrical resistance of a superconducting Nb film, usually
interpreted as field-induced variations of the critical
temperature, were experimentally studied by Pannetier {\it et al.}
\cite{Pannetier-mohograph-95}, Otani {\it et al.}
\cite{Otani-JMMM-93}, Geoffroy {\it et al.}
\cite{Geoffroy-JMMM-93}. The oscillatory dependence of the
resistivity $\rho$ on the perpendicularly oriented external field
$H_{ext}$ with a period $\Delta H_{ext}$ close to $\Phi_0/a^2$ was
observed at $T<T_{c0}$ only when the dots had been magnetized
before cooling (here $a$ is the period of the magnetic dot array).
The appearance of minima in the $\rho(H_{ext})$ dependence, which
are reminiscent of the Little-Parks oscillations for
multiply-connected superconductors and superconducting networks
(see the monograph of Berger and Rubinstein \cite{Berger-book} and
references therein), were attributed to the variation of the
critical temperature, $T_c=T_c(H_{ext})$ due to the fluxoid
quantization in each magnetic unit cell.

\vspace*{0.3cm}

\noindent  {\it Tunable field-induced superconductivity for arrays
of out-of-plane magnetized dots}

\begin{figure}[t!]
    \begin{center}
    \epsfxsize=80mm \epsfbox{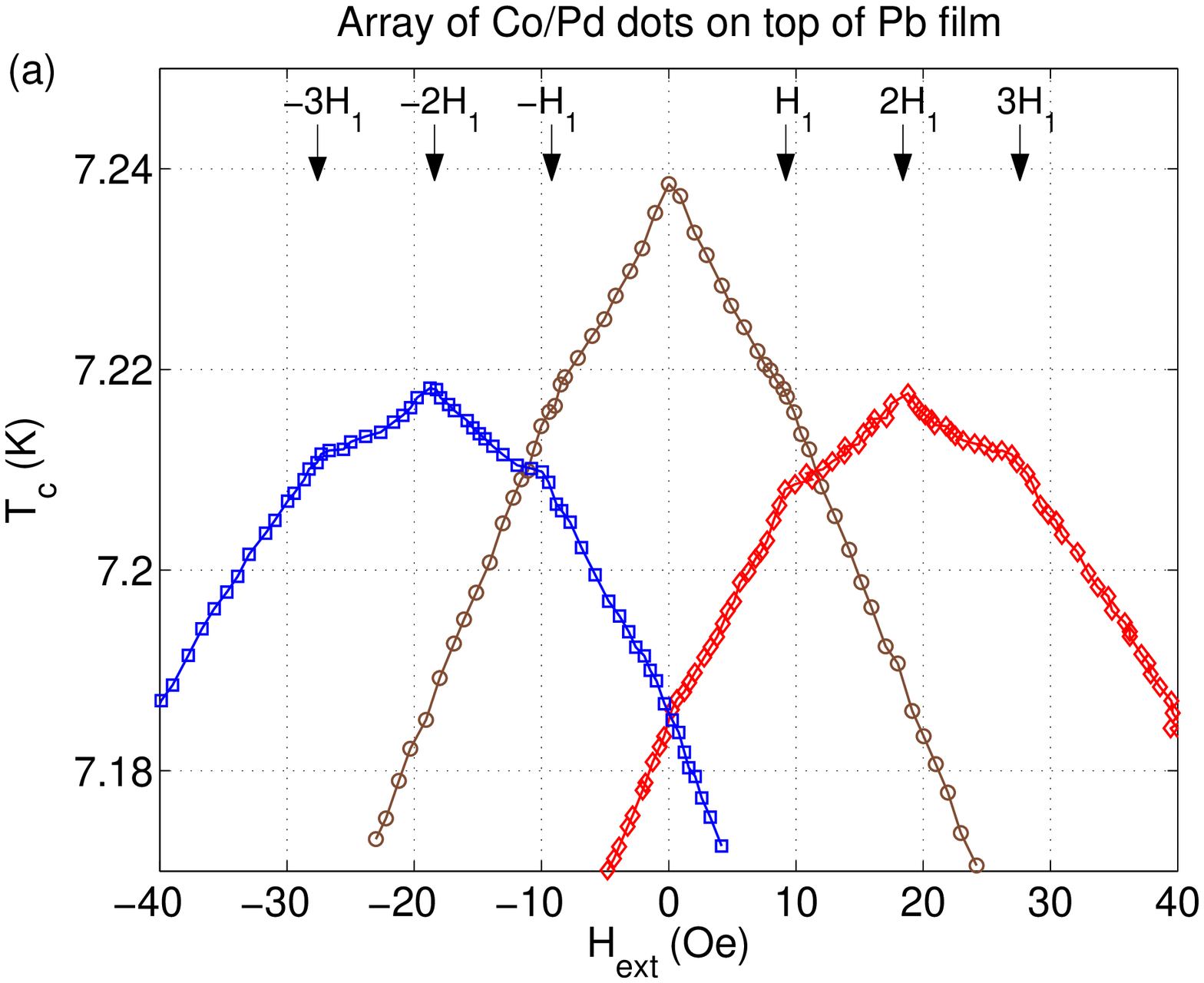} \epsfxsize=80mm
    \epsfbox{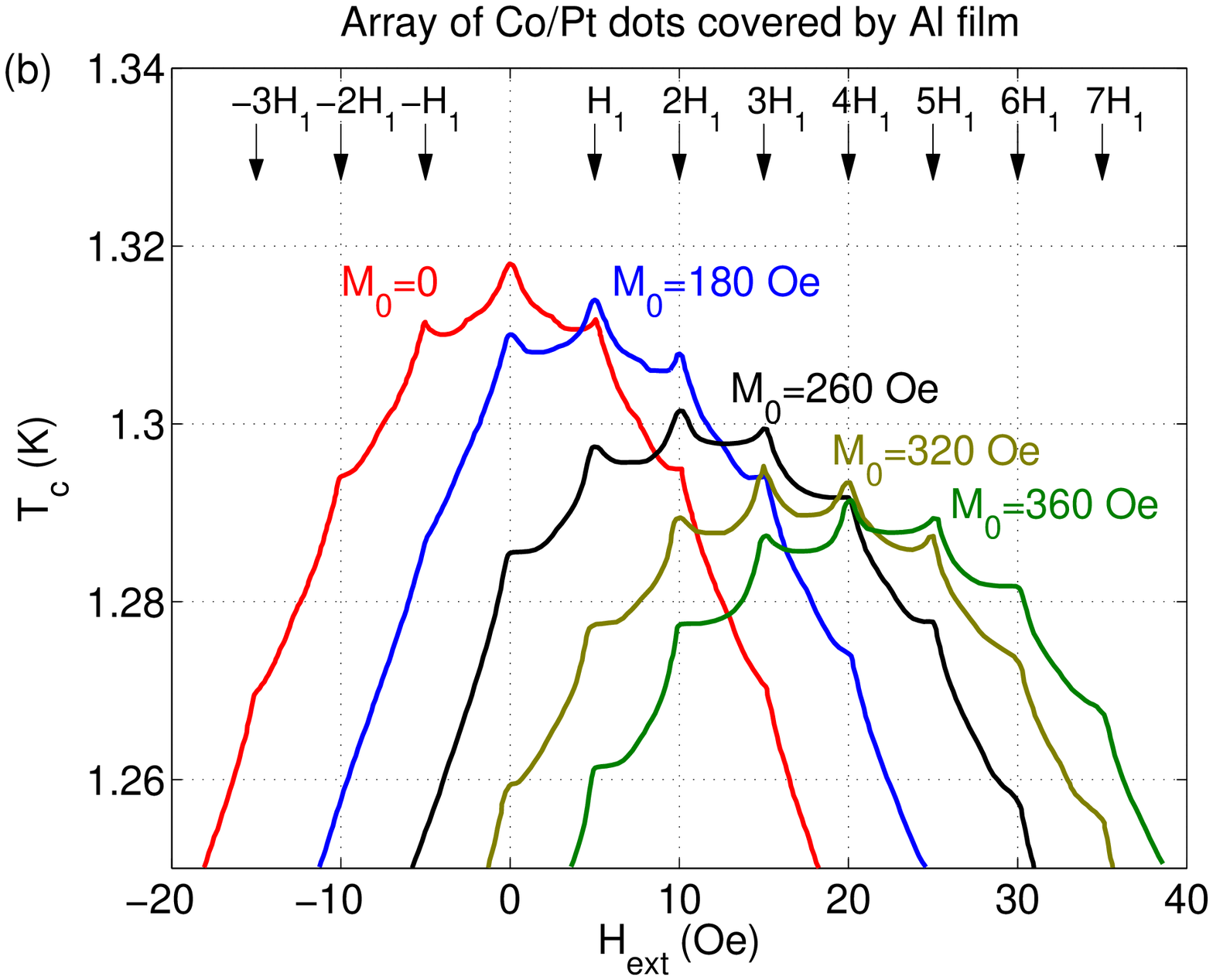}
    \caption{(color online)
    Field-induced superconductivity in a superconducting film with an
    array of magnetic dots: \newline
    (a) The $T_c(H_{ext})$ dependences
    obtained for a Pb film after demagnetization of the ferromagnetic
    dot array (the brown central curve marked by circles), saturation of the
    dots in a large positive $H_{ext}$ (the red right curve marked by
    diamonds), and saturation in a large negative $H_{ext}$ (the
    blue left
    curve marked by squares), adapted from   Lange {\it et al.}
    \cite{Lange-PRL-03}. The period of the lattice was 1.5 $\mu$m.
    The arrows depict the corresponding matching
    fields. \newline
    (b)
    Superconducting transition $T_c(H)$ of an Al film for different
    magnetic states of the square array of the ferromagnetic dots of
    the period 2 $\mu$m, adapted from Gillijns {\it et al.}
    \cite{Gillijns-PRB-06}. By increasing the magnetization a clear
    shift of $T_c(H_{ext})$ and a decrease of $T_c^{max}$ is observed.
    } \label{Fig-QuantizedDisplacement}
    \end{center}
    \end{figure}

\noindent The fabrication of periodic arrays of multilayered Co/Pd
and Co/Pt magnetic dots with out-of-plane magnetic moments allows
one to observe both the matching phenomena and the modified OP
nucleation due to the field-compensation effect. The stray field
induced by positively magnetized dots has a positive $z-$component
of the magnetic field under the dots and a negative one in the
area in between the dots. Thus, the magnetic field in the region
in between the dots will be effectively compensated at
$H_{ext}>0$, stimulating the appearance of superconductivity at
nonzero $H_{ext}$ values (magnetic field-induced
superconductivity). Lange {\it et al.} \cite{Lange-PRL-03}
demonstrated that the $T_c$ maximum is located at $H_{ext}=0$ for
the demagnetized magnetic dots and it is shifted towards a certain
$H_{ext,n}$ which depends on the dot's magnetization [see the
panel (a) in Fig.~\ref{Fig-QuantizedDisplacement}]. This quantized
shift of the $T_c$ was attributed to the field compensation in the
interdot areas accompanied by an annihilation of the interstitial
antivortices under the action of the external field, since (i) the
number of antivortices is determined by the magnetic moment of the
dots; (ii) the interstitial antivortices can be fully cancelled
only at the matching fields $H_{ext,n}=n\Phi_0/a^2$
(Milo\v{s}evi\'{c} and Peeters \cite{Milosevic-EPL-05}). Thus, the
appearance of periodic kinks in the $T_c(H_{ext})$ phase boundary
with a period coinciding with the first matching field $H_1$ can
be associated with the fluxoid quantization, confirming that
superconductivity indeed nucleates in multiply connected regions
of the film.

The results of further investigations on similar hybrid systems
(an array of micron-sized Co/Pt dots on top of an Al film) was
presented by Gillijns {\it et al.}
\cite{Gillijns-PRB-06,Gillijns-PRB-07b,Gillijns-PhysC-08}. As a
consequence of the rather large diameter of the dots, the
demagnetized dot's state microscopically corresponds to a magnetic
multidomain state with very weak stray field.  As it was
demonstrated in Ref. \cite{Gillijns-PRB-06}, the remanent
magnetization of the dots, which were initially demagnetized,
depends monotonously on the maximal applied field (excursion
field) $H_{ret}$. Thus the total remanent magnetic moment of the
dot becomes variable and tunable, hereby changing the influence of
the ferromagnet on the superconductor in a continuous way. It was
found that a gradual increase of the dot's magnetization from zero
to a certain saturated value results (i) in a quantized
displacement of the main $T_c$ maximum towards $nH_1$ ($n$ is
integer) due to the quantized character of the field-induced
superconductivity [the panel (b) in
Fig.~\ref{Fig-QuantizedDisplacement}]; (ii) in an enhancement of
the local $T_c$ maxima, attributed to the formation of a
commensurate vortex phase at discrete matching fields, which
becomes more pronounced as compared to Ref.~\cite{Lange-PRL-03}.

The effect of changing the average remanent magnetization
$M_{rem}$ and the radius $R_f$ of the magnetic dots on the
superconducting properties of an Al film deposited on top of a
periodic array of such dots was studied by Gillijns {\it et al.}
\cite{Gillijns-PRB-07b,Gillijns-PhysC-08}. Indeed, once the dot's
magnetization becomes saturated, the only way to further increase
the magnetic flux from each magnet can be achieved by increasing
the lateral size of the dots. It was experimentally found that the
larger the $R_f$ value, the smaller the necessary $\Delta M$
needed to shift the main $T_c(H_{ext})$ maximum by one matching
field.

\begin{figure}[t!]
    \begin{center}
    \epsfxsize=80mm \epsfbox{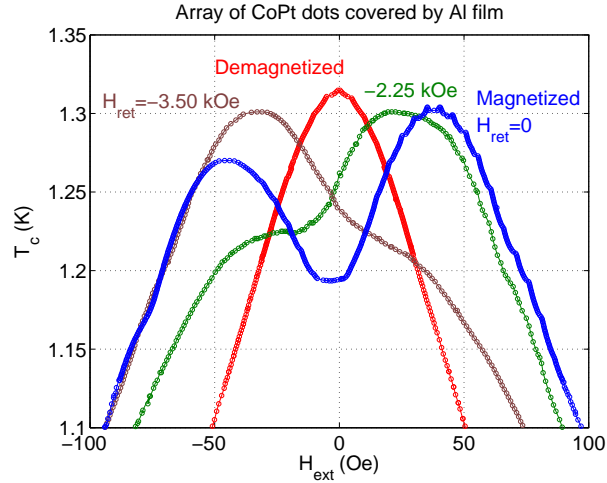}
    \caption{(color
    online) The phase boundaries $T_c(H_{ext})$ for an S/F
    hybrid, consisting of an Al film and an array of magnetic dots, in
    the demagnetized state, in the completely magnetized state in
    positive direction as well as in several intermediate magnetic
    states, adapted from  Gillijns {\it
    et al.} \cite{Gillijns-PRB-07a} and Aladyshkin {\it et al.}
    \cite{Aladyshkin-PhysC-08}. The period $\Delta H_{ext}$ of the $T_c$ oscillations,
    which are distinctly seen in the curve corresponding to the
    magnetized states, is equal to 5.1 Oe and it exactly coincides
    with the matching field, i.e., $\Delta H_{ext}=\Phi_0/S$, where
    $\Phi_0=2\cdot10^{-7}$~Oe$\cdot$cm$^2$ is the flux quantum and
    $S=4~\mu$m$^2$ is the area of the unit cell. Note that field and temperature intervals shown here are much
    broader than those presented in Fig. \ref{Fig-QuantizedDisplacement}
    (b). } \label{Fig-Gillijns-PRB-2007-b}
    \end{center}
    \end{figure}

Both the field compensation and matching effects in plain Al films
with a square array of ferromagnetic Co/Pt disks were investigated
by Gillijns {\it et al.} \cite{Gillijns-PRB-07a} and Aladyshkin
{\it et al.} \cite{Aladyshkin-PhysC-08}. Due to the presence of
the out-of-plane magnetized dots, there are {\it three} different
areas, where the OP can potentially nucleate: above the positive
or negative domains, inside the magnetic dot, and in between the
dots. In the demagnetized state the interdot field is close to
zero, therefore superconductivity starts to nucleate at this
position at relatively low magnetic fields, resulting in an almost
linear phase boundary centered at $H_{ext}=0$. By magnetizing the
dots positively (i) the amplitude of the field between the dots
increases negatively and (ii) the typical width of the positive
domains becomes larger than that for negative domains. Therefore
the peak, associated with the OP localization between the dots,
shifts towards positive fields. In addition, a second local $T_c$
maximum, corresponding to the appearance of superconductivity
above the broader positive domains, appears, while the OP
nucleation above narrower negative domains is still suppressed.
For negatively magnetized dots the reversed effect occurs. It is
important to note, that the amplitude of the main $T_c$ peak,
corresponding to the OP nucleation between the dots, remains
almost constant, since the mentioned area of the OP localization
is almost independent of the dot's magnetic state (Fig.
\ref{Fig-Gillijns-PRB-2007-b}).

\vspace*{0.3cm}

\noindent  {\it Individual ferromagnetic dots above/inside
superconducting films}

\begin{figure}[b!]
    \begin{center}
    \epsfxsize=40mm \epsfbox{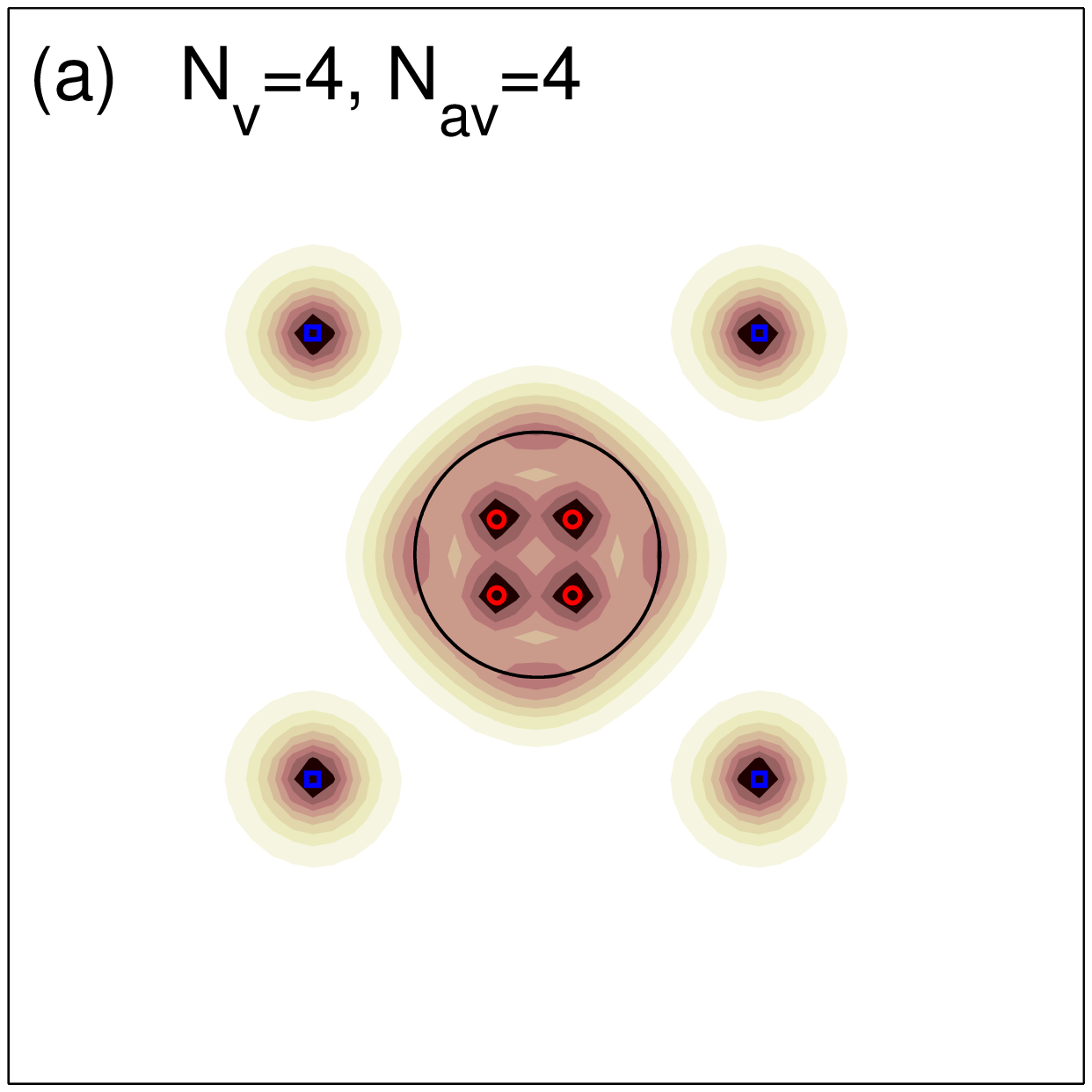}
    \epsfxsize=40mm \epsfbox{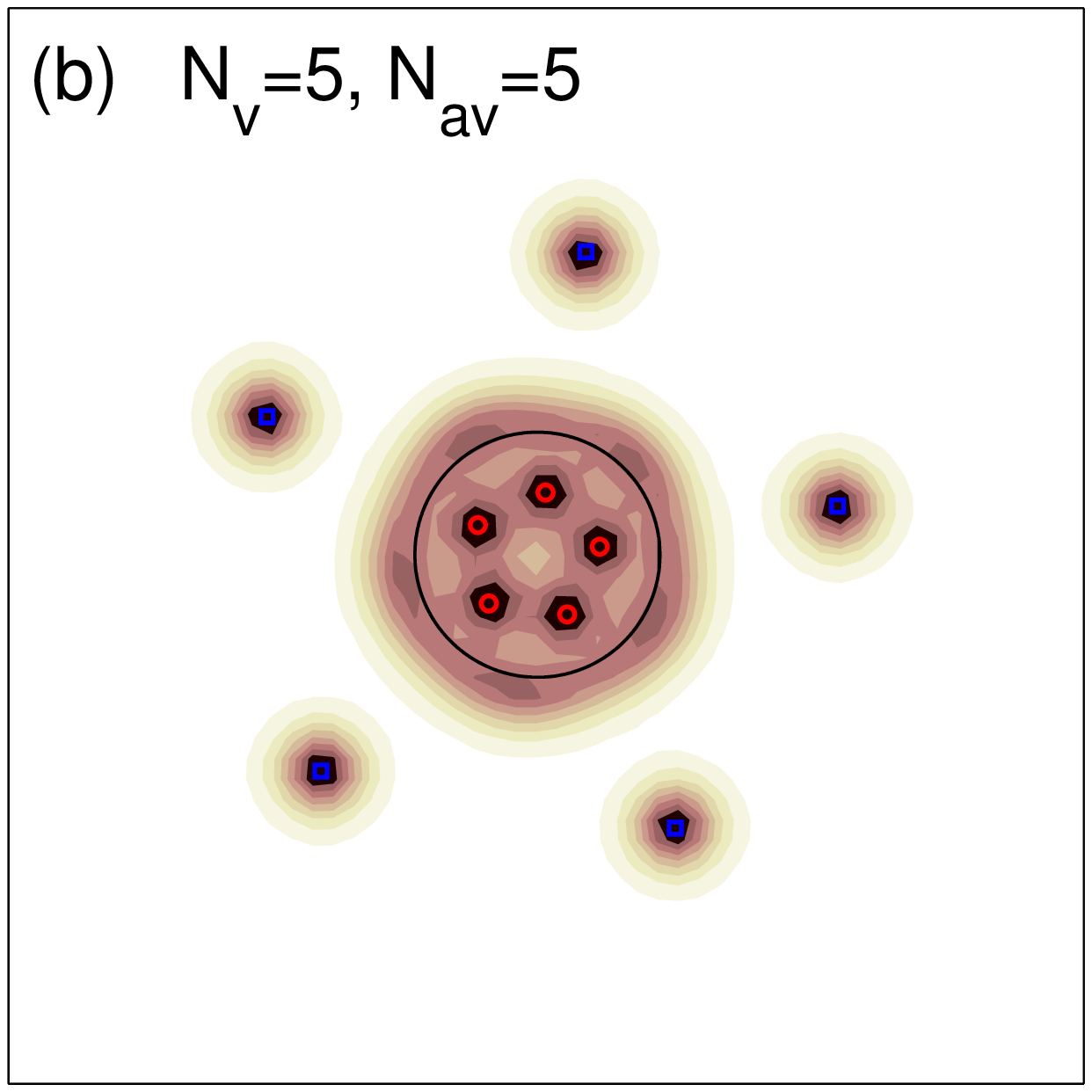}
    \epsfxsize=40mm \epsfbox{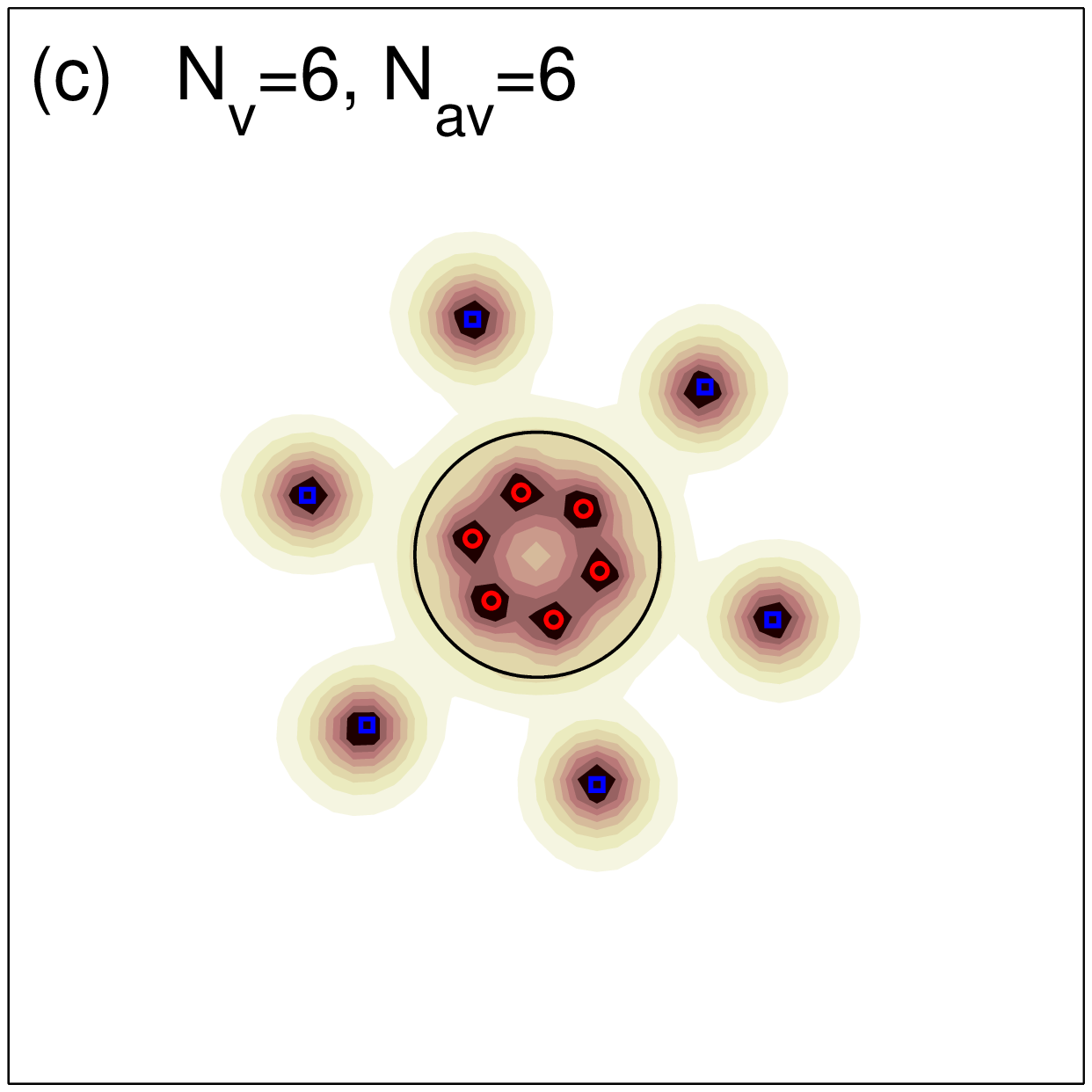}
    \epsfxsize=40mm \epsfbox{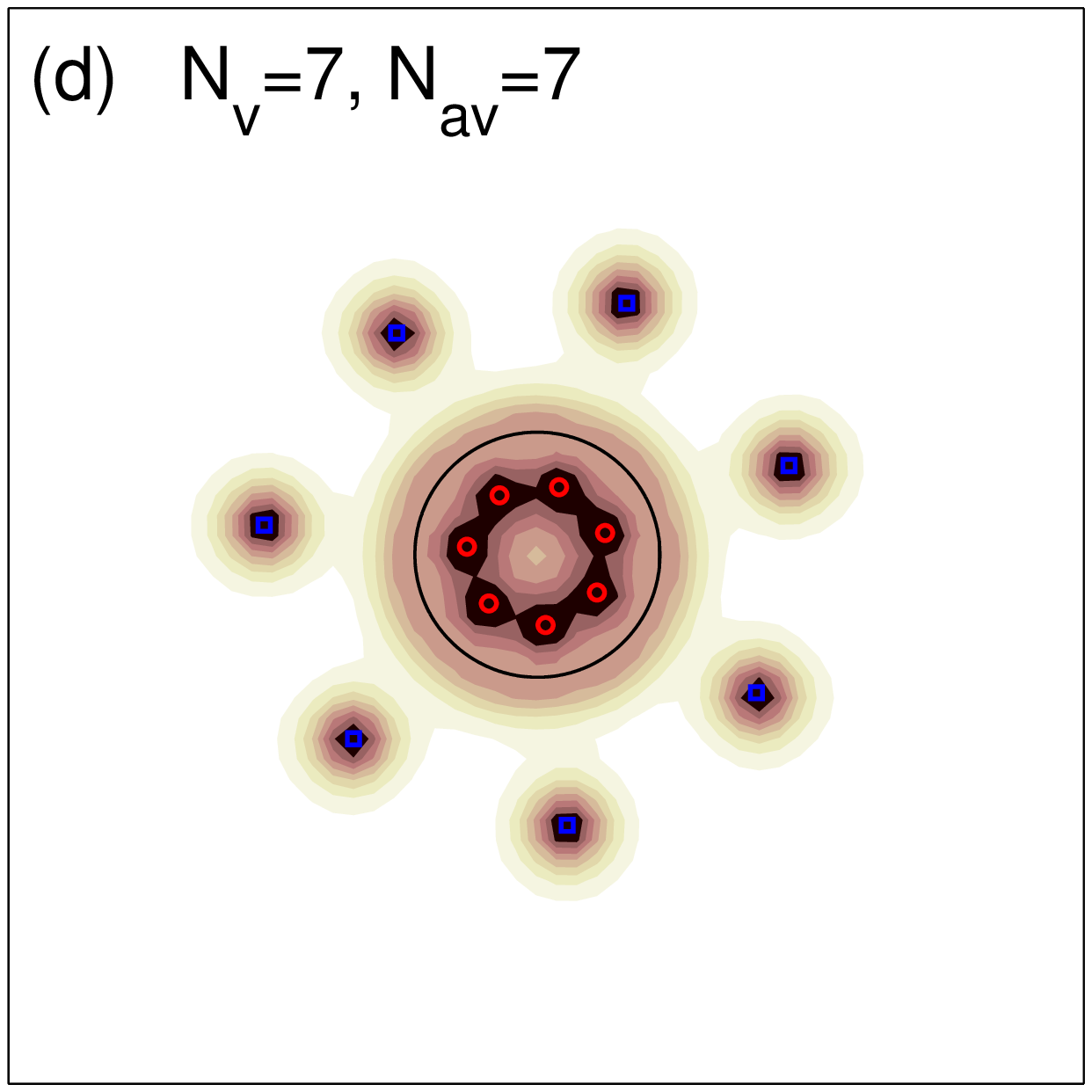}
    \caption{(color online) Contour-plots of the Cooper pair density
    $|\psi|^2$, illustrating the appearance of vortex-antivortex
    shell structures in large-area superconducting film
    in the field of perpendicularly magnetized disk for different magnetic
    moments $m$ of the ferromagnetic particle: $m/m_0=25$ (a), 29 (b), 35 (c) and 38
    (d), where $m_0=H_{c2}\,\xi^3$, by courtesy of M.V. Milo\v{s}evi\'{c}.
    The highest $|\psi|$ values are shown in lighter shades
    and the lowest densities in darker shades. It is important to note
    that only the central part
    of the superconducting film (45$\,\xi\times$ 45$\,\xi$) is shown here,
    black solid line schematically depicts the edge of the magnetic
    dot. Red circles correspond to the vortex cores, while blue
    squares mark the position of antivortices. The number of
    vortices $N_v$ and antivortices $N_{av}$ are indicated on the
    plots. The simulations were performed for the following
    parameters: $\kappa=\lambda/\xi=1.2$,
    the lateral size of superconducting sample is 256$\,\xi\times$ 256$\,\xi$,
    the radius of ferromagnetic disk is 4.53 $\xi$,
    and the thicknesses of superconductor and ferromagnet are equal to 0.1
    $\xi$.} \label{Fig-Milosevic}
    \end{center}
    \end{figure}

Marmorkos {\it et al.} \cite{Marmorkos-PRB-96} studied a
possibility to create giant vortices by a ferromagnetic disk with
out-of-plane magnetization embedded in thin superconducting film
within full nonlinear, self-consistent Ginzburg-Landau equations.
Later using the same model Milo\v{s}evi\'{c} and Peeters
\cite{Milosevic-PRB-03a,Milosevic-JLTP-03a,Milosevic-PhysC-04b}
considered the formation of vortex-antivortex structures in plain
superconducting films, infinite in the lateral direction, in the
field of an isolated ferromagnetic disk with out-of-plane
magnetization within the full nonlinear Ginzburg-Landau theory.
Antivortices were shown to be stabilized in shells around a
central core of vortices (or a giant vortex) with
magnetization-controlled "magic numbers" (Fig.
\ref{Fig-Milosevic}). The transition between the different vortex
phases while varying the parameters of the ferromagnetic dot
(namely, the radius and the magnetization) occurs through the
creation of a vortex-antivortex pair under the magnetic disk edge.

\subsection{Mesoscopic S/F hybrids: theory and experiments}

In all the description of nucleation of superconductivity so far,
we have ignored the effects of the sample's borders. It is well
known that the OP patterns in mesoscopic superconducting samples
with lateral dimensions comparable to the superconducting
coherence length and magnetic penetration depth, is substantially
influenced by the geometry of the superconductor (see review of
Chibotaru {\it et al.} \cite{Chibotaru-JMP-05} and references
therein). As a result, the presence of the sample's boundaries
allows the appearance of exotic states (giant vortex states,
vortex clusters, shell configurations etc.), otherwise forbidden
for bulk superconductors and non-patterned plain superconducting
films. Since, as we have pointed out above, a nonuniform magnetic
field is an alternative way to confine the superconducting OP in a
certain $H_{ext}$ and $T$ range, mesoscopic S/F hybrids seem to be
of interest for studying the interplay between different
mechanisms of confinement of the superconducting condensate.

It is important to note, that the screening effects can still be
omitted provided that the lateral size of the thin superconducting
sample is smaller than the effective penetration depth
$\lambda_{2D}=\lambda^2/D_s$. In this case the self-interaction of
the superconducting condensate can be taken into account solving
the nonlinear decoupled GL equation
    \begin{eqnarray}
    \label{NonlinearizedGLEquation-1} -\xi^2\left[\nabla +
    \frac{2\pi i}{\Phi_0}\,{\bf A}({\bf r})\right]^2 \psi - \psi + |\psi|^2\psi=0,
    \end{eqnarray}
where the vector potential distribution is given by the external
sources and the ferromagnet only [see
Eq.~(\ref{LinearizedGLEquation-2})].

\vspace*{0.3cm}

\noindent  {\it Interplay between different regimes of the OP
nucleation}

\noindent As we anticipated above, a very interesting phenomenon
in mesoscopic S/F hybrids is the interplay between competing
regimes of the OP nucleation, which can be clearly seen in the
case of a small-sized magnetic dot of radius $R_f$ placed above a
mesoscopic superconducting disk, \mbox{$R_f\ll R_s$}. Indeed, the
$|\psi|$ maximum can be generally located either at the central
part of the superconducting disk, close to the magnetic dot
(magnetic-dot-assisted superconductivity) or at the outer
perimeter of the superconducting disk (surface superconductivity).
For a positively magnetized dot  the regime of the
magnetic-dot-assisted superconductivity, associated with the
appearance of superconductivity in the region with compensated
magnetic field, can be realized only for $H_{ext}<0$. At the
compensation field ($|H_{ext}|\simeq B_0$) and provided that
$\sqrt{\Phi_0/(2\pi B_0)}\ll R_s$ the enhancement of the
$z-$component of the field near the disk edge acts as a magnetic
barrier for the superconducting condensate and it prevents the
edge nucleation of superconductivity even in small-sized
superconductors ($B_0$ being the maximum of the self-field of the
magnetized dot). The OP nucleation near the magnetic dot becomes
possible, if the critical temperature $T_c^{(0)}$ of the formation
of localized superconductivity with the OP maximum at the
superconducting disk center
    \begin{eqnarray}
    \nonumber
    \label{Eq-Bulk-Supercond}
    1-\frac{T^{(0)}_c}{T_{c0}} \simeq \frac{2\pi
    \xi_0^2}{\Phi_0}\big|H_{ext}+B_0\big|.
    \end{eqnarray}
exceeds the the critical temperature for the edge nucleation regime
    \begin{eqnarray}
    \nonumber
    \label{Eq-Surface-Supercond}
    1-\frac{T_{c3}}{T_{c0}} \simeq 0.59 \frac{2\pi
    \xi_0^2}{\Phi_0}\big|H_{ext}\big| \ ,
    \end{eqnarray}
corresponding to the critical field of surface superconductivity
$H_{c3}=1.69\,H_{c2}$
\cite{Abrikosov-book,Schmidt-book,Tinkham-book}. Due to the
different slopes $dT^{(0)}_c/dH_{ext}$ and $dT_{c3}/dH_{ext}$ and
the offsets, one can conclude that the edge OP nucleation regime
apparently dominates both for positive and large negative
$H_{ext}$ values. Only in the intermediate field range the highest
critical temperature corresponds to the formation of
superconductivity near the magnetic particle.

\vspace*{0.3cm}

\noindent  {\it Little-Parks oscillations in mesoscopic samples}

\noindent The nucleation of superconductivity in
axially-symmetrical mesoscopic S/F structures (e.g.,
superconducting disks or rings in the field of a perpendicularly
magnetized ferromagnetic circular dot) were studied theoretically
by Aladyshkin {\it et al.} \cite{Aladyshkin-PRB-07}, Cheng and
Fertig \cite{Cheng-PRB-99}, Milo\v{s}evi\'{c} {\it et al.}
\cite{Milosevic-PRB-02a,Milosevic-PRB-07}, and experimentally by
Golubovi\'{c} {\it et al.}
\cite{Golubovic-APL-03,Golubovic-EPL-04,Golubovic-PRB-03,Golubovic-PRL-04}
and Schildermans {\it et al.} \cite{Schildermans-PRB-08}. Due to
the cylindrical symmetry of the problem, superconductivity was
found to appear only in the form of giant vortices
$\psi(r,\theta)=f_L(r)\exp(iL\theta)$, where $L$ is the angular
momentum $L$ of the Cooper-pairs (vorticity). The appearance of
vortex-antivortex configurations in superconducting disks of
finite radius at temperatures close to $T_c$ is possible, although
these states were predicted to be metastable states.

\begin{figure}[t!]
    \begin{center}
    \epsfxsize=80mm \epsfbox{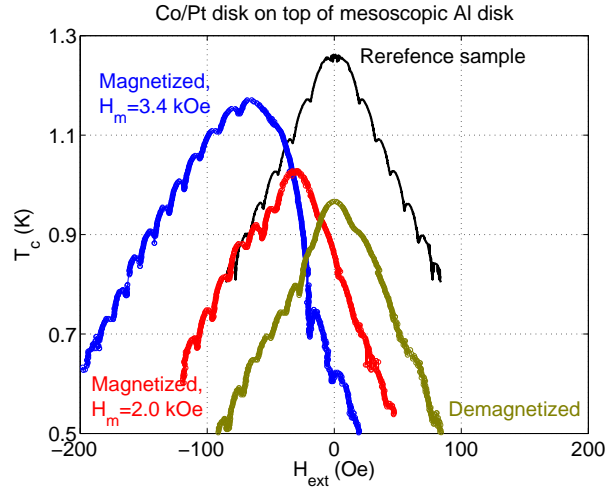}
    \caption{(color online) The phase transition lines $T_c(H_{ext})$, obtained
    experimentally for the mesoscopic S/F hybrid system with a $0.1\,R_n$ criterion
    for three different magnetic states (completely magnetized,
    partly magnetized and demagnetized states), $R_n$ and $H_m$ being the normal-state
    resistance and the magnetizing field, adapted from Schildermans
    {\it et al.} \cite{Schildermans-PRB-08}.
    The considered S/F system consists of
    superconducting Al disk covered by ferromagnetic Co/Pt multilayered
    film of the same radius $R_s=R_f=0.825$ $\mu$m.
    Black solid line represents the $T_c(H_{ext})$ dependence for the
    reference Al disk of the same lateral size.}
    \label{Figure-Schildermans-PRB-08}
    \end{center}
    \end{figure}

The observed periodic cusp-like behavior of the $T_c(H_{ext})$
dependence was attributed to the field-induced transition between
states with different vorticity similar to that of mesoscopic
superconductors in a uniform magnetic field. However, the stray
field, induced by the magnetized dot, was shown to be responsible
for a peculiar asymmetry of the oscillatory $T_c(H_{ext})$ phase
boundary and a shift of the main $T_c$ maximum towards nonzero
$H_{ext}$ values
\cite{Aladyshkin-PRB-07,Golubovic-APL-03,Golubovic-EPL-04,Golubovic-PRB-03}.
The mentioned abrupt modification of the preferable nucleation
regime when sweeping $H_{ext}$ can lead to a double change in the
slope of the $T_c(H_{ext})$ envelope from $T_{c0}/H_{c3}^{(0)}$ to
$T_{c0}/H_{c2}^{(0)}$ \cite{Aladyshkin-PRB-07,Schildermans-PRB-08}
(see Fig. \ref{Figure-Schildermans-PRB-08}). The restoration of
the slope close to $T_{c0}/H_{c2}^{(0)}$  can be interpreted as an
effective elimination of the boundary effects in mesoscopic S/F
samples at the compensation field (near the main $T_c$ maximum).
Interestingly, the nonuniform magnetic field can be used to
control the shift in the field dependence of the maximal critical
current $I_c(H_{ext})$ for a bias current flowing through the
superconducting loop, which allows one to tune the internal phase
shift in superconducting networks (Golubovic {\it et al.}
\cite{Golubovic-PRL-04}).

It is important to note that the periodicity $\Delta H_{ext}$ of
the Little-Parks oscillations in the $T_c(H_{ext})$ dependence is
explicitly given by the area where the superconducting OP is
confined. For edge nucleation only the area enclosed by the
superconductor determines the period of the oscillations, which
can be roughly estimated as $\Delta H_{ext}\simeq \Phi_0/R_s^2$.
However in the case of the magnetic-dot-assisted nucleation (in
the vicinity of the compensation field) the area of the OP
localization is determined by the spatial characteristics of the
nonuniform magnetic field (either the dot' radius $R_f$  or the
vertical separation between the dot and the superconductor $Z_d$),
therefore $\Delta H_{ext}\simeq \Phi_0/\max\{R_f^2,Z_d^2\}$. As a
consequence, the change of the nucleation regimes manifests itself
as an abrupt modification of the oscillatory $T_c(H_{ext})$
dependence. In particular, both the amplitude and the period of
the Little-Parks oscillations become much larger, provided that
\mbox{$(Z_d,R_f)\ll R_s$} and the OP wave function localizes far
from the sample's edges (Aladyshkin {\it et al.}
\cite{Aladyshkin-PRB-07}, Carballeira {\it et al.}
\cite{Carballeira-PRL-05}).

\vspace*{0.3cm}

\noindent  {\it Symmetry-induced vortex-antivortex patterns}

\begin{figure}[t!]
    \begin{center}
    \epsfxsize=80mm \epsfbox{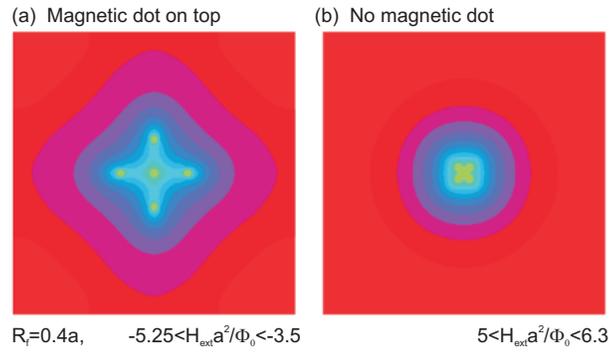} \caption{(color
    online) A comparison between the vortex-antivortex patterns that
    can be observed for $L=3$ (a) with and (b) without the magnetic
    dot on top of the superconducting square, $a$ is the lateral size of the sample, adapted from
    Carballeira {\it et al}. \cite{Carballeira-PRL-05}.
    As can be seen, the vortex-antivortex pattern rotates 45$^{\circ}$ and expands
    dramatically in the presence of the magnetic dot.}
    \label{Figure-Carballeira-PRL-05}
    \end{center}
    \end{figure}

\noindent Mesoscopic S/F hybrids of a reduced symmetry (e.g.,
structures consisting of a superconductor/ferromagnet disks and/or
regular polygons) represent nice model systems for studying
symmetry-induced phenomena. The formation of different
vortex-antivortex configurations was studied theoretically by
Carballeira {\it et al.} \cite{Carballeira-PRL-05} and Chen {\it
et al.} \cite{Chen-PRB-06b} for mesoscopic superconducting squares
with a circular ferromagnetic dot magnetized perpendicularly, and
experimentally by Golubovic {\it et al.} \cite{Golubovic-PRB-05}
for a superconducting Al disk with a magnetic triangle of Co/Pt on
top. It was shown that the symmetry-consistent solutions of the
Ginzburg-Landau equations\footnote[1]{By symmetry-consistent
    solutions we mean those vortex configurations reflecting the symmetry of the problem.
    For instance, in a
    mesoscopic superconducting square with
    vorticity $L=3$, the state consisting of four vortices and a central
    antivortex may have lower energy than the configuration of three
    equidistant vortices, which breaks the square symmetry
    (Chibotaru {\it et al.} \cite{Chibotaru-Nature-00}).},
earlier predicted for mesoscopic superconducting polygons by
Chibotaru {\it et al.} \cite{Chibotaru-JMP-05}, are preserved for
regular superconducting polygons in the stray field of
ferromagnetic disk. However, since spontaneously formed vortices
and antivortices interact with the magnetic dot in a different
way, it leads to a modification of the symmetry-induced vortex
patterns (see Fig. \ref{Figure-Carballeira-PRL-05}). In
particular, the dot can be used to enlarge these vortex-antivortex
patterns, thus facilitating their experimental observation with
local vortex-imaging techniques (``magnetic lensing").

\vspace*{0.3cm}

\noindent  {\it Embedded magnetic particles}

\noindent Doria \cite{Doria-PhysC-04a,Doria-PhysC-04b} and Doria
{\it et al.} \cite{Doria-EPL-07,Doria-PhysC-08} studied
theoretically the formation of vortex patterns induced by magnetic
inclusions embedded in a superconducting material but electrically
insulated from the multiply-connected superconductor. Since in the
absence of an external field, flux lines should be closed, vortex
lines are expected to start and end at the magnetic inclusions.
The calculations, performed in the framework of GL theory for a
mesoscopic superconducting sphere with a single magnetic point
like particle in its center,  reveal that the confined vortex
loops arise in triplets from the normal core when the magnetic
moment reaches the scale defined by $m_0=\Phi_0\xi/2\pi$.
Therefore a vortex pattern is made of confined vortices, loops and
also broken loops that spring to the surface in the form of pairs.
This vortex state provides a spontaneous vortex phase scenario for
bulk superconductors with magnetic inclusions, where the growth of
the vortex loops interconnects neighboring magnetic inclusions.

    \begin{figure}[b!]
    \begin{center}
    \epsfxsize=40mm \epsfbox{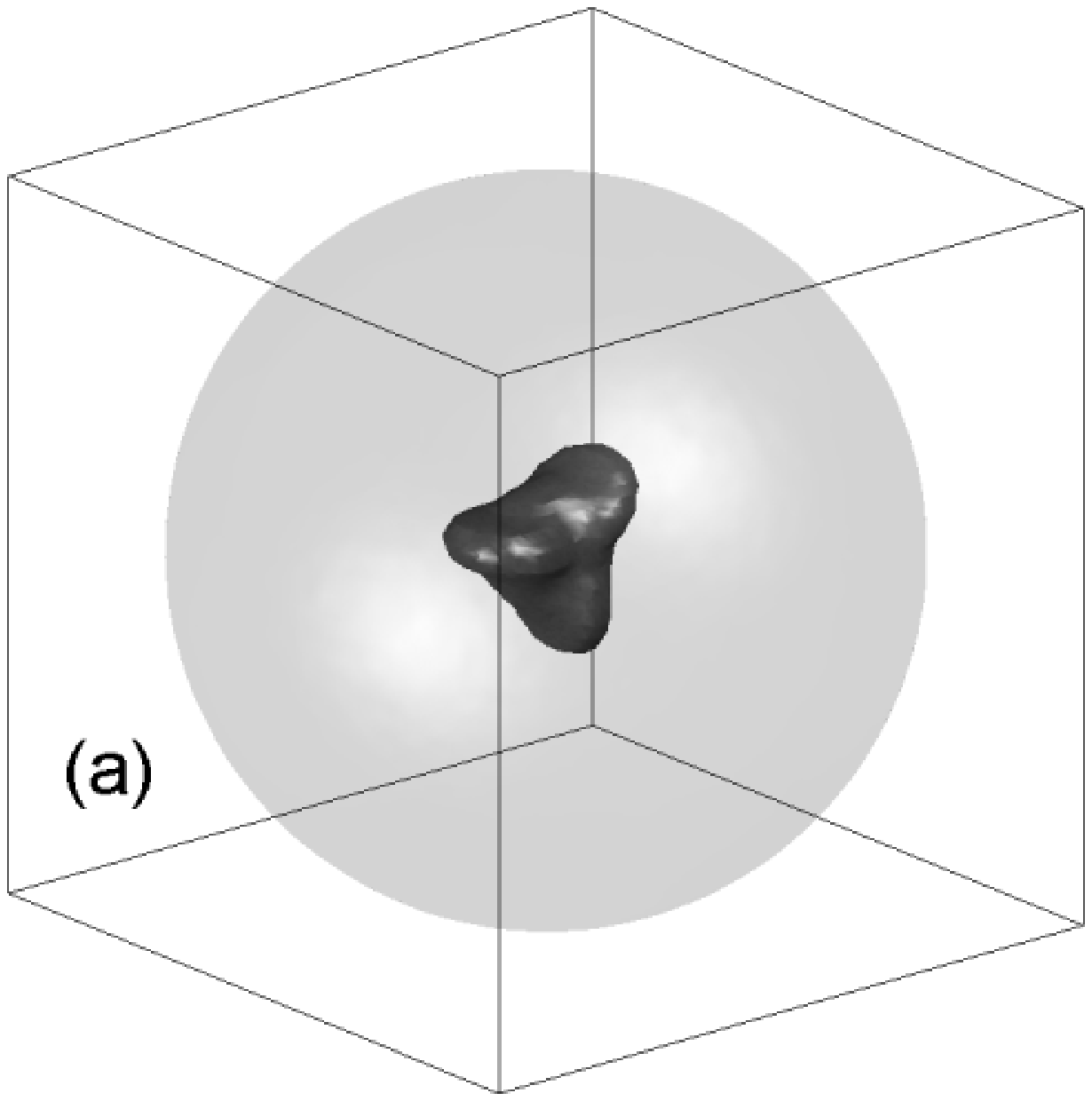}
    \epsfxsize=40mm \epsfbox{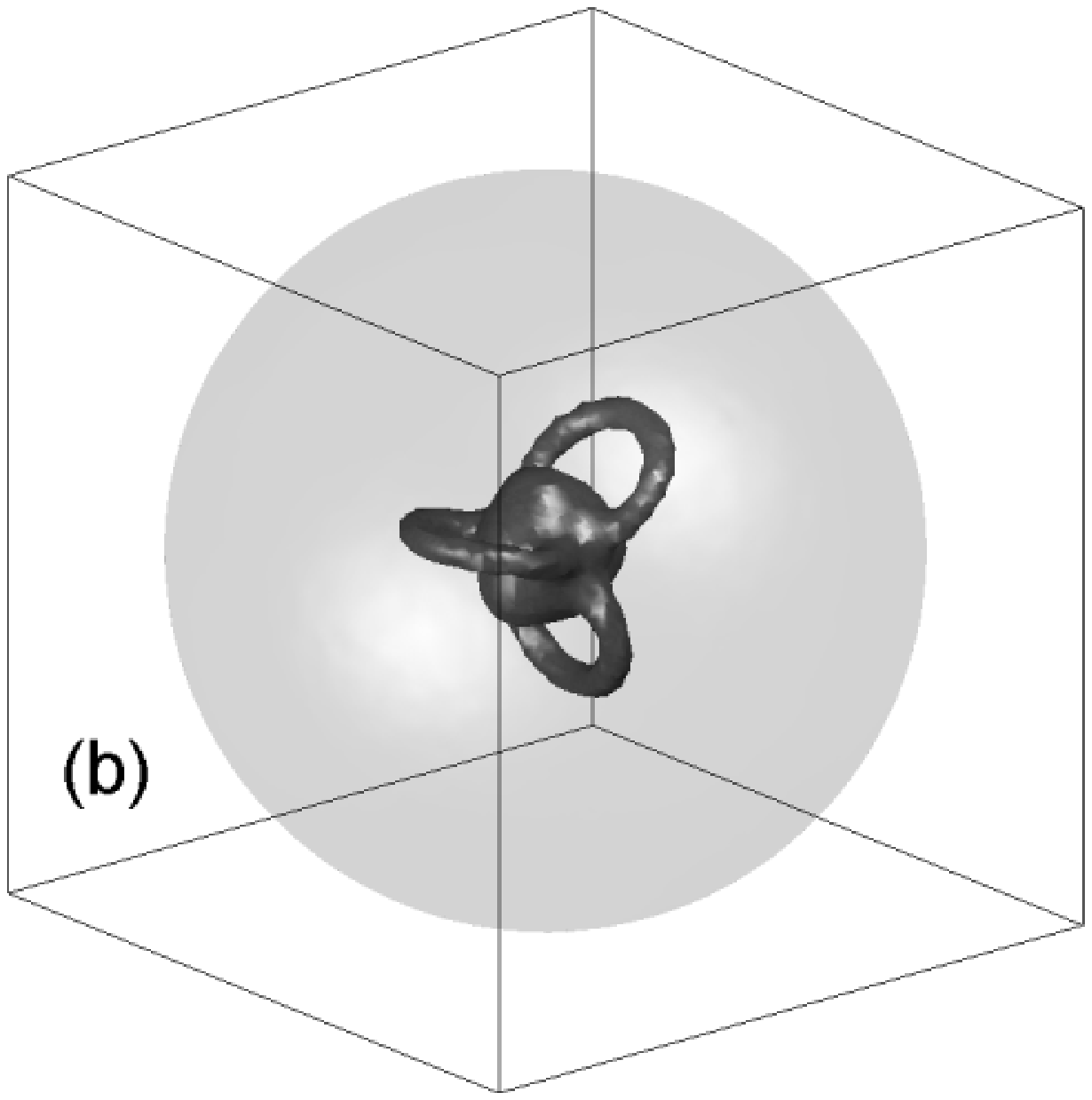}
    \caption{(color online) Confined vortex loops arise in sets of threes
    inside a mesoscopic superconducting sphere (radius 15 $\xi$), as
    shown here for two consecutive values of the point-like magnetic
    moment that occupies its center: (a) $m/m_0=15$; (b)
    $m/m_0=20$, by the courtesy of M. M. Doria.}
    \label{Figure-Doria}
    \end{center}
    \end{figure}

\vspace*{0.3cm}

\noindent  {\it Effect of the finite thickness of the
superconducting film}

\noindent The effect of the finite thickness of the
superconducting film on the $T_c(H_{ext})$ dependence was studied
theoretically by Aladyshkin {\it et al.} \cite{Aladyshkin-PRB-07},
Aladyshkin and Moshchalkov \cite{Aladyshkin-PRB-06},  and
Schildermans {\it et al.} \cite{Schildermans-PRB-08}. It follows
from Maxwell's equations that the faster the spatial variation of
the magnetic field in the lateral $(x,y)-$directions, the faster
the decay of the field in the transverse $z-$direction. However
the OP variation along the sample's thickness can be effectively
suppressed by a requirement of vanishing the normal derivative
$\partial \psi/\partial n$ of the superconducting OP at the top
and bottom surfaces of the superconductor. Indeed, the external
field applied along the $z-$direction makes the OP variations over
the superconducting film thickness more energetically unfavorable
than in the lateral direction, especially for $H_{ext}$ values of
the order of the compensation field or higher. The field
dependence $T_c(H_{ext})$ for rather thick superconducting films
in the high-field limit is similar to the phase transition line
described by Eq.~(\ref{Hc2}). This behavior results from the
effective averaging of the inhomogeneous magnetic field by the
quasi--uniform OP wave function over the sample thickness, which
substantially weakens the effect arising from the field modulation
in the lateral direction. In other words, only the lateral
inhomogeneity leads to the anomalous $T_c(H_{ext})$ dependence and
reentrant superconductivity in particular, while the transverse
field non-uniformity masks this effect. Interestingly, the
location where superconductivity starts to nucleate shifts towards
the point with zero total magnetic field (not only $z-$component
of the magnetic field) as the thickness of superconductor
increases (Aladyshkin {\it et al.} \cite{Aladyshkin-PRB-07}).

\section{Vortex matter in non-uniform magnetic fields at low temperatures}
\label{LondonSection}

\subsection{London description of a magnetically coupled S/F hybrid system}

In section \ref{Part-I} we have reviewed the theoretical modelling
within the GL formalism and the experimental data concerning the
superconducting properties of S/F hybrids rather close to the
phase transition line $T_c(H_{ext})$. In this section we consider
the properties of S/F hybrids for a fully developed
superconducting OP wave function, i.e. for $T\ll T_{c0}$. In this
limit, the superconducting properties of the S/F hybrids can be
correctly described by the London theory, omitting any spatial
variations of the OP. As before, we assume that the coercive field
of the ferromagnet is much higher than the upper critical field of
the superconductor. This guarantees that, during the investigation
of the superconducting properties, no changes in the ferromagnetic
element(s) will occur.

At rather low temperatures one should take into account that the
magnetic field, induced by screening (Meissner) currents or by
vortices, can no longer be neglected and will strongly interact
with the ferromagnet. The free energy functional of the S/F hybrid
can be written in the following form
\cite{Schmidt-book,Tinkham-book}
    \begin{eqnarray}
    \label{GibbsEnergyLondon} G_{sf}=G_{s0} + G_{m}+\int\limits_{V}
    \left\{\frac{\lambda^2}{8\pi}\, \big({\rm rot}\,{\bf
    B}\big)^2 + \frac{{\bf B}^2}{8\pi}- {\bf
    B}\cdot{\bf M} - \frac{{\bf
    B}\cdot{\bf H}_{ext}}{4\pi}  \right\} \, dV.
    \end{eqnarray}
As before, $G_{s0}$ is the self-energy of superconductor, while
the term $G_{m}$ depends on the magnetization of ferromagnet only.
Assuming that $G_m$ is constant for a given distribution of
magnetization and minimizing this functional $G_{sf}$ with respect
to the vector potential ${\bf A}$, one can obtain the
London-Maxwell equation:
    \begin{eqnarray}
    \label{LondonEquation-2}
    {\bf B} + \lambda^2\, {\rm rot~}{\rm rot~}{\bf B} = \Phi_0\,\sum_i\delta({\bf r}-{\bf R}_{v,i})\,{\bf z}_0 +
    4\pi\lambda^2\, {\rm rot}\,{\rm rot~}{\bf M},
    \end{eqnarray}
where summation should be done over all vortices, positioned at
points $\{{\bf R}_{v,i}\}$ and $\delta(r)$ is the Dirac
delta-function\footnote[1]{
    We follow the standard definition of the $\delta-$function: $\delta(x,y)=0$
    for any $x\neq 0$,  $y\neq 0$, and $\int\limits_{-\infty}^{-\infty}\int\limits_{-\infty}^{-\infty}
    \delta(x,y)\,dxdy=1$.}.
In Eq. (\ref{LondonEquation-2}) we assumed that each vortex line,
parallel to the $z-$axis and carrying one flux quantum, generates
a phase distribution $\Theta=\varphi$ with ${\rm rot}(\nabla
\Theta)={\rm rot}~([{\bf z}_0\times{\bf r}_0]/r)=\delta(r)\,{\bf
z}_0$ [here ($r,\varphi,z$) are cylindrical coordinates with the
origin chosen at the vortex]. It should be noticed that the
solution of Eq.~(\ref{LondonEquation-2}) gives the magnetic field
distribution for a given vortex configuration, and in order to
find the field pattern one should find the minimum of the total
free energy $G_{sf}$ with respect to the vortex positions $\{{\bf
R}_{v,i}\}$.

\subsection{Interaction of a point magnetic dipole with a superconductor}

Next we consider the generic problem of the interaction of a
superconductor with a point magnetic dipole positioned at a height
$Z_d$ above the superconducting sample. Introducing the OP phase,
which determines the density of supercurrents ${\bf j}_s=
c/(4\pi\lambda^2)\,[\Phi_0/(2\pi)\nabla\Theta-{\bf A}]$, the free
energy functional Eq. (\ref{GibbsEnergyLondon}) at $H_{ext}=0$ can
be rewritten as $G_{sf}=G_{s0}+G_{m}+G_s$, where
    \begin{eqnarray}
    \label{ErdinFunctional}
    G_s = \int \left\{\frac{\Phi_0^2}{32\pi^3\lambda^2} (\nabla\Theta\cdot\nabla\Theta) -
    \frac{\Phi_0}{16\pi^2\lambda^2} (\nabla\Theta\cdot{\bf A}) -\frac{1}{2}\,{\bf B}\cdot{\bf M}\right\}\,dV
    \end{eqnarray}
similar to that obtained by Erdin {\it et al.}
\cite{Erdin-PRB-02}. Due to the linearity of Eq.
(\ref{LondonEquation-2}), we can introduce the field ${\bf
B}_v={\rm rot}~{\bf A}_v$, generated by a single vortex line
positioned at ${\bf r}={\bf R}_v$, and the field ${\bf B}_m={\rm
rot}~{\bf A}_m$, corresponding to the screening (Meissner) current
distribution induced by the magnetic dipole\footnote[2]
    {These fields ${\bf B}_v$ and ${\bf B}_m$ are the solutions of the
    following differential
    equations
    \begin{eqnarray}
    \nonumber
    {\bf B}_v + \lambda^2\, {\rm rot~}{\rm rot~}{\bf B}_v = \Phi_0\,\delta({\bf r}-{\bf R}_v)\,{\bf z}_0, \\
    \nonumber
    {\bf B}_m + \lambda^2\, {\rm rot~}{\rm rot~}{\bf B}_m = 4\pi\lambda^2\, {\rm rot}\,{\rm rot~}{\bf
    M}.
    \end{eqnarray}
    }.
As a result, for the particular case of a point magnetic dipole,
${\bf M}={\bf m}_0\,\delta({\bf r}-{\bf R}_d)$, the energy of the
system $G_s$ depending on the supercurrents can be represented as
a sum of the self-energy of the vortex
$\epsilon^{(0)}_v=(\Phi_0/4\pi\lambda)^2\,\ln \lambda/\xi$, which
is independent on the dipolar moment, and a term $G_{int}$,
responsible for the interaction between the dipole and
superconductor (e.g., Wei {\it et al.} \cite{Wei-PRB-96}, Carneiro
\cite{Carneiro-PRB-04}):
    \begin{eqnarray}
    \label{InteractionEnergy}
    G_{int}=-\frac{1}{2}\,{\bf m}_0\cdot {\bf B}_m({\bf R}_d)-{\bf m}_0\cdot {\bf B}_v({\bf
    R}_d).
    \end{eqnarray}
The first term in Eq. (\ref{InteractionEnergy}) corresponds to the
interaction between the magnetic dipole and the local field at the
dipole's position ${\bf B}_{m}({\bf R}_d)$ due to the screening
currents. The second term in Eq. (\ref{InteractionEnergy}) gives
the energy of the interaction between the vortex and the magnetic
dipole, which consists of two parts: (i) the ``hydrodynamical"
interaction between circulating supercurrents due to the vortex on
one hand and the dipole on the other hand; (ii) the interaction
between the stray field of the vortex with the magnetic moment.
Interestingly both contributions turn out to be equal to
$(-1/2)\,{\bf m}_0\cdot {\bf B}_v({\bf R}_d)$.

\vspace*{0.2cm} \noindent {\it Interaction between a point
magnetic dipole and the Meissner currents}

\noindent In the case of a superconductor without vortices [the
so-called Meissner state, ${\rm rot}(\nabla \Theta)=0$], the
interaction between the dipole and the superconductor is given by
$U_{m}=(-1/2)\,{\bf m}_0\cdot{\bf B}_m({\bf R}_d)$. It can easily
be seen that this energy contribution is always positive, since
the screening currents induced by a dipole generate a magnetic
field which is opposite to the orientation of the dipole. This is
in principle just a reformulation of the fact that a magnetic
dipole will be repelled by a superconducting film with a force
${\bf f}=-\partial U_{m}/\partial {\bf R}_d$. The related problem
of levitating ferromagnetic particle over various superconducting
structures was considered theoretically by Wei {\it et al.}
\cite{Wei-PRB-96}, Xu {\it et al.} \cite{Xu-PRB-95}, Coffey
\cite{Coffey-PRB-95,Coffey-PRB-02} and Haley
\cite{Haley-PRB-96a,Haley-PRB-96b}. In particular, in the limiting
case $Z_d/\lambda\gg 1$ the magnetic levitation force was found to
vary linearly with $\lambda$, regardless of the shape of the
magnet. The fact that $\lambda$ (as well as the vertical component
of the force) shows an exponential temperature dependence for
$s-$wave superconductors, linear$-T$ for $d-$wave superconductors,
and quadratic$-T^2$ dependence in a wide low-temperature range for
materials with $s+id$ symmetry of the gap, can assist in
distinguishing the type of pairing in real samples based on
magnetic force microscopy (MFM) measurements \cite{Xu-PRB-95}.

\vspace*{0.2cm}
\noindent {\it Creation of vortices by the stray
field of a magnetic dipole}

\noindent In reality, magnetic dipoles which have a high dipolar
moment or are localized rather close to the superconductor can
suppress the Meissner state due to the dipole's stray field. The
possible appearance of a vortex would change the total energy of
the S/F hybrid system by $G_{s}=\epsilon^{(0)}_v + \epsilon_{mv}$,
where $\epsilon_{mv} \propto -m_0$ accounts for the dipole-vortex
interaction. Clearly, one can make $G_{s}<0$ by increasing $m_0$
and therefore it will become energetically favorable to generate
vortices in the system. Such a scenario for the formation of the
mixed state was considered, e.g., by Wei {\it et al.}
\cite{Wei-PRB-96,Wei-PhysC-97} for a vertically magnetized dipole
placed above a thin superconducting film. As a consequence of
Eq.~(\ref{InteractionEnergy}), the appearance of a vortex line in
the superconductor drastically changes the resulting force acting
on the magnetic dipole (Xu {\it et al.} \cite{Xu-PRB-95}, Wei {\it
et al.} \cite{Wei-PRB-96,Wei-PhysC-97}).

It is important to notice that in the case of superconducting
films infinite in the lateral direction and cooled in zero
magnetic field, a spontaneously induced vortex should be
accompanied by an antivortex in order to provide flux conservation
imposed by Maxwell's equations\footnote[1]{In
    mesoscopic superconducting systems, the
    returning flux lines generated by an out-of-plane magnetized dipole can
    bypass the border of the sample and the existence of an
    antivortex is not required.}.
The destruction of the Meissner state in a zero-field-cooled
superconducting film was theoretically studied by Melnikov {\it et
al.} \cite{Melnikov-PRB-98} and Aladyshkin {\it et al.}
\cite{Aladyshkin-JETP-99}. In this case the formation of the mixed
state occurs via the penetration of vortex semi-loops which split
into vortex-antivortex pairs. The local suppression of the
Bean-Livingston energy barrier \cite{Bean-PRL-64}, which controls
the process of vortex penetration, takes place when the screening
current density will be of the order of the depairing current
density. The threshold distance $Z_d^*=Z_d^*(T)$, corresponding to
the suppression of the energy barrier can be experimentally
detected by measuring a nonzero remanent magnetization as long as
pinning is relevant. Interestingly, such non-contact technique
allows one to estimate the depairing current density and its
temperature dependence in thin superconducting YBa$_2$Cu$_3$O$_7$
films \cite{Melnikov-PRB-98,Aladyshkin-JETP-99,Nozdrin-IEEE-99}.
Since the surface energy barrier in superconductors with a flat
surface is known to be suppressed by applying an external field on
the order of the critical thermodynamical field
$H_{cm}=\Phi_0/(2\sqrt{2}\pi\lambda\xi)$ \cite{Abrikosov-book},
one can get a rough criterion for the persistence of the
vortex-free state in the presence of a magnetic dipole:
$m_0/Z_d^3\le H_{cm}$. A different criterion should be obtained in
case the dipole is already present close to the superconductor and
the whole system is cooled down below the superconducting
temperature. Under this field-cooled condition a lower critical
magnetization to induce a vortex-antivortex pair is expected.

The problem of the formation of vortex-antivortex pairs in
superconducting films in the presence of vertically and
horizontally magnetized dipole at $H_{ext}=0$ was considered
theoretically by Milo\v{s}evi\'{c} {\it et al.}
\cite{Milosevic-PRB-02b,Milosevic-JLTP-03b}, Carneiro
\cite{Carneiro-PRB-04}. It was shown that an equilibrium vortex
pattern could consist of spatially separated vortices and
antivortices (for rather thin superconducting films) or, when the
superconducting film thickness increases and becomes comparable to
the London penetration depth, curved vortex lines, which start and
terminate at the surface of the superconductor.

\vspace*{0.2cm} \noindent {\it Magnetic pinning}

\noindent Now we will discuss the properties of S/F hybrids where
the magnetization of the ferromagnet and the distance between
ferromagnet and superconductor are assumed to be fixed resulting
in a constancy of the interaction energy with the screening
currents induced by the ferromagnet. In this case, any variation
of the free energy of the S/F hybrid in the presence of vortices
(either induced by the ferromagnetic element or by external
sources) can be attributed to the rearrangements of the vortex
pattern. The part of the interaction energy $G_{int}$,
proportional to the magnetization of the ferromagnet and sensitive
to the vortex' positions, is usually called the magnetic pinning
energy $U_p$.

\begin{figure}[b!]
    \begin{center}
    \epsfxsize=75mm \epsfbox{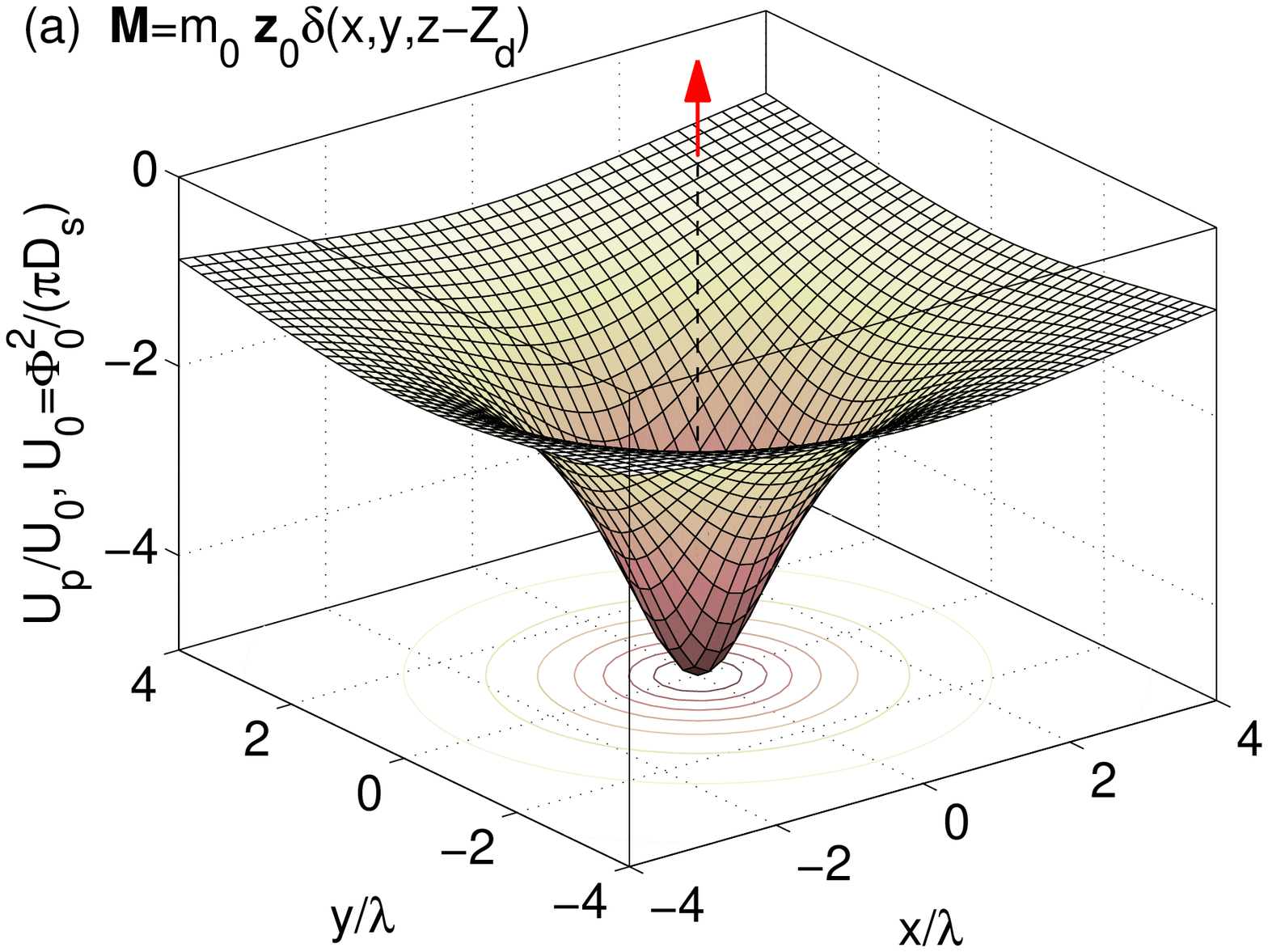}
    \epsfxsize=75mm \epsfbox{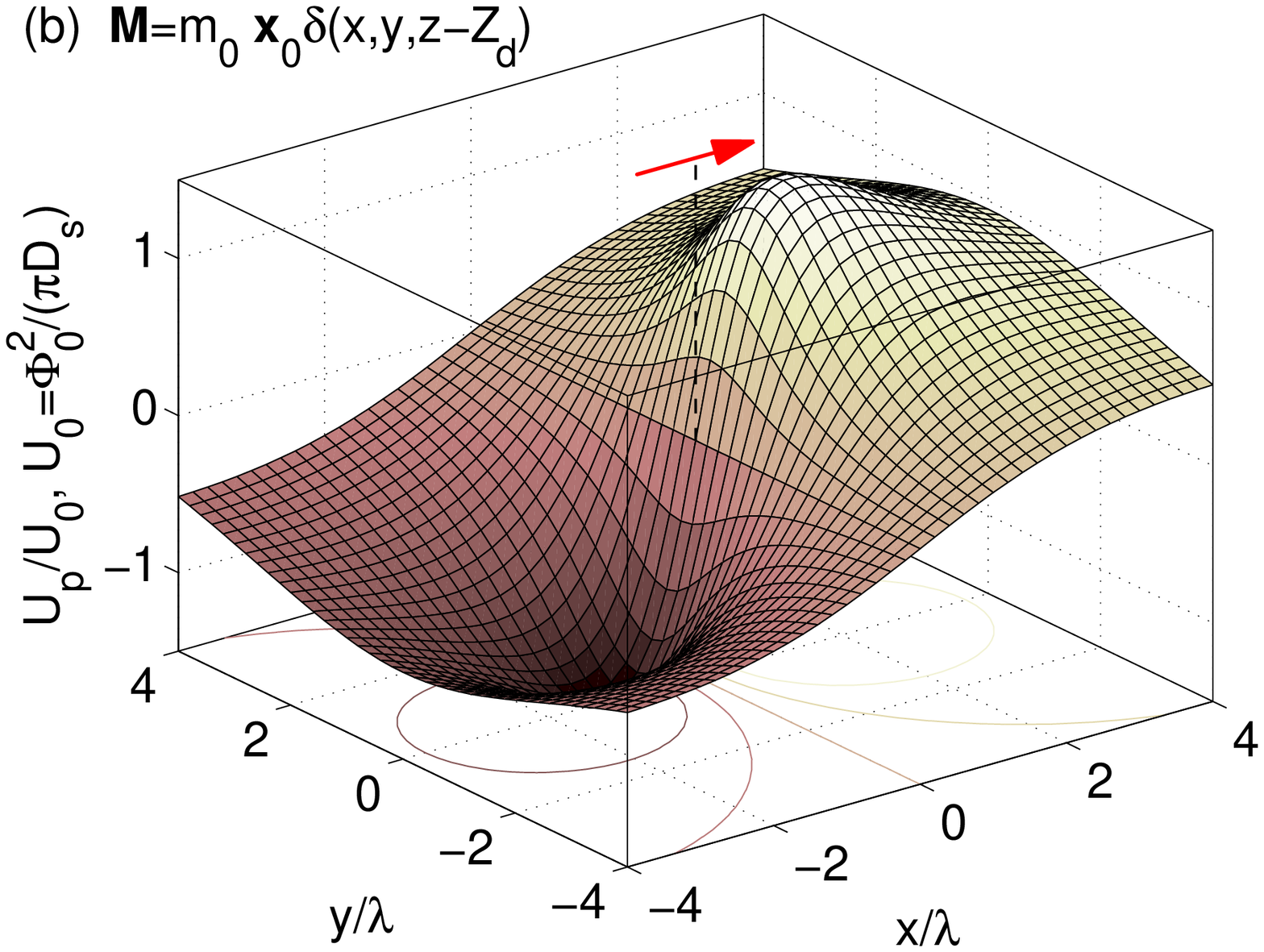}
    \end{center}
    \caption{(color online) The spatial dependence of the energy of a single
    vortex in a field of point magnetic dipole, ${\bf M}={\bf m}_0 \delta(x,y,z-Z_d)$,
    calculated according to Eq. (\ref{Uvm}) for the height $Z_d/\lambda=1$ and the dipole strength
    $m_0/(\Phi_0\lambda)=10$: \newline
    (a) vertically magnetized dipole, ${\bf m}_0=m_0{\bf
    z}_0$, \newline
    (b) horizontally magnetized dipole, ${\bf m}_0=m_0{\bf x}_0$.}
    \label{Fig:MagneticPinning}
    \end{figure}

In order to illustrate the angular dependence of the interaction
between a point magnetic dipole of fixed magnetization and a
vortex, we refer to the following expression (Carneiro
\cite{Carneiro-PRB-04}, Milo\v{s}evi\'{c} {\it et al.}
\cite{Milosevic-PRB-02b,Milosevic-JLTP-03b})
    \begin{equation}
    U_{p}(r_{\perp},0)=\frac{\Phi_0^2}{2\pi\lambda}\,\left[-\frac{1}{2}\left(\frac{m_{0,z}}{\Phi_0\lambda}\right)
    \,\frac{D_s}
    {(r_{\perp}^2+Z_d^2)^{1/2}} + \frac{D_s}{2r_{\perp}}
    \left(\frac{m_{0,z}}{\Phi_0\lambda}\right)\left(1-\frac{Z_d}{(r_{\perp}^2+Z_d^2)^{1/2}}
    \right)\,  \cos\phi\right],
    \label{Uvm}
    \end{equation}
\noindent obtained for a thin superconducting film $D_s\ll
\lambda$ and for vortex-dipole separations smaller than the
effective penetration depth $\lambda_{2D}=\lambda^2/D_s$. Here
$m_{0,x}$ and $m_{0,z}$ are the in-plane and out-of-plane
components of the dipolar moment, $r_{\perp}=\sqrt{x^2+y^2}$,
${\bf R}_d=(0,0,Z_d)$ is the position of the dipole, $\phi$ is the
angle between the $x$-axis and the vector position of the vortex
in the plane of the superconductor ${\bf r}_{\perp}$. The
resulting pinning potentials for an in-plane and out-of-plane
magnetized dipole, derived from Eq.~(\ref{Uvm}), are shown in
Fig.~\ref{Fig:MagneticPinning}. The pinning of vortices (and
antivortices) in the superconducting films of a finite thickness
on the magnetic dipole at $H_{ext}=0$ was analyzed by
Milo\v{s}evi\'{c} {\it et al.}
\cite{Milosevic-PRB-02b,Milosevic-JLTP-03b} and Carneiro
\cite{Carneiro-PRB-04,Carneiro-PhysC-04}. In addition, Carneiro
considered the magnetic pinning for the case of a dipole able to
rotate freely in the presence of the external magnetic field
\cite{Carneiro-PRB-05,Carneiro-EPL-05,Carneiro-PhysC-06} or
external current \cite{Carneiro-PRB-05}.

In Ref. \cite{Carneiro-PRB-04} it was shown that this magnetic
pinning potential has a depth of $m_0/(4\pi\lambda)$, penetrates a
distance $\lambda$ into the film whereas its range parallel to the
film surface is a few times $\lambda$. This finding points out the
relevance of the penetration depth $\lambda$ to characterize the
purely magnetic pinning potential. Since typical $\lambda$ values
are similar for a broad spectrum of superconducting materials,
this suggests that magnetic pinning represents a promising way of
increasing the critical current not only in conventional
superconductors but also in high-$T_c$ superconductors. Notice
that the fact that the magnetic pinning range is determined by
$\lambda$ sets a limit for the minimum distance between magnetic
particles, beyond which the vortex lines cannot resolve the field
modulation and therefore the pinning efficiency decreases. Having
in mind that many practical applications typically involve
high-$T_c$ superconductors with $\xi\ll\lambda$, it becomes clear
that this maximum density of pinning sites to trap individual
vortices is much lower than that limited due to core pinning.

It is important to mention that real nanoengineered ferromagnetic
elements are quite far from point dipoles but rather consist of
extended volumes of a magnetic material, such as magnetic dots and
stripes. In this case, due to the principle of superposition, the
magnetic pinning energy needs to be integrated over the volume of
the ferromagnet $V_f$ to determine the interaction between a
vortex line and a finite size permanent magnet
    \begin{equation}
    U_{p}({\bf r})=-\int\limits_{V_f} {\bf M}({\bf r}^{\prime})\cdot{\bf B}_v({\bf r}-
    {\bf r}^{\prime})~d^3 r^{\prime}. \label{Uvm-dot}
    \end{equation}

The magnetic pinning in an infinite superconducting film produced
by individual ferromagnetic objects of finite size was considered
for the following cases: a ferromagnetic sphere magnetized
out-of-plane (Tokman \cite{Tokman-PLA-92}), magnetic disks, rings,
rectangles, and triangles magnetized out-of-plane
(Milo\v{s}evi\'{c} {\it et al.}
\cite{Milosevic-JLTP-03b,Milosevic-PRB-03b}), magnetic bars and
rectangular dots magnetized in-plane (Milo\v{s}evi\'{c} {\it et
al.} \cite{Milosevic-PRB-04}), circular dots magnetized either
in-plane and out-of-plane (Erdin {\it et al.} \cite{Erdin-PRB-02},
Erdin \cite{Erdin-PRB-05}), circular and elliptic dots and rings
magnetized out-of-plane (Kayali
\cite{Kayali-PLA-02,Kayali-PRB-04a}, Helseth
\cite{Helseth-PLA-03}), columnar ferromagnetic rod (Kayali
\cite{Kayali-PRB-05}). The analysis of the pinning properties of a
vortex in a semi-infinite superconducting film due to magnetic
dots was done by Erdin \cite{Erdin-PRB-04}. Similar to the case of
a point magnetic dipole, the extended magnetic objects are able to
generate vortex-antivortex pairs provided that the size and the
magnetization are large enough.

We would like to note that the attraction of a superconducting
vortex to a source of inhomogeneous magnetic field (e.g., coil on
a superconducting quantum interference device or magnetized tip of
a magnetic force microscope) makes it possible to precisely
manipulate the position of an individual vortex. Such experiments
were performed by Moser {\it et al.} \cite{Moser-JMMM-98}, Gardner
{\it et al.} \cite{Gardner-APL-02}, Auslaender {\it et al.}
\cite{Auslaender-NatPhys-09} for high$-T_c$ superconducting thin
films and single crystals in intermediate temperatures when
intrinsic pinning become weaker. This technique apparently opens
unprecedent opportunities, e.g., for a direct measuring of the
interaction of a moving vortex with the local disorder potential
and for preparing exotic vortex states like entangled vortices
(Reichhardt \cite{Reichhardt-NatPhys-09}).

Equation (\ref{Uvm-dot}) shows that the pinning potential does not
only depend on the size and the shape of the ferromagnetic
elements but also on the particular distribution of the
magnetization (i.e. on their exact magnetic state). It is expected
that the average pinning energy is less efficient for magnetic
dots in a multidomain state whereas it should reach a maximum for
single domain structures. In other words, if the size of the
magnetic domains is small in comparison with $\lambda$ or the
separation distance $Z_d$, then a vortex line would feel the
average field emanating from the domains and the magnetic pinning
should be strongly suppressed. This flexibility of magnetic
pinning centers makes it possible to tune the effective pinning
strength, as we shall discuss below.

The question now arises whether the pinning potential produced by
a magnetic dipole will remain efficient when a vortex-antivortex
pair is induced by the magnetic dipole. In order to answer this
question it is necessary to minimize the mutual interaction
between the induced vortex-antivortex pair, the magnetic element
and the test vortex generated by the external field. The magnetic
moment -- test vortex interaction is a linear function of $m_0$
always favoring the test vortex to sit on the positive pole of the
magnet, whereas the induced currents -- test vortex interaction
will follow the Little-Parks oscillations (due to the creation of
vortices by the dipole). For the case of an in-plane dipole,
Milo\v{s}evi\'{c} {\it et al.} \cite{Milosevic-PRB-04}
demonstrated that all these terms produce a subtle balance of
forces which lead to a switching of the optimum pinning site from
the positive to the negative magnetic pole, as $m_0$ is increased.
For the case of out-of-plane magnetic dipole, once a
vortex-antivortex pair is induced, the test vortex will annihilate
the antivortex leaving a single vortex on top of the magnetic dot
(Gillijns {\it et al}. \cite{Gillijns-PRB-06}).

\subsection{Magnetic dots in the vicinity of a plain superconducting film}

The rich variety of possibilities considered theoretically in the
previous section for an individual ferromagnetic element in close
proximity to a superconducting film represents an experimental
challenge in part due to the difficulties associated with
recording small induced signals. A very successful way to overcome
this limitation consists of studying the average effect of a
periodic array of dots at the expense of blurring sharp
transitions, such as the vortex-antivortex generation, or inducing
new collective phenomena associated with the periodicity of the
magnetic templates. An important experimental condition that
should be satisfied in order to justify the analogies drawn
between individual dipoles and arrays of dipoles, is the lack of
magnetostatic interactions between neighboring dipoles. In other
words, it is necessary to ensure that the field generated by a
dipole at the position of its nearest neighbors lies below the
coercive field of the chosen ferromagnetic materials (Cowburn {\it
et al.} \cite{Cowburn-PRL-99,Cowburn-NJP-99}, Novosad {\it et al.}
\cite{Novosad-PRB-02,Novosad-APL-03}).

\vspace*{0.2cm} \noindent {\it Commensurability effects in S/F
hybrids with periodic arrays of magnetic dots}

\noindent In the early 1970's, Autler
\cite{Autler-JLTP-72,Autler-HPA-72} proposed that a periodic array
of ferromagnetic particles should give rise to an enhancement of
the critical current of the superconducting material. Recent
developments on lithographic techniques have made it possible to
prepare superconducting structures containing a regular array of
magnetic dots (Co, Ni, Fe) at submicrometer scale of desirable
symmetry in a controlled way (Otani {\it et al.}
\cite{Otani-JMMM-93}, Geoffroy {\it et al.}
\cite{Geoffroy-JMMM-93}, Mart\'{\i}n {\it et al.}
\cite{Martin-PRL-97,Martin-PRL-99,Martin-JMMM-98,Martin-PRB-00},
Morgan and Ketterson \cite{Morgan-PRL-98,Morgan-JLTP-01}, Hoffmann
{\it et al.} \cite{Hoffmann-PRB-00}, Jaccard {\it et al.}
\cite{Jaccard-PRB-98}, Villegas {\it et al.}
\cite{Villegas-PRB-05b,Villegas-PRB-03,Villegas-PRB-05c}, Stoll
{\it et al.} \cite{Stoll-PRB-02}, van Bael {\it et al.}
\cite{vanBael-PRB-99,vanBael-PRL-01,vanBael-PhysC-00a,vanBael-PhysC-00b,vanBael-JC-01,vanBael-PhysC-02},
van Look {\it et al.} \cite{vanLook-PhysC-00}). The S/F hybrids
containing periodic arrays of magnetic elements with out-of-plane
magnetization (multilayered Co/Pt and Co/Pd structures) were
fabricated and investigated  by van Bael {\it et al.},
\cite{vanBael-PhysC-00a,vanBael-PhysC-00b,vanBael-JC-01,vanBael-PhysC-02,vanBael-PhysC-01,vanBael-PRB-03}
and Lange {\it et al.} \cite{Lange-JLTP-05,Lange-PRB-05}. In all
these works, it was found that the presence of the lattice of
magnetic dots leads either to (i) a resonant change in the
magnetoresistance $\rho(H_{ext})$ and the appearance of pronounced
equidistant minima of resistivity with the period $H_1=\Phi_0/S_0$
determined by the size of the magnetic unit cell $S_0$
\cite{Otani-JMMM-93,Geoffroy-JMMM-93,Martin-PRL-97,Martin-PRL-99,Martin-JMMM-98,Martin-PRB-00,Morgan-JLTP-01,Hoffmann-PRB-00,Jaccard-PRB-98,Villegas-PRB-03,Villegas-PRB-05b,Villegas-PRB-05c,Stoll-PRB-02};
(ii) or to the presence of distinct features as peaks or plateaus
in the field dependence of the critical current $I_c(H_{ext})$ or
in the magnetization curve $M(H_{ext})$
\cite{Morgan-PRL-98,Morgan-JLTP-01,Hoffmann-PRB-00,vanBael-PRB-99,vanBael-PhysC-00a,vanBael-PhysC-00b,vanBael-JC-01,vanBael-PhysC-02,vanLook-PhysC-00,vanBael-PhysC-01,vanBael-PRB-03,Lange-JLTP-05,Lange-PRB-05}.

Such matching effects are also very well known for
spatially-modulated superconducting systems with antidot lattices
without ferromagnetic constituents (e.g., Baert {\it et al.}
\cite{Baert-PRL-95}, Moshchalkov {\it et al.}
\cite{Moshchalkov-PRB-98}, Hebard {\it et al.}
\cite{Hebard-IEEE-77}, Rosseel {\it et al.} \cite{Rosseel-PRB-96},
Harada {\it et al.} \cite{Harada-Science-96}, Metlushko {\it et
al.} \cite{Metlushko-PRB-99}). These periodic anomalies are
commonly explained in terms of commensurability effects between
the vortex lattice governed by the external field and the
artificially-introduced pinning potential at an integer number of
vortices per unit cell at $H_{ext}=\pm nH_1$ (with $n$ integer).
Typically these lithographically defined arrays are limited to a
minimal period of the unit cell of the order of a few hundred
nanometers, giving rise to a maximum matching field $H_1\simeq
10-10^2$ Oe. Alternative methods for introducing more closely
packed particles and thus larger $H_1$ values  can be achieved by
Bitter decoration (Fasano {\it et al.}
\cite{Fasano-PRB-99,Fasano-PRB-00}, Fasano and Menghini
\cite{Fasano-SuST-08}) or a diversity of self-assembled techniques
(Goyal {\it et al.} \cite{Goyal-SuST-05}, Villegas {\it et al.}
\cite{Villegas-PRL-07}, Welp {\it et al.}
\cite{Welp-PRB-02,Welp-PRB-05}, Vinckx {\it et al.}
\cite{Vinckx-PhysC-07}, Vanacken {\it et al.}
\cite{Vanacken-PhysC-08}).

    \begin{figure}[t!]
    \begin{center}
    \epsfxsize=80mm \epsfbox{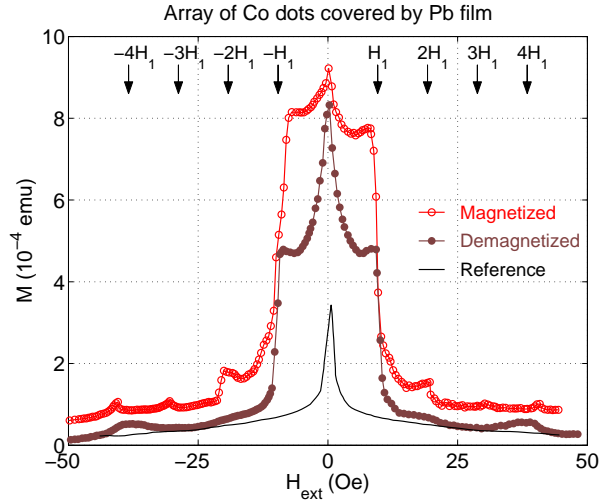}
    \end{center}
    \caption{(color online) Upper half of the magnetization loop $M$ vs. $H_{ext}$ at $T/T_{c0}=0.97$, for a
    superconducting Pb film (50 nm thickness) on top of a triangular lattice
    of Au/Co/Au dots (period 1.5 $\mu$m corresponding to a first matching field of 9.6 Oe)
    before and after magnetizing the dots
    (filled and open symbols, respectively), and for a reference Pb
    50 film (solid line), adapted from
    van Bael {\it et al.} \cite{vanBael-PRB-99}.
    The curve for the magnetized dots is slightly shifted upwards for clarity.}
    \label{Fig:vanBael-PRB-99}
    \end{figure}

\begin{figure}[t!]
    \begin{center}
    \epsfxsize=80mm \epsfbox{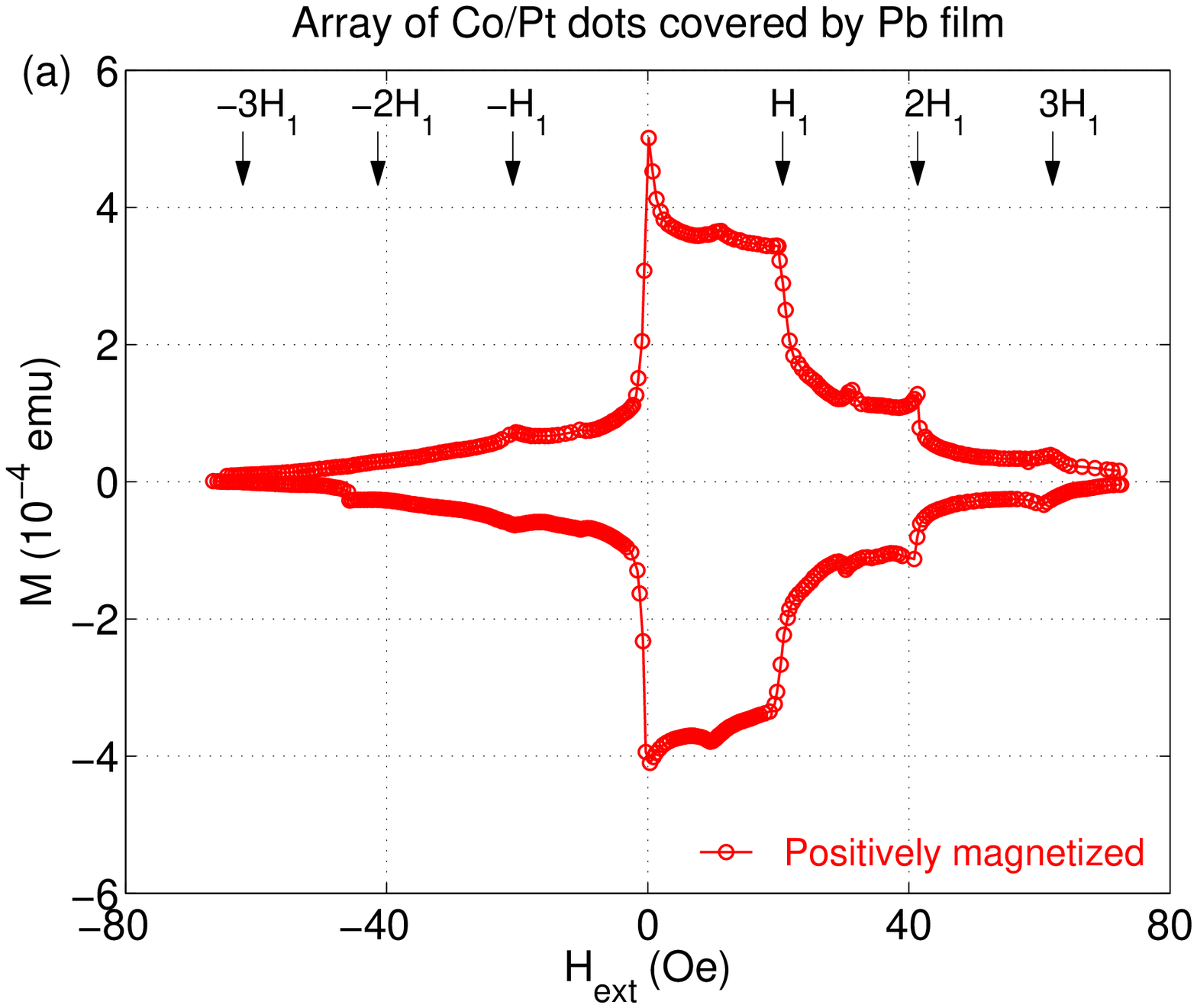}
    \epsfxsize=80mm \epsfbox{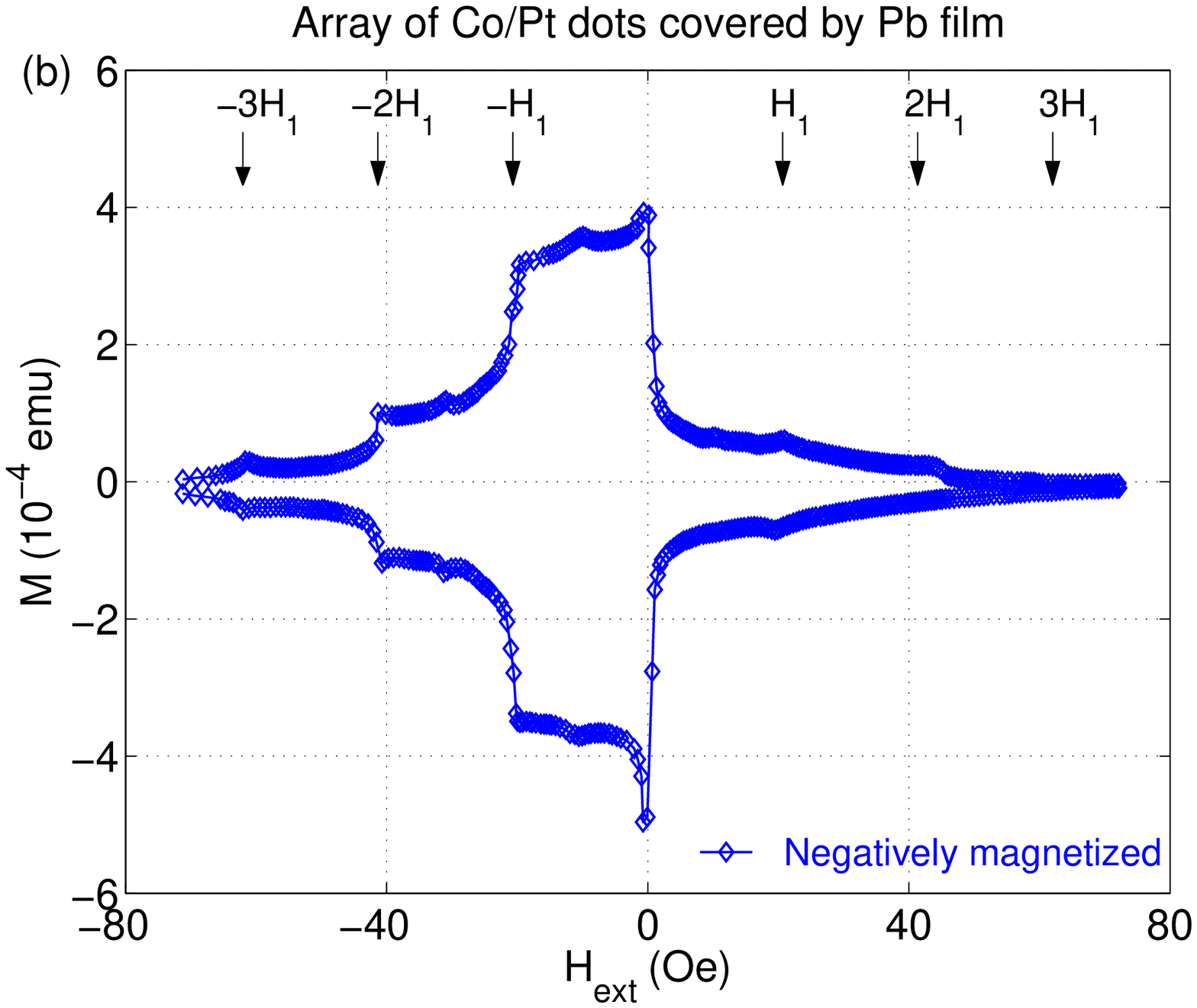}
    \end{center}
    \caption{(color online) Magnetization curves  $M$ vs. $H_{ext}$ at
    $T=7$ K ($T_{c0}=7.17$ K, $T/T_{c0}=0.976$) for a
    superconducting Pb film (50 nm thickness) on top of a Co/Pt dot
    array (the period is 1.0 $\mu$m, the first matching field is
    20.68 Oe) with all dots aligned in the positive (a) and
    negative (b) direction, adapted from
    van Bael {\it et al.} \cite{vanBael-PRB-03}.}
    \label{Fig:vanBael-PRB-03}
    \end{figure}

\begin{figure}[b!]
    \begin{center}
    \epsfxsize=80mm \epsfbox{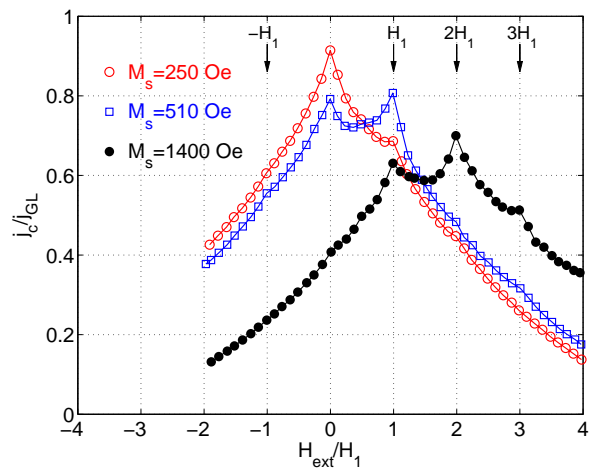}
    \end{center}
    \caption{(color online) Field dependence of
    the critical depinning current $j_c$, calculated for a periodic array of out-of-plane magnetized dots.
    The values of the dot's magnetization are indicated in the
    plot.  $j_{GL}$ is the density
    of the depairing (Ginzburg-Landau) current, $T/T_{c0}=0.9$, adapted from
    Milo\v{s}evi\'c and Peeters \cite{Milosevic-EPL-05}. It is worth noting that (i)
    $j_c(H_{ext})$ is asymmetric similar to that shown in Fig.
    \ref{Fig:vanBael-PRB-03}, and
    (ii) the quantized displacement of the $j_c$ maximum
    toward nonzero $H_{ext}$ value is sensitive to the magnetization of the dots.}
    \label{Fig:Milosevic-EPL-2005}
    \end{figure}

The field dependence of the critical current $I_c(H_{ext})$ and
magnetization $M(H_{ext})$, can be symmetrical or asymmetrical
with respect to $H_{ext}=0$ depending on the dot's net
magnetization (compare Figs. \ref{Fig:vanBael-PRB-99} and
\ref{Fig:vanBael-PRB-03}). The latter takes place for arrays of
magnetic dots with out-of-plane magnetization
\cite{vanBael-PhysC-00a,vanBael-PhysC-00b,vanBael-JC-01,vanBael-PhysC-02,vanBael-PhysC-01,vanBael-PRB-03,Lange-JLTP-05,Lange-PRB-05}.
Indeed, vertically magnetized dots with average magnetic moment
$\langle{\bf m}\rangle_z>0$, similar to point magnetic dipoles,
produce a stronger pinning potential for vortices (at $H_{ext}>0$
when \mbox{$\langle{\bf m}\rangle \parallel {\bf H}_{ext}$}) than
for antivortices ($H_{ext}<0$). In contrast to that, in-plane
magnetized dots are able to pin vortices and antivortices at the
magnetic poles equally well [see Eq. (\ref{Uvm}) and Fig.
\ref{Fig:MagneticPinning}]. This explains the experimentally
observed field polarity-dependent (asymmetric) pinning for arrays
of out-of-plane magnetized particles (Fig.
\ref{Fig:vanBael-PRB-03} and Fig. \ref{Fig:Milosevic-EPL-2005}).

The interaction between vortices and a periodic array of hard
magnetic dots on top or underneath a plain superconducting film
within the London approximation\footnote[1]{This issue
    seems to be part of a more general problem of the interaction of vortex matter
    with a periodic potential regardless the nature of the pinning in
    the superconducting system (see, e.g., Reichhardt {\it et al.}
    \cite{Reichhardt-PRB-98,Reichhardt-PRL-00,Reichhardt-PRB-00,Reichhardt-PRB-01a,Reichhardt-PRB-01b}
     and references therein).
    In this review we discuss only the results obtained for the S/F hybrids,
    keeping in mind that similar effects can be observed for non-magnetic patterned superconductors as well.}
was theoretically analyzed by Helseth \cite{Helseth-PRB-02b},
Lyuksyutov and Pokrovsky \cite{Lyuksyutov-PRL-98},
\v{S}\'{a}\v{s}ik and Hwa \cite{Sasik-CondMat-00}, Erdin
\cite{Erdin-PhysC-03}, Wei \cite{Wei-PRB-05,Wei-PhysC-06}, Chen
{\it et al.} \cite{Chen-PRB-06a}. These calculations show that at
$H_{ext}=0$ for out-of-plane magnetized dots, vortex-antivortex
pairs can be created in thin-film superconductors with the
vortices always sitting on top of the magnetic dot and the
antivortices located in between the dots. For in-plane magnetized
dots (or magnetic bars), the vortex and antivortex will be located
at opposite sides of the magnetic dots as described above for
individual magnetic dipoles. Unlike the case of an isolated
dipole, the threshold magnetization value needed to create a
vortex-antivortex pair is also a function of the period of the
lattice (Milo\v{s}evi\'{c} and Peeters \cite{Milosevic-PRL-04}).
Direct visualization of vortex-antivortex pairs via scanning Hall
probe microscopy was achieved for a square array of in-plane dots
by van Bael {\it et al.} \cite{vanBael-PRL-01} and for
out-of-plane dots by van Bael {\it et al.} \cite{vanBael-PRB-03}
and Neal {\it et al.} \cite{Neal-PRL-07}.

It is known that the preferred vortex configuration in a
homogeneous defect-free superconducting film should be close to a
triangular lattice because of the repulsive vortex-vortex
interaction \cite{Abrikosov-book,Schmidt-book,Tinkham-book}. The
artificially-introduced pinning appears to be the most effective
provided that each vortex is trapped by a pinning center, i.e.
when the symmetry of the pinned vortex lattice coincides with that
imposed by the topology of the internal magnetic field. The
transition between square and distorted triangular vortex lattice,
induced by variation of the strength of the periodic pinning
potential and the characteristic length scale of this interaction,
was considered by Pogosov {\it et al.} \cite{Pogosov-PRB-03} for
superconductors with a square array of pinning centers.
Experimentally the field-induced reconfiguration of the vortex
lattice (from rectangular to square) for superconducting Nb films
and rectangular arrays of circular magnetic Ni and Co dots was
reported by Mart\'{\i}n {\it et al.} \cite{Martin-PRL-99} and
Stoll {\it et al.} \cite{Stoll-PRB-02} as an abrupt increase of
the period of the oscillation in the $\rho(H_{ext})$ dependence
(resulting from the shrinkage of the period of the vortex lattice)
and decrease of the amplitude of such oscillations (due to a
weakening of the effective pinning) while increasing $|H_{ext}|$.

\begin{figure}[t!]
    \begin{center}
    \epsfxsize=80mm \epsfbox{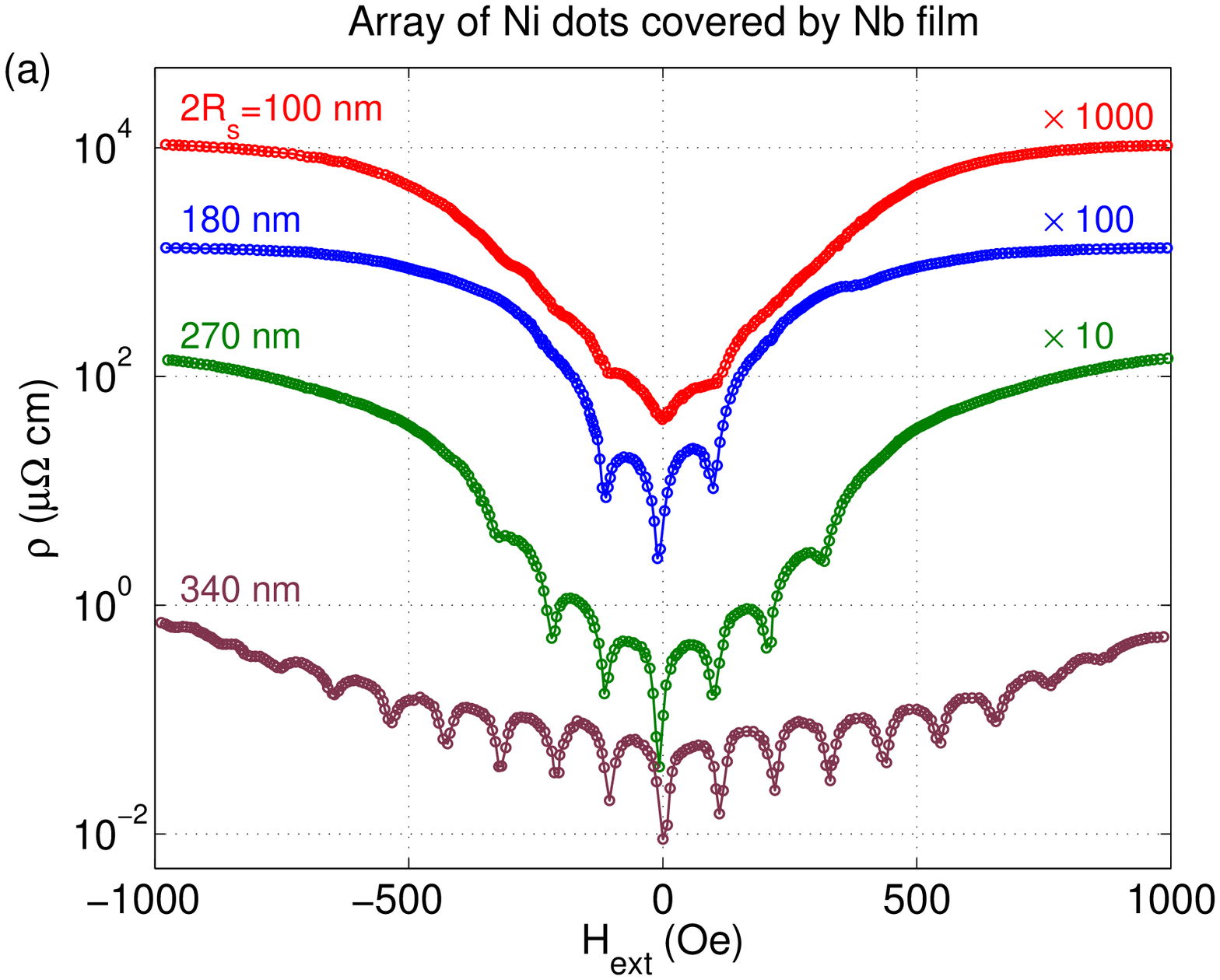}
    \epsfxsize=80mm \epsfbox{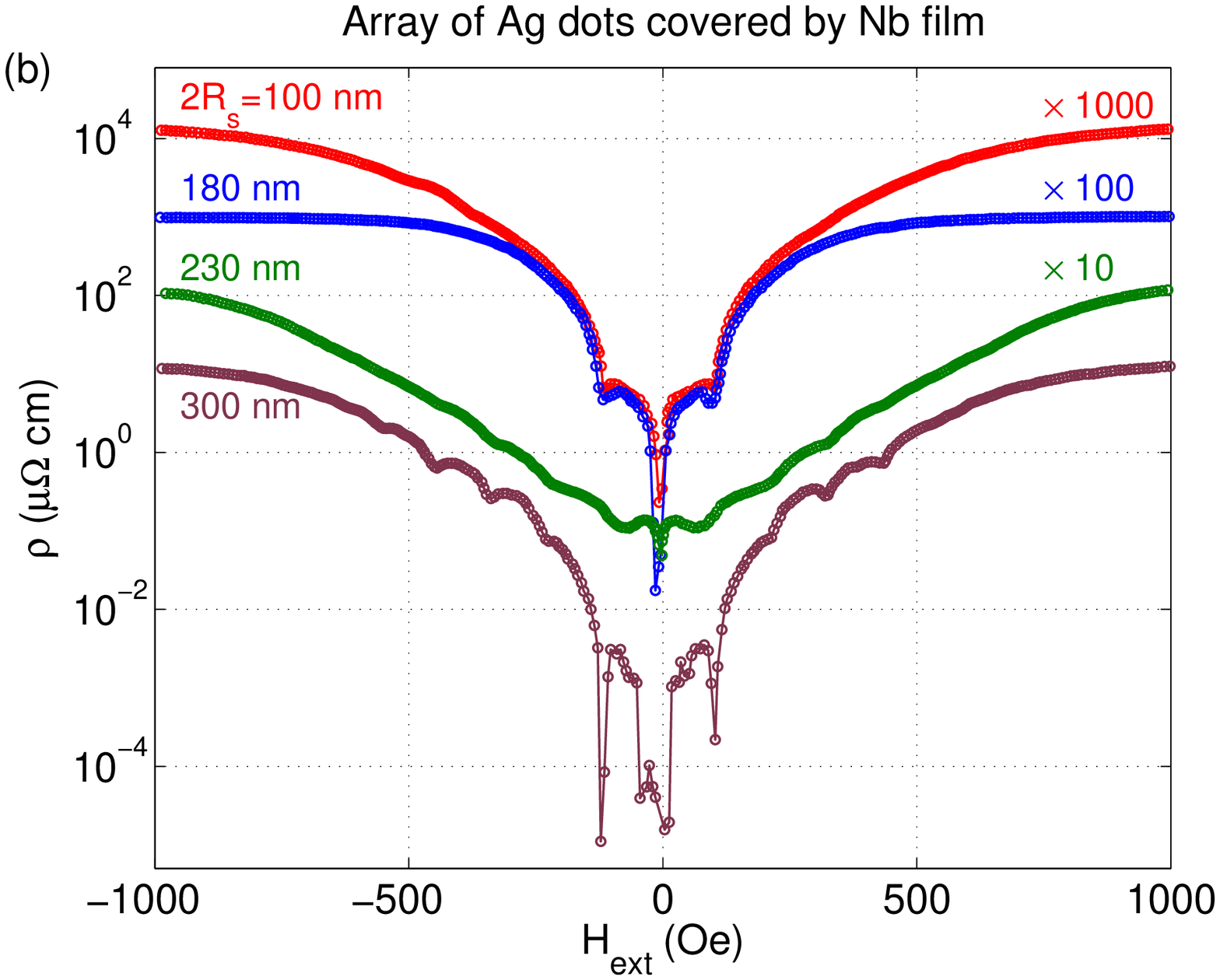}
    \end{center}
    \caption{(color online) (a) Field dependence of the electrical resistance
    $\rho$ for a superconducting Nb films covering periodic arrays of the magnetic Ni dots with different
    dot's diameter (indicated in the plot),
    but for the same lattice constant 400 nm, adapted from
    Hoffmann {\it et al.} \cite{Hoffmann-PRB-00}. The curves are shifted by factors of 10 from each other for
    clarity.\newline
    (b) Dependences $\rho(H_{ext})$ for samples with different diameters of non-magnetic Ag dots, adapted from
    Hoffmann {\it et al.} \cite{Hoffmann-PRB-00}.}
    \label{Fig:Hoffmann-PRB-00}
    \end{figure}

The dependence of the magnetic pinning in superconducting Nb films
on the diameter of the Ni dots was studied by Hoffmann {\it et
al.} \cite{Hoffmann-PRB-00}. They found that more minima appear in
the magnetoresistance (or maxima in the critical current) as the
lateral dot's size increases, indicating thus an enhanced pinning
(the panel (a) in Fig. \ref{Fig:Hoffmann-PRB-00}). This effect can
be caused by the two parameters which increase with the dot size:
the total magnetic moment $\langle{\bf m}\rangle$ (proportional to
the dot's volume $V_f=\pi R_f^2 D_f$) and the area on the order of
$\pi R_f^2$ where superconductivity might be locally suppressed
due to the high stray field or proximity effect. In addition,
larger magnetic dots can stabilize giant vortices carrying more
than one flux quantum.

\vspace*{0.3cm} \noindent {\it Periodic arrays of magnetic
antidots}

\noindent The antipode of arrays of magnetic dots is a perforated
ferromagnetic film (so-called magnetic antidots), which also
produces a periodic magnetic field. This system can be regarded as
the limiting case of big magnetic dots with a diameter larger than
the period of the periodic lattice.

\begin{figure}[t!]
    \begin{center}
    \epsfxsize=80mm \epsfbox{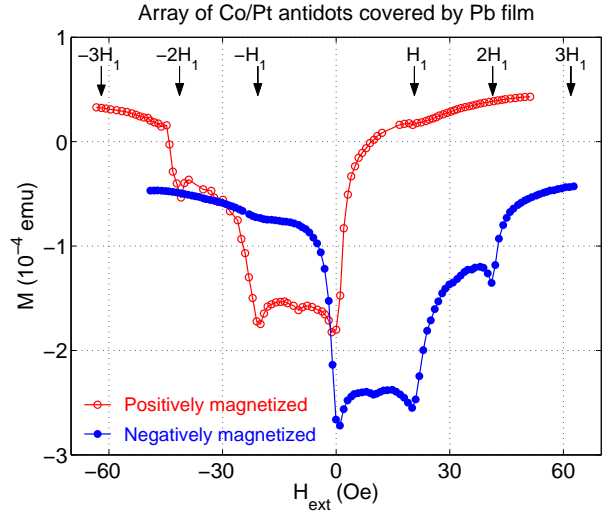}
    \end{center}
    \caption{(color online) Magnetization curves $M(H_{ext})$ at $T=7.05$ K
    ($T_{c0}=7.20$ K, $T/T_{c0}=0.972$) of a superconducting Pb
    film on top of a magnetic Co/Pt antidot lattice (the period is 1.0 $\mu$m, the first matching field is
    20.68 Oe) after saturation in a
    positive field ($M_z>0$, open circles) and after saturation in a negative
    field ($M_z<0$, filled circles), adapted from Lange {\it et
    al.} \cite{Lange-EPL-01,Lange-EPL-02}.}
    \label{Fig:Lange-EPL-02}
    \end{figure}

Magnetic antidots in multilayered Co/Pt films, characterized by an
out-of-plane remanent magnetization, and their influence on the
superconducting properties of Pb films were studied by Lange {\it
et al.}
\cite{Lange-JLTP-05,Lange-EPL-01,Lange-EPL-02,Lange-JMMM-02}. From
magnetostatic considerations, such submicron holes in a
ferromagnetic thin film generates a very similar field pattern as
an array of magnetic dots of the same geometry, but with opposite
sign. As a consequence, the enhanced magnetic pinning and the
pronounced commensurability peaks in the $M(H_{ext})$ dependence
are observed for the opposite polarity of the external field (i.e.
at $H_{ext}<0$ for positively magnetized film and vice versa), see
Fig. \ref{Fig:Lange-EPL-02}. However the matching effects are
considerably weakened in the demagnetized state of the Co/Pt
multilayer with holes as compared with the demagnetized array of
magnetic dots, thus indicating that an irregular domain structure
effectively destroys a long-range periodicity
\cite{Lange-JLTP-05}.

Van Bael {\it et al.} \cite{vanBael-JAP-02} and Raedts {\it et
al.} \cite{Raedts-PhysC-02} explored perforated Co film with
in-plane anisotropy. In this case the magnetic field distribution
becomes non-trivial since such magnetic antidots effectively pin
magnetic domain walls which generate a rather strong magnetic
field. As a result, neither matching effects nor pronounced
asymmetry were observed in the magnetization curves of the
superconducting layer, but only an overall enhancement of the
critical current after the sample was magnetized along the
in-plane easy axis, in comparison with the demagnetized state.

\vspace*{0.3cm} \noindent {\it Anisotropy of the transport characteristics and guidance of vortices}

\noindent In any periodic array of pinning centra, transport
properties such as magnetoresistance $\rho$($H$) and the critical
current $J_c$($H$), exhibit a dependence not only on the absolute
value of $H_{ext}$, but also on the direction of the applied
transport current with respect to the principal translation
vectors of the periodic pinning array (Villegas {\it et al.}
\cite{Villegas-PRB-03,Villegas-PRB-05c}, Soroka and Huth
\cite{Soroka-LTP-02}, V\'elez {\it et al.} \cite{Velez-PRB-02b},
Silhanek {\it et al.} \cite{Silhanek-PRB-03}, W\"{o}rdenweber {\it
et al.} \cite{Wordenweber-PhysC-04}). Interestingly, the direction
of the Lorentz force \mbox{${\bf f}_L=c^{-1} [{\bf j}\times {\bf
B}]$} and the drift velocity of the vortex lattice do not
generally coincide. It was demonstrated that for rectangular
arrays of magnetic dots the minimum of resistivity corresponds to
a motion of the vortex lattice along the long side of the array
cell. Such behavior was predicted by Reichhardt {\it et al.}
\cite{Reichhardt-PRB-01a} by numerical simulations indicating that
a rectangular array of pinning centers induces an easy direction
of motion for the vortex lattice (and larger dissipation as well)
along the short side of the array cell.

\begin{figure}[t!]
    \begin{center}
    \epsfxsize=80mm \epsfbox{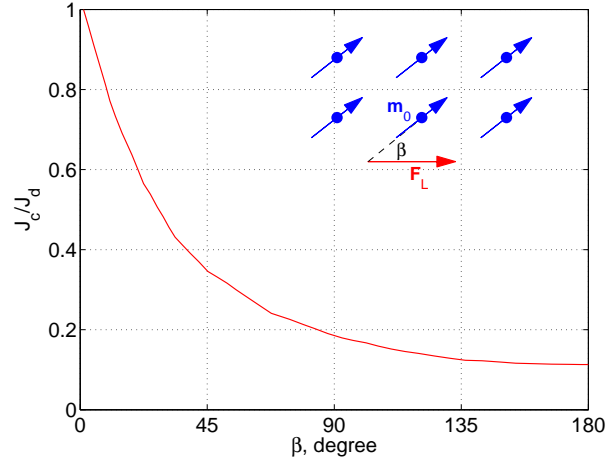}
    \end{center}
    \caption{(color online) Critical current, $J_c$, for a vortex pinned by a dipole
    array as a function of the angle $\beta$ between the magnetic moments and the
    driving force, adapted from Carneiro \cite{Carneiro-PhysC-05}. Here $J_d$
    is the depairing current.}
    \label{Fig:Carneiro-PhysicaC-05}
    \end{figure}

\begin{figure*}[t!]
    \begin{center}
    \epsfxsize=140mm \epsfbox{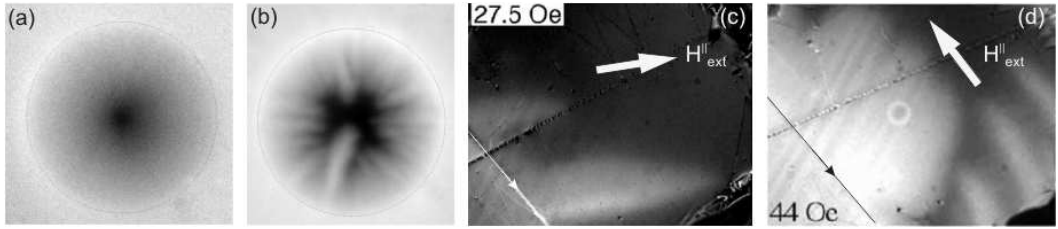}
    \end{center}
    \caption{(a) Magneto-optical image of a non-patterned superconducting Pb
    disk
    at $T=2$ K and $H_{ext}=50$ Oe, demonstrating an isotropic flux penetration, after Gheorghe {\it et al.}
    \cite{Gheorghe-PRB-08}. White corresponds to a high local magnetic field and black to zero local field.
    (b) Magneto-optical image of a circular Pb sample decorated by fully magnetized Co/ Pt dots,
    obtained at $T=2$ K and $H_{ext}=72$ Oe, after Gheorghe {\it et al.}
    \cite{Gheorghe-PRB-08}. The external field is applied parallel to the dot's magnetic moment. \newline
    (c) Magneto-optical image of flux entry in a superconducting MoGe
    film at $H^{||}_{ext}=16.5$ Oe, $T=4.5$ K following the preparation of
    stripe domain structures in a permalloy film by turning on and
    off an in-plane field of $H^{\parallel}_{ext}=1$ kOe at an angle of
    45$^{\circ}$ with respect to the sample edge at $T>T_{c0}$, after Vlasko-Vlasov {\it et al.}
    \cite{Vlasko-Vlasov-PRB-08a}. The  brightness of the magneto-optical contrast corresponds to the
    vortex density. The large yellow arrow
    shows the preferential flux entry direction coinciding with the
    direction of the stripe domains in the Py film. The thin solid line with arrow
    marks the sample edge. (d) Same as in panel (c) after application and switching off of
    $H^{\parallel}_{ext}=1$ kOe along the sample edge at $T>T_{c0}$, after
    Vlasko-Vlasov {\it et al.} \cite{Vlasko-Vlasov-PRB-08a}.}
    \label{Fig:Gheorghe-Vlasko-MO}
    \end{figure*}

Similar anisotropic transport properties was studied by Carneiro
\cite{Carneiro-PhysC-05} for the case of a periodic array of
in-plane magnetic dipoles. In order to illustrate the angular
dependence of the critical depinning current on the angle $\beta$
between the direction of the injected current and the magnetic
moment of in-plane oriented dipoles we refer to Fig.
\ref{Fig:Carneiro-PhysicaC-05}. Interestingly, Verellen {\it et
al.} \cite{Verellen-APL-08} showed that this resulting guided
vortex motion in square arrays of magnetic rings can be rerouted
by 90$^\circ$ simply by changing the dipole orientation or can
even be suppressed by inducing a flux-closure magnetic vortex
state with very low stray fields in the rings. Similar anisotropic
vortex motion was recently observed in Nb films with a periodic
array of one-dimensional Ni lines underneath by Jaque {\it et al.}
\cite{Jaque-APL-02}. The mentioned channelling of vortices lead to
an anisotropic vortex penetration that has been directly
visualized by means of magnetooptics experiments [Gheorghe {\it et
al.} \cite{Gheorghe-PRB-08}, see Fig.
\ref{Fig:Gheorghe-Vlasko-MO}(b)].

\vspace*{0.2cm} \noindent {\it Mechanisms of pinning in S/F hybrids}

\noindent It should be noticed that the magnetic pinning
originating from the spatial modulation of the ``internal"
magnetic field generally competes with so-called core pinning
resulting from structural inhomogeneities in real samples (either
regular or random defects). In addition to random intrinsic
pinning, the fabrication of an array of magnetic particles
naturally leads to an alteration of the local properties of the
superconducting film (e.g., due to proximity effects, corrugation
of the superconducting layer or local suppression of the critical
temperature). As a consequence, both magnetic and structural
modulation share the same periodicity, and a clear identification
of the actual pinning type becomes difficult.

A direct comparison of the pinning properties of arrays of
magnetic vs. nonmagnetic dots have been addressed by Hoffmann {\it
et al.} \cite{Hoffmann-PRB-00} and Jaccard {\it et al.}
\cite{Jaccard-PRB-98}. These reports show that even though both
systems display commensurability features, the pinning produced by
magnetic arrays of Ni dots is substantially stronger than that
produced by non-magnetic Ag particles (Fig.
\ref{Fig:Hoffmann-PRB-00}). In our opinion, the main issue whether
the enhanced pinning for the sample with ferromagnetic Ni dots
actually arises from purely magnetic interactions and not from an
additional suppression of the local critical temperature, e.g. due
to the enhanced magnetic field near magnetic dots, remains
unclear. In principle, the most straightforward way to distinguish
the two competing pinning mechanisms is the mentioned
field-polarity of the magnetic pinning for the S/F hybrids with
dots magnetized perpendicularly, i.e. exploring the broken
field-polarity symmetry.

\begin{figure}[b!]
    \begin{center}
    \epsfxsize=80mm \epsfbox{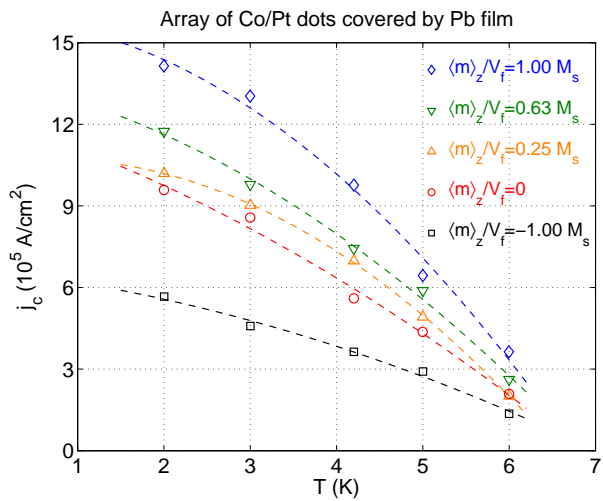}
    \end{center}
    \caption{(color online) Temperature dependence of the critical
    current density $j_c$, estimated from magneto-optical images, for
    a superconducting Pb film with square array of the ferromagnetic Co/Pt dots on top, in
    various magnetic states of the dots:
    demagnetized ($\circ$), fully magnetized parallel configuration ($\diamond$), fully magnetized
    antiparallel configuration (square), partially magnetized
    parallel, $\langle m \rangle_z=0.25\, M_s V_f$ ($\Delta$) and
    $\langle m \rangle_z=0.63\, M_s V_f$
    ($\nabla$), adapted from   Gheorghe {\it et al.}
    \cite{Gheorghe-PRB-08}. The dashed lines are guides to the eye.}
    \label{Fig:Gheorghe-PRB-07}
    \end{figure}

Clear evidence of the field-polarity dependent pinning properties
has been reported by Gheorghe {\it et al.} \cite{Gheorghe-PRB-08}
in Pb films on top of a square array of [Co/Pt]$_{10}$ dots with a
well defined out-of-plane magnetic moment. In this work the
authors show that the critical current of the hybrid system can be
increased by a factor of 2 when the magnetic dots are switched
from low stray field in the demagnetized state (disordered
magnetic moment) to high stray field in the magnetized state
(nearly single domain state) at temperatures as low as
\mbox{$T\simeq0.3\,T_{c0}$} (see Fig. \ref{Fig:Gheorghe-PRB-07}).
Additional evidence of an increase of the critical current at low
temperatures (far from the superconducting/normal phase boundary)
produced by magnetic dots was reported by Terentiev {\it et al.}
\cite{Terentiev-PhysC-99,Terentiev-PhysC-00,Terentiev-PRB-00}.

\vspace*{0.3cm} \noindent {\it Tunable pinning centers}

\begin{figure}[t!]
    \begin{center}
    \epsfxsize=80mm \epsfbox{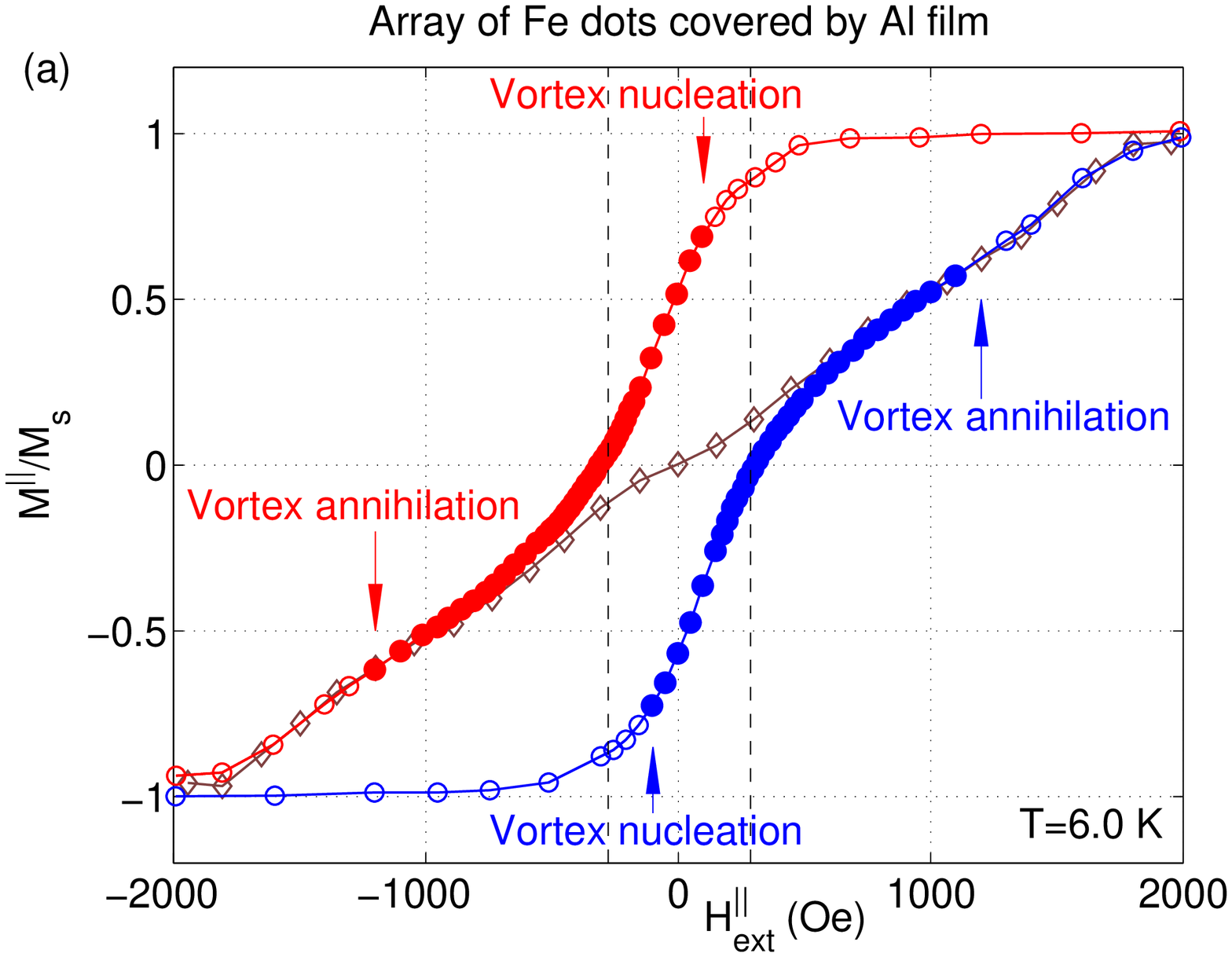}
    \epsfxsize=80mm \epsfbox{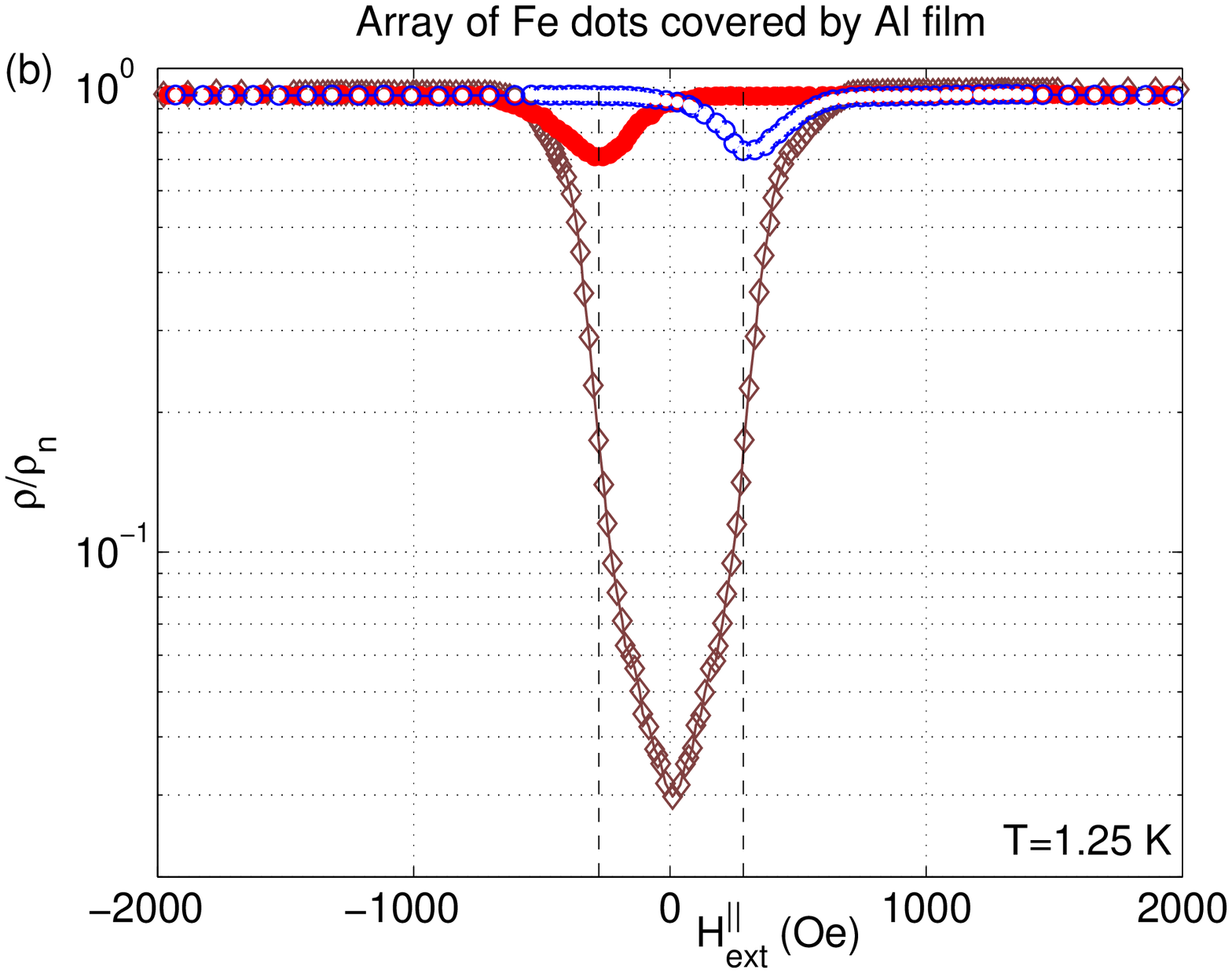}
    \end{center}
    \caption{(color online) (a) Normalized magnetization $M^{\parallel}/M_s$ vs. in-plane
    applied field $H^{\parallel}_{ext}$ ($M_s$ is the saturated magnetization)
    for an array of Fe dots with average diameter of 140 nm and
    average interdot distance of 180 nm measured at $T=6$ K (above the critical
    temperature of superconducting Al film), adapted from
    Villegas {\it et al.} \cite{Villegas-PRL-07}. Brown diamonds correspond to a virgin state of the Fe dots,
    the magnetic state depicted by red (blue) circles obtained after saturation in positive (descending branch)
    and negative (ascending branch) magnetic fields, respectively.
    The filled circles schematically show the regions where the
    magnetic vortex is expected to take place. Vertical dashed lines mark the coercive fields. \newline
    (b) Normalized resistivity $\rho$ vs. in-plane applied field $H^{\parallel}_{ext}$ for the same
    sample at $T=1.25$ K (below the critical temperature of superconducting Al film), $\rho_n$ is the normal-state
    resistance. Open (blue) and filled (red) circles mark the curves measured from negative and positive saturation
    respectively, while brown diamonds correspond to the virgin
    state, adapted from
    Villegas {\it et al.} \cite{Villegas-PRL-07}.}
    \label{Fig:Villegas-PRL-07}
    \end{figure}

\noindent An apparent advantage of using magnetic pinning centra
is their flexibility (tunability) in contrast to core pinning on
structural inhomogeneities. Indeed, according to Eq.
(\ref{Uvm-dot}) the magnetic pinning should be sensitive to the
particular distribution of magnetization inside the ferromagnetic
elements. Depending on the geometrical details of the dot and the
magnetic anisotropy of the chosen material a huge variety of
magnetic states can be found. For instance, domain formation is
expected to be suppressed for structures with lateral dimensions
smaller than tens of nm (Raabe {\it et al.} \cite{Raabe-JAP-00}),
whereas for larger sizes the magnetic sample breaks into domains
of different orientation (Seynaeve {\it et al.}
\cite{Seynaeve-JAP-01}). The exact transition from single domain
to multi-domain structures depends on the shape, dimensions,
temperature and particular material, among other parameters. More
recently, Villegas {\it et al.}
\cite{Villegas-PRL-07,Villegas-PRB-08} and Hoffmann {\it et al.}
\cite{Hoffmann-PRB-08} experimentally investigated the switching
of the ferromagetic dots from single domain to magnetic vortex
state while sweeping the external field and the influence of their
stray fields on the resistivity of the S/F hybrid sample (Fig.
\ref{Fig:Villegas-PRL-07}). The interaction between a vortex in a
superconducting film and a magnetic nanodisk in the magnetic
vortex state was studied theoretically by Carneiro
\cite{Carneiro-PRB-07}. For magnetic dots big enough to host a
multi-domain state it is possible to tune the average magnetic
moment by partially magnetizing the sample in a field lower than
the saturation field or even recover the virgin state by
performing a careful degaussing procedure similar to that shown in
Fig. \ref{Fig:Gillijns-PRB-07-abc} (Gillijns {\it et al.}
\cite{Gillijns-PRB-07a,Gillijns-PRB-06}, Lange {\it et al.}
\cite{Lange-APL-02}). Interestingly, recently Cowburn {\it et al.}
\cite{Cowburn-PRL-99} showed that small disks of radius about 50
nm made of supermalloy (Ni-80\%, Fe-14\%, Mo-5\%) lie in a single
domain state with the magnetization parallel to the disk plane and
with the property that their direction can be reoriented by small
applied fields. This systems represents the closest experimental
realization of in-plane free-rotating dipoles, which was
theoretically analyzed by Carneiro
\cite{Carneiro-PRB-05,Carneiro-EPL-05,Carneiro-PhysC-06} within
the London formalism.

Whatever the mechanism of pinning produced by the magnetic dots,
either core or electromagnetic, it is now a clearly established
fact that changing the domain distribution in each dot has
profound effects on the superconducting pinning properties as
demonstrated, for example, by van Bael {\it et al.}
\cite{vanBael-PRB-99}, van Look {\it et al.}
\cite{vanLook-PhysC-00}. This result points out the importance of
performing a careful study of the magnetic properties of the dots
in order to identify the domain size, shape, distribution, and
stable states. Van Bael {\it et al.} \cite{vanBael-PRB-99}
presented the first report directly linking changes in the
hysteresis loop of a superconducting Pb film when the underlying
submicron Co islands are switched from \mbox{$2\times 2$}
checkerboard magnetic domain pattern to single domain structures.

    \begin{figure*}[t!]
    \begin{center}
    \epsfxsize=120mm \epsfbox{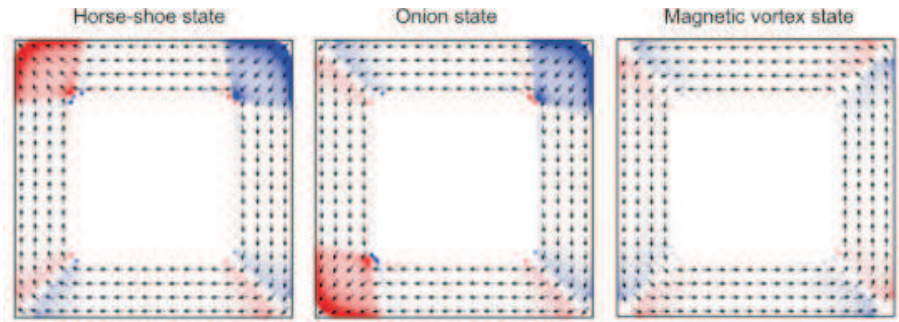}
    \end{center}
    \caption{(color online) Different magnetic states realized in square permalloy micro-loops depending on
    the direction of the applied magnetic field, by courtesy of Metlushko {\it et
    al.}}
    \label{Fig:Metlushko-private}
    \end{figure*}

As we pointed out above, unfortunately both, the multidomain state
and the magnetic vortex state, still involve a sizable component
of the magnetic stray field which eventually influences the
response of the superconducting properties by locally suppressing
the order parameter. In other words, it is actually not possible
to completely switch off the magnetic pinning using singly
connected structures. It has been recently shown that a way to
partially circumvent this limitation can be achieved by using
multiply connected ring-like magnetic structures. In this case, if
a flux-closure state is induced in the magnetic ring, in
principle, there is almost no stray field present, besides small
fields due to domain walls at the sharp corners of the ring.
Indeed, a two-dimensional magnetic material of ring like shape of
group symmetry $C_n$ can be set in two flux closure states of
opposite chirality and $n(n-1)$ different polarized states. In a
square loop, for instance, 12 states corresponding to six
different dipole directions with two opposite dipole orientations
are expected. If the net dipolar moment is parallel to the side of
the square, the final state is named horseshoe state whereas if
the dipole is along the diagonal of the square, it is called onion
state. Figure \ref{Fig:Metlushko-private} shows the different
topologically non-equivalent magnetic states for a square ring of
magnetic material with in-plane magnetization. Clear experimental
evidence of ON/OFF magnetic pinning potentials induced by these
type of multiply-connected ferromagnetic structures have been
demonstrated by Silhanek {\it et al.}
\cite{Silhanek-APL-06,Silhanek-APL-07,Silhanek-PhysC-08}. It is
worth mentioning that the S/F structures investigated in Ref.
\cite{Silhanek-APL-07} exhibit two well distinguished phases
corresponding to a disordered phase when the sample is in the
as-grown state and an ordered phase when the sample is magnetized
with an in-plane field. These order-disorder transitions manifest
themselves as an enhancement of submatching features in the field
dependence of the critical current which cannot be explained from
a simple rescaling of the response corresponding to the disordered
phase.

\vspace*{0.3cm} \noindent {\it Random (disordered) magnetic inclusions}

Early studies of the influence of ferromagnet on the
superconducting state were performed in the sixties by Strongin
{\it et al.} \cite{Strongin-RMP-64}, Alden and Livingston
\cite{Alden-APL-66,Alden-JAP-66} and Koch and Love
\cite{Koch-JAP-69} for a dispersion of fine ferromagnetic
particles (Fe, Gd, Y) in a superconducting matrix. These reports
motivated further experimental and theoretical investigations of
the influence of randomly distributed particles
on/underneath/inside superconducting materials, which continue
nowadays (Sikora and Makiej \cite{Sikora-PSS-82,Sikora-PSS-85},
Wang {\it et al.} \cite{Wang-IEEE-97}, Lyuksyutov and Naugle
\cite{Lyuksyutov-MPLB-99,Lyuksyutov-IJMPB-03a,Lyuksyutov-IJMPB-03b},
Santos {\it et al.} \cite{Santos-PRB-01}, Kuroda {\it et al.}
\cite{Kuroda-SuST-06}, Togoulev {\it et al.}
\cite{Togoulev-PhysC-06}, Kruchinin {\it et al.}
\cite{Kruchinin-SuST-06}, Palau {\it et al.}
\cite{Palau-SuST-07,Palau-PRL-07}, Haindl {\it et al.}
\cite{Haindl-PhysC-07,Haindl-SuST-08}, Snezhko {\it et al.}
\cite{Snezhko-PRB-05}, Rizzo {\it et al.} \cite{Rizzo-APL-96},
Stamopoulos {\it et al.}
\cite{Stamopoulos-PRB-04,Stamopoulos-SuST-04,Stamopoulos-PRB-05,Stamopoulos-SuST-05,Stamopoulos-PhysC-06},
Suleimanov  {\it et al.} \cite{Suleimanov-PhysC-04}, Xing {\it et
al.} \cite{Xing-PRB-08,Xing-CondMat-08,Xing-CondMat-09}). In most
of these investigations no precautions were taken to electrically
isolate the magnetic particles from the superconducting material
which presumably results in a substantial core pinning due to
proximity effects.


Xing {\it et al.} \cite{Xing-CondMat-08} reported on controlled
switching between paramagnetic\,\footnote[1]{The paramagnetic
Meissner
    effect in various superconducting systems is discussed in the review of Li \cite{Li-PhysRep-03}.}
and diamagnetic Meissner effect in S/F nanocomposites consisting
of Pb films with embedded single-domain Co particles. These
authors argue that in this particular system the paramagnetic
Meissner effect attributed to the superconducting part only,
originates from the spontaneous formation of vortices induced by
the ferromagnetic inclusions. Therefore, the different
contributions of the external field and the spontaneous vortices
to the resulting magnetization of the sample, make it possible to
manipulate the sign of Meissner effect by changing the orientation
of the magnetic moments embedded in the supercondicting
matrix\,\footnote[2]{Previously, Monton {\it et al.}
    \cite{Monton-PRB-07} reported on an experimental observation of
    the paramagnetic Meissner effect in Nb/Co superlattices in
    field-cooled measurements, however the origin of this effect
    remains unclarified.}.

\vspace*{0.3cm} \noindent  {\it Vortex dynamics in a periodic
magnetic field}

\noindent Here we want to briefly discuss the peculiarities of
low-frequency vortex dynamics in nonuniform magnetic fields.
Magnetic templates placed in the vicinity of a superconducting
film not only induces changes in the static pinning properties but
also in the overall dynamic response of the system. Lange {\it et
al.}~\cite{Lange-PRB-05} demonstrated that the vortex-antivortex
pairs induced by an array of out-of-plane magnetized dots lead to
a strong field polarity dependent vortex creep as evidenced in the
current-voltage characteristics. This result shows that in S/F
hybrids with perpendicular magnetized dots vortices and
antivortices experience a different pinning strength. A
theoretical study of the dynamic evolution of these interleaved
lattices of vortices and antivortices in the case of in-plane
point like dipoles has been recently addressed by Carneiro
\cite{Carneiro-arxiv-08} and Lima and de Souza Silva
\cite{Lima-arxiv-08}.

A more subtle effect, namely magnetic-dipole-induced voltage
rectification was predicted by Carneiro \cite{Carneiro-PhysC-05}.
Unlike conventional ratchet systems\,\footnote[3]{Early
theoretical studies showed that
    a vortex lattice submitted to an oscillatory excitation in the
    presence of a non-centro-symmetric pinning potential gives rise to
    a net drift $\langle{\bf v}\rangle$ of the vortex lattice which in
    turn generates a dc voltage signal $V_{dc}=\int \left[\langle{\bf
    v}\rangle\times{\bf H}_{ext}\right]\cdot {\bf dl}$ along the direction
    of bias current (Zapata {\it et al.} \cite{Zapata-PRL-96}, Lee
    {\it et al.} \cite{Lee-Nature-99}, Wambaugh {\it et al.}
    \cite{Wambaugh-PRL-99}). These predictions are in agreement with
    recent experimental results obtained for purely superconducting
    systems (Villegas {\it et al.} \cite{Villegas-Science-03},
    W\"{o}rdenweber {\it et al.} \cite{Wordenweber-PRB-04}, van de
    Vondel {\it et al.} \cite{vandeVondel-PRL-05}, Togawa {\it et al.}
    \cite{Togawa-PRL-05}, de Souza Silva {\it et al.}
    \cite{deSouzaSilva-Nature-06},  Wu {\it et al.}
    \cite{Wu-PhysC-06}, Aladyshkin {\it et al.}
    \cite{Aladyshkin-APL-08}). }, in the particular case of magnetic
ratchet, induced by in-plane magnetized dots, the motion of
vortices is in the opposite direction than the motion of
antivortices, thus giving rise to a field-polarity independent
rectification (Silhanek {\it et al.}
\cite{Silhanek-APL-07,Silhanek-PhysC-08}, de Souza Silva {\it et
al.} \cite{deSouzaSilva-PRL-07}). This magnetic dipole-induced
ratchet motion depends on the mutual orientation and strength of
the local magnetic moments thus allowing one to control the
direction of the vortex drift. In some cases, a nonzero rectified
signal is observed even at $H_{ext}=0$ resulting from the
interaction between the induced vortex-antivortex pairs by the
magnetic dipoles \cite{deSouzaSilva-PRL-07}. It is worth
emphasizing that in the case of in-plane magnetic dipoles treated
by Carneiro \cite{Carneiro-PhysC-05}, the inversion symmetry is
broken by the stray field of the dipoles, thus giving rise to
different depinning forces parallel and anti-parallel to the
dipoles orientations, as shown in Fig.
\ref{Fig:Carneiro-PhysicaC-05}.

    \begin{figure}[t!]
    \begin{center}
    \epsfxsize=80mm \epsfbox{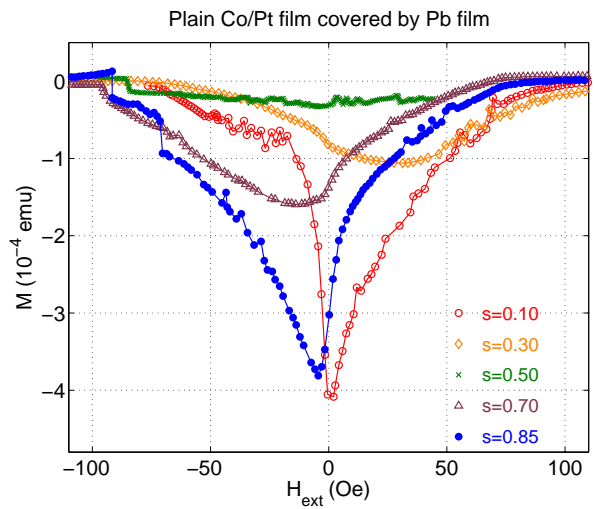}
    \end{center}
    \caption{(color online) Bottom parts of the magnetization curves $M$ vs. $H_{ext}$ for a superconducting Pb
    film covering a Co/Pt multilayer.
    The curves $M(H_{ext})$ corresponding to the different values of the parameters $s$,
    which is defined as the fraction of magnetic moments that are
    pointing up ($m>0$) relative to the total number of magnetic
    moments: $s=0.1$ (open circles), $s=0.3$ (diamonds), $s=0.5$ (crosses),
    $s=0.7$ (triangles) and $s=0.85$ (filled circles), adapted from  Lange {\it et
    al.} \cite{Lange-APL-02}.}
    \label{Fig:Lange-APL-02}
    \end{figure}

\subsection{Planar S/F bilayer hybrids}

In this section we shall discuss the properties of continuous
planar S/F structures which have macroscopically large lateral
dimensions. As before, the superconducting and ferromagnetic films
are assumed to be electrically insulated from each other.

\vspace*{0.3cm} \noindent  {\it Appearance of vortices in planar
S/F structures}

\noindent The interaction of the Meissner currents and the
currents induced by vortex lines with a one-dimensional
distribution of the magnetization (both single domain walls,
periodic domain structures and magnetic bars) in the London
approximation was considered by Sonin \cite{Sonin-PRB-02b}, Genkin
{\it et al.} \cite{Genkin-JMMM-94}, Bespyatykh and Wasilevski
\cite{Bespyatykh-PSS-01a}, Bespyatykh {\it et al.}
\cite{Bespyatykh-PSS-01b}, Helseth {\it et al.}
\cite{Helseth-PRB-02a}, Laiho {\it et al.} \cite{Laiho-PRB-03},
Traito {\it et al.} \cite{Traito-PhysC-03}, Erdin
\cite{Erdin-PRB-06}, Bulaevskii and Chudnovsky
\cite{Bulaevskii-PRB-00,Bulaevskii-PRB-02}, Kayali and Pokrovsky
\cite{Kayali-PRB-04b}, Burmistrov and Chtchelkatchev
\cite{Burmistrov-PRB-05}, Ainbinder and Maksimov
\cite{Ainbinder-SuST-07}, Maksimova {\it et al.}
\cite{Maksimova-PRB-06,Maksimova-PRB-08}. It was found that in
order to create vortex-antivortex pairs in the S/F bilayer with
out-of-plane magnetization at $H_{ext}=0$ (and thus keeping the
total flux through the superconducting film zero) the amplitude of
the magnetization $M_s$ should overcome the following threshold
value \cite{Bespyatykh-PSS-01a,Laiho-PRB-03,Traito-PhysC-03}
    \begin{eqnarray}
    \label{BW-1}
    \nonumber
    M_{v-av}^{\perp}=\frac{H_{c1}}{4\alpha}\, \frac{D_s}{w} \propto \frac{\Phi_0 D_s}{\lambda^2 w}\,\ln \lambda/\xi,
    \end{eqnarray}
where $\alpha$ is a numerical factor of the order of unity and the
period $w$ of the domain structure is assumed to be fixed (the
hard-magnet approximation). This estimate corresponds to the case
when the width of the domain walls is much smaller than other
relevant length scales. The critical magnetization
$M_{v-av}^{\perp}$ decreases monotonically with decreasing
superconducting film thickness $D_s$. The equilibrium vortex
pattern appearing in the superconducting film at $H_{ext}=0$ and
$M_s>M_{v-av}^{\perp}$ consists of straight vortices, arranged in
one-dimensional chains, with alternating vorticities corresponding
to the direction of the magnetization in the ferromagnetic domains
\cite{Bespyatykh-PSS-01a,Laiho-PRB-03,Traito-PhysC-03}. The
parameters of such vortex configuration with one or two vortex
chains per half-period was analyzed by Erdin \cite{Erdin-PRB-06}.
It was shown that in equilibrium the vortices in the neighboring
domains are halfway shifted, while they are next to each other in
the same domain. Alternatively, as the thickness $D_s$ increases,
the vortex configuration, consisting of vortex semi-loops between
the ferromagnetic domains with opposite directions of the
magnetization, becomes energetically favorable
\cite{Laiho-PRB-03,Traito-PhysC-03} provided that
$M_s>M_{loops}^{\perp}$, where
    \begin{eqnarray}
    \label{Laiho-1}
    \nonumber
    M_{loops}^{\perp}=\frac{H_{c1}}{8 \ln(w/\pi\lambda)} \propto \frac{\Phi_0}{\lambda^2}\,\frac{\ln
    \lambda/\xi}{\ln(w/\pi\lambda)}.
    \end{eqnarray}

The destruction of the Meissner state in the S/F bilayer with
in-plane magnetization was considered by Burmistrov and
Chtchelkatchev \cite{Burmistrov-PRB-05}. Since the out-of-plane
component of the field, which is responsible for the generation of
the vortex, is maximal near the domain wall (unlike from the
previous case) and goes to zero in the center of magnetic domains,
one can consider only a single domain wall. At $H_{ext}=0$ a
creation of a single vortex near the Bloch-type domain wall of
width $\delta$ corresponds to the condition
    \begin{eqnarray}
    \label{Burm-1}
    \nonumber
    M_{v}^{\parallel}\simeq\frac{H_{c1}}{4\pi}\,\frac{\lambda}{D_f}\,\times\, \left\{ 2\lambda/\delta
    ,
    \atop 1-32\lambda/(\pi^2\delta) , \right.
    \left.  \pi\delta/(4\lambda) \ll 1 \atop  \pi\delta/(4\lambda) \gg 1.  \right.
    \end{eqnarray}

\vspace*{0.3cm}

\vspace*{0.3cm} \noindent  {\it Magnetic pinning and guidance of
vortices in planar S/F structures}

\noindent Irrespective of whether the domain structure in the
ferromagnetic layer is spontaneously created or was present
beforehand, the spatial variation of the magnetization will lead
to an effective vortex pinning (Bespyatykh {\it et al.}
\cite{Bespyatykh-PSS-01b}, Bulaevskii {\it et al.}
\cite{Bulaevskii-APL-00}). However, there are discrepancies in the
estimates concerning the pinning effectiveness. Indeed, Bulaevskii
{\it et al.} \cite{Bulaevskii-APL-00} argued that
superconductor/ferromagnet multilayers of nanoscale period can
exhibit strong pinning of vortices by the magnetic domain
structure in magnetic fields below the coercive field when the
ferromagnetic layers exhibit strong perpendicular magnetic
anisotropy. The estimated maximum magnetic pinning energy for a
single vortex in such a system is about 100 times larger than the
core pinning energy produced by columnar defects. In contrast to
that, Bespyatykh {\it et al.} \cite{Bespyatykh-PSS-01b} have shown
that the effectiveness of magnetic pinning of vortices in a
layered system formed by an uniaxial ferromagnet, does not
considerably exceed the energy of artificial pinning by a
column-type defect, regardless the saturation magnetization of the
ferromagnet. The limitation of the pinning energy is caused by the
interaction of external vortices with the spontaneous vortex
lattice formed in the superconducting film when the magnetization
of the ferromagnetic film exceeds the critical value (see Eq.
\ref{BW-1}).

There have been numerous experimental investigations corroborating
the enhancement of the critical current in planar S/F hybrids. It
was shown that the presence of a bubble domain structure in Co/Pt
ferromagnetic films with out-of-plane magnetization modifies the
vortex pinning in superconducting Pb films (Lange {\it et al.}
\cite{Lange-JLTP-05,Lange-APL-02,Lange-MPLB-03,Lange-PhysC-04}),
leading to an increase of the width of the magnetization loop
$M(H_{ext})$ as compared with a uniformly magnetized S/F sample
(Fig. \ref{Fig:Lange-APL-02}). The crossover between an enhanced
magnetic pinning on bubble magnetic domains observed at low
temperatures and a suppressed magnetic pinning at temperatures
close to $T_c$ for a demagnetized S/F bilayer can be possibly
associated with an increase of an effective penetration length
$\lambda^2/D_s$, characterizing the vortex size, and an effective
averaging on the small-scale variation of the nonuniform magnetic
field provided that $\lambda^2/D_s$ considerably exceed the period
of the magnetic field (Lange {\it et al.} \cite{Lange-PhysC-04}).
Interestingly, the parameters of the bubble domain structure (the
size and the density of domains of both signs of magnetization)
can be controlled by demagnetization similar to that reported in
Refs. \cite{Gillijns-PRB-07a,Gillijns-PRB-06}. A three-fold
enhancement of the critical depinning current in Nb films
fabricated on top of ferromagnetic Co/Pt multilayers was observed
by Cieplak {\it et al.} \cite{Cieplak-APP-04,Cieplak-JAP-05} based
on magnetization measurements and on the analysis of the magnetic
field distribution obtained by using a 1D array of Hall sensors.
The mentioned enhancement of the magnetic pinning takes place in
the final stages of the magnetization reversal process, and it can
attributed to residual un-inverted dendrite-shaped magnetic
domains.

High-resolution magneto-optical imaging performed by Goa {\it et
al.} \cite{Goa-APL-03} in superconducting NbSe$_2$ single crystals
and ferrite-garnet films demonstrates that the stray field of
Bloch domain walls can be used to manipulate vortices. Indeed,
depending on the thickness of the sample, the vortices are either
swept away or merely bent by the Bloch wall.

Vlasko-Vlasov {\it et al.}
\cite{Vlasko-Vlasov-PRB-08a,Vlasko-Vlasov-PRB-08b} and Belkin {\it
et al.} \cite{Belkin-APL-08,Belkin-PRB-08} studied the anisotropic
transport properties of superconducting MoGe and Pb films and
NbSe$_2$ single crystals which are in the vicinity of a
ferromagnetic permalloy film. In these works a
quasi-one-dimensional distribution of magnetization can be
achieved by applying a strong enough in-plane field
$H_{ext}>300\,$Oe, which aligns the domain walls in a desired
direction. Such a domain structure was maintained even after
switching off the external magnetic field. Magneto-optical
measurements directly display the preferential direction of the
vortex entry in the presence of the perpendicular magnetic field
along the domain walls [the panel (c) and (d) in Fig.
\ref{Fig:Gheorghe-Vlasko-MO}]. By reorienting the magnetic domains
using combinations of dc and ac fields, it is possible to
rearrange current patterns in the S/F bilayer and thus manipulate
its conductivity. The presence of this rotatable periodic
stripe-like magnetic domain structure with alternating
out-of-plane component of magnetization results in a difference in
the critical depinning current density between cases when the
magnetic domain stripes are oriented parallel and perpendicular to
the superconducting current: $J_c^{\parallel}>J_c^{\perp}$. For
planar thin-film Pb/Py structures Vlasko-Vlasov {\it et al.}
\cite{Vlasko-Vlasov-PRB-08b} observed a pronounced
magnetoresistance effect yielding four orders of magnitude
resistivity change in a few millitesla in-plane field. In
addition, the S/F bilayer exhibits commensurability features that
are related to the matching of the Abrikosov vortex lattice and
the magnetic stripe domains (Belkin {\it et al.}
\cite{Belkin-PRB-08}). The matching effects are less apparent than
for S/F hybrids with magnetic dots, although commensurability
becomes more pronounced as temperature is lowered. This result can
be explained by the gradual decrease in the $\lambda$ value, which
leads to stronger modulation of the magnetic field in the
superconductor at lower temperatures and consequently, to more
prominent magnetic interaction with ferromagnetic domain
structure.

It is interesting to note that the effect of magnetic domains on
the pinning of vortices, was also observed in high$-T_c$
superconductors such as YBa$_2$Cu$_3$O$_{7-\delta}$
(Garc\'{i}a--Santiago {\it et al.} \cite{Garsia-APL-00}, Jan {\it
et al.} \cite{Jan-APL-03}, Zhang {\it et al.} \cite{Zhang-EPL-01},
Laviano {\it et al.} \cite{Laviano-PRB-07}). At the same time the
influence of the ferromagnet on the nucleation in the high$-T_c$
superconductors should be rather small due to extremely small
coherence length (of the order of few nanometers).

\vspace*{0.2cm}

\noindent  {\it Current compensation effect and field-polarity
dependent critical current}

A superconducting square with in-plane magnetized ferromagnet on
top was proposed by Milo\v{s}evi\'{c} {\it et al.}
\cite{Milosevic-PRL-05b} as a potential field and current
compensator, allowing to improve the critical parameters of
superconductors. Indeed, such a magnet generates stray fields of
the same amplitude but opposite signs at the poles of the magnet,
therefore the field-compensation effect leads to the enhancement
of the upper critical field equally for both polarities of the
external field. The superconducting state was shown to resist much
higher applied magnetic fields for both perpendicular polarities.
In addition, such ferromagnet induces two opposite screening
currents inside the superconducting film plane (in the
perpendicular direction to its magnetization), which effectively
compensates the bias current, and therefore superconductivity
should persist up to higher applied currents and fields. These
effects have been recently studied experimentally by Schildermans
{\it et al.} \cite{Schildermans-JAP-09} in an Al/Py hybrid disk of
1.7 $\mu$m diameter where a finite dipolar moment lying in the
plane of the structure was achieved by pinning magnetic domains
with the contact leads used for electrical measurements.

\begin{figure}[t!]
    \begin{center}
    \epsfxsize=80mm \epsfbox{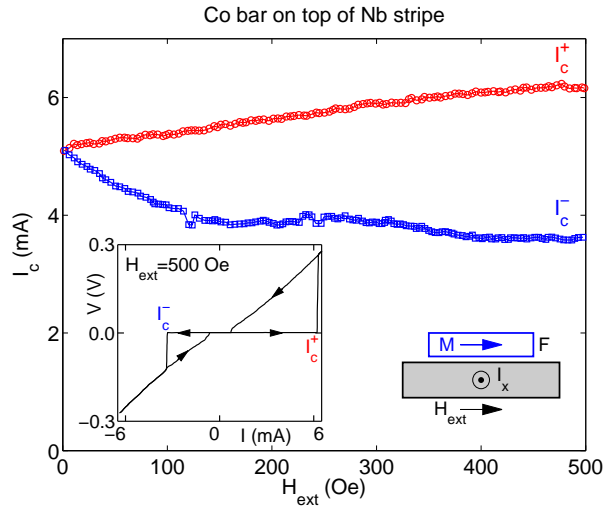}
    \end{center}
    \caption{(color online) Diode effect in the Nb bridge (2 $\mu$m width) with
    the in-plane magnetized Co stripe on top:
    Experimental dependence of $I_c^{+}$ (the critical current in the $x-$direction,
    see the geometry of the S/F system on the inset)
    and $I_c^{-}$ (the critical current in the opposite direction) on
    the external magnetic field $H_{ext}$ applied in the $y-$direction ($T=4.2$ K, $T_{c0}=9.2$ K),
    adapted from Vodolazov {\it et al.} \cite{Vodolazov-PRB-05b}.
    In the inset the dc
    $I-V$ characteristic of our hybrid system $H_{ext}=500$ Oe is presented,
    showing a pronounced diode effect.}
    \label{Fig:Vodolazov-PRB-2005}
    \end{figure}

Vodolazov {\it et al.} \cite{Vodolazov-PRB-05b} and Touitou {\it
et al.} \cite{Touitou-APL-04}  considered an alternative
experimental realization of the current compensator, consisting of
a superconducting bridge and a ferromagnetic bar magnetized
in-plane and perpendicularly to the direction of the bias current.
Such geometry allows one to weaken the self-field of the
superconducting bridge near its edge and thus to enhance the total
critical current corresponding to the dissipation-free current
flow. Since the self-field compensation occurs only for a certain
direction of the current (for fixed magnetization), the presence
of magnetized coating leads to a diode effect -- the
current-voltage $I-V$ dependence becomes asymmetrical (Fig.
\ref{Fig:Vodolazov-PRB-2005}). Later the similar difference in
critical currents flowing in opposite directions was studied
experimentally by Morelle and Moshchalkov \cite{Morelle-APL-06}
for a system consisting of a superconducting Al strip placed close
to a perpendicularly magnetized Co/Pd rectangle and Vodolazov {\it
et al.} \cite{Vodolazov-APL-09} for Nb/Co bilayer in the presence
of titled external magnetic field.


\subsection{Stray field-induced Josephson junctions}

Josephson junctions consist of weak links between two
superconducting reservoirs of paired electrons. Commonly, these
junctions are predefined static tunnel barriers that, once
constructed, can no longer be modified/tuned. In contrast to that,
a new concept of the Josephson junctions with a weak link
generated by the local depletion of the superconducting condensate
by a ``magnetic barrier'' from a micro/nano-patterned ferromagnet
can be realized (Sonin \cite{Sonin-JTP-88}). Interestingly, this
type of devices offer an unprecedented degree of flexibility as it
can be readily switched ON/OFF by simply changing between
different magnetic states using an in-plane field. This switching
process is fully reversible, and non-volatile since does not
require energy to keep one of the magnetic states.

\begin{figure}[t!]
    \begin{center}
    \epsfxsize=80mm \epsfbox{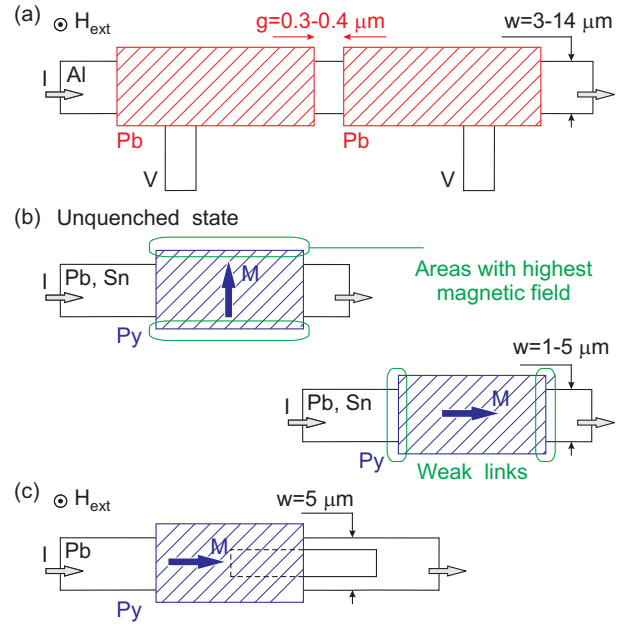}
    \end{center}
    \caption{(color online) (a) Sample layout investigated by Dolan and Lukens
    \cite{Dolan-IEEE-1977}: An uniform Al bridge was covered with a
    superconducting
    Pb strips (dashed rectangles) everywhere but in a small region near the center of
    the bridge. Due to the flux expulsion from the Pb strips the local magnetic field is primarily confined to
    this gap. \newline (b) Sample configuration investigated by
    Clinton and Johnson \cite{Clinton-APL-97,Clinton-JAP-99}: a Pb
    (or Sn)
    transport bridge is partially covered with a ferromagnetic Py strip
    with in-plane magnetic moment $M$. When $M$ is parallel to the
    bridge a strong stray field depletes the superconducting order
    parameter in a small region near the border of the Py bar
    (quenched state) thus inducing a weak link. \newline (c) Schematic presentation of
    magnetoquenched superconducting quantum interferometer,
    consisting of two superconducting Pb bridges connected in
    parallel and permalloy film on top, after
    Eom and Johnson \cite{Eom-APL-01}.}
    \label{Fig-Dolan-Clinton}
    \end{figure}

A pioneer investigation of the properties of superconducting weak
links achieved by local intense magnetic fields was performed by
Dolan and Lukens \cite{Dolan-IEEE-1977}. The sample layout used by
these authors and their typical dimensions is schematically shown
in Fig. \ref{Fig-Dolan-Clinton} (a) and it consists of an Al
bridge locally covered by a plain Pb film which has a thin gap of
width $g$ and spans the width of the Al strip at its center. By
applying an external magnetic field, the Pb film screens the
magnetic field due to the Meissner effect in the whole Al bridge
but magnify its intensity at the gap position. This effect leads
to a local region of suppressed superconductivity which gives rise
to dc and ac Josephson effect as evidenced by a finite critical
current and the presence of Shapiro steps in the current-voltage
characteristics at $V_n=n\hbar\omega/(2e)$ when the system was
irradiated with rf-excitations with frequency $\omega=2\pi f$, $n$
is integer. Interestingly, the Josephson-like features appear for
applied fields in the shield gap approximately equal to the upper
critical field of the Al film.

\begin{figure}[b!]
    \begin{center}
    \epsfxsize=80mm \epsfbox{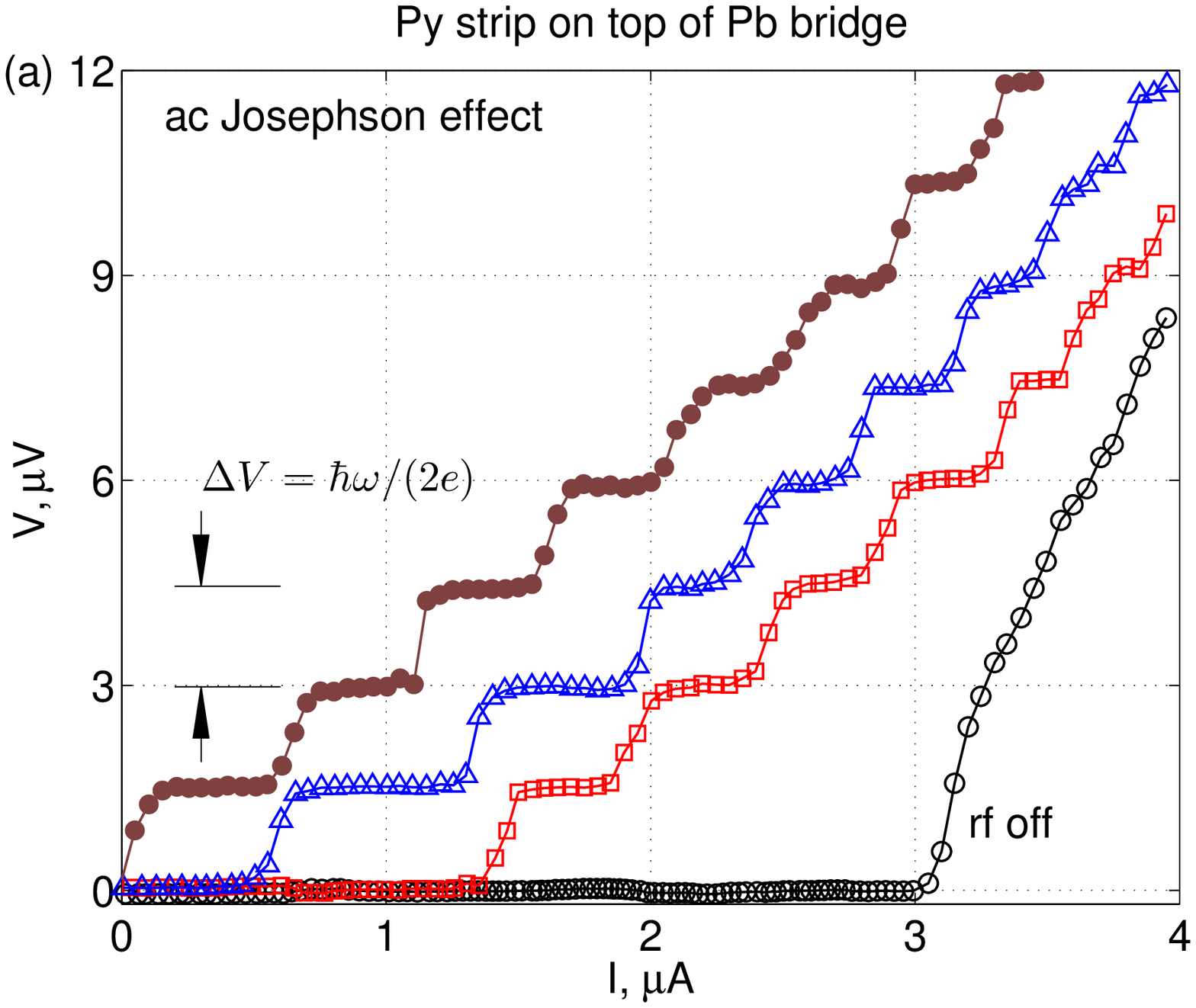}
    \epsfxsize=80mm \epsfbox{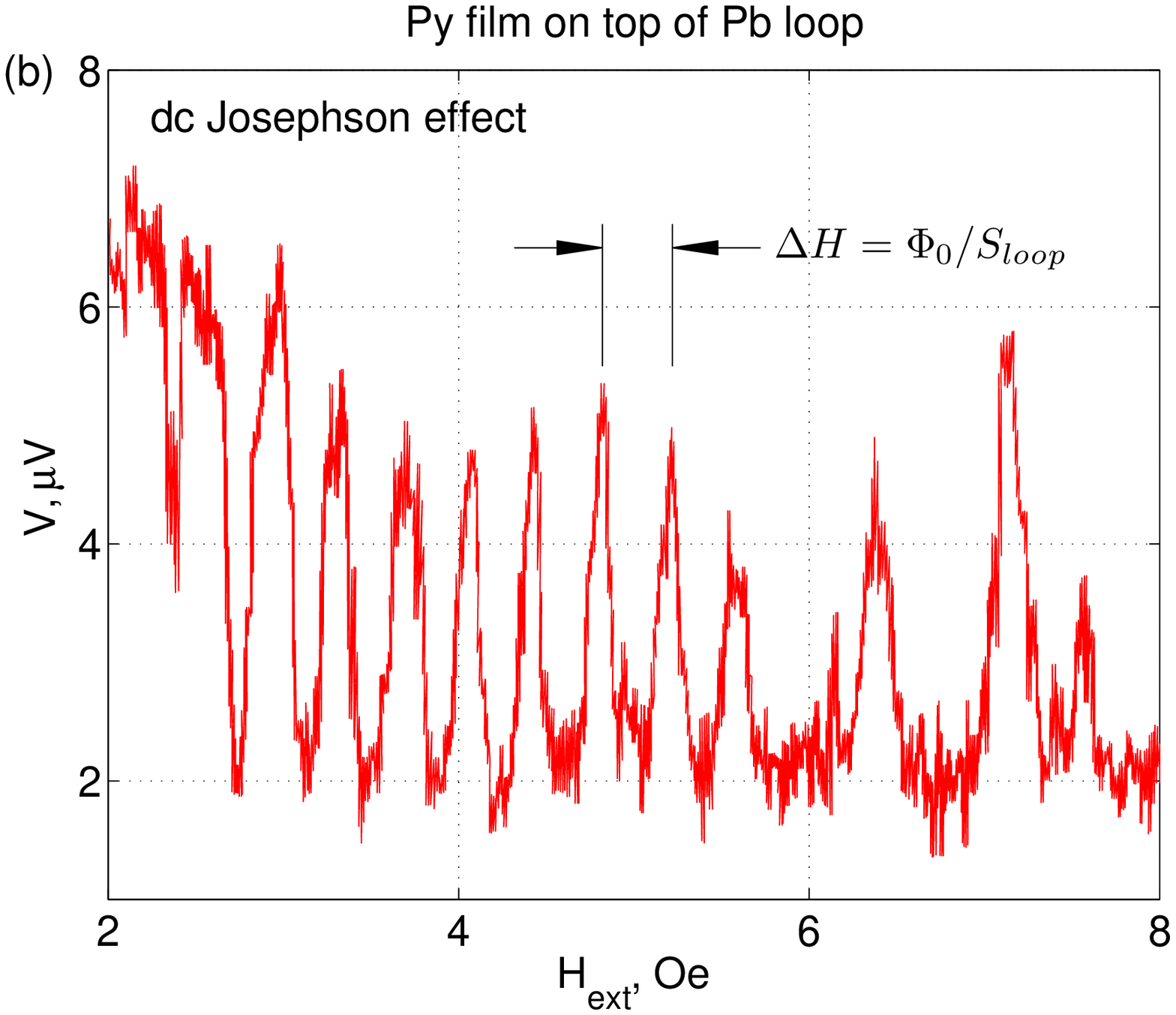}
    \end{center}
    \caption{(color online)
    (a) The $I-V$ curves obtained for a plain superconducting Pb
    bridge (2 $\mu$m wide) subjected in the inhomogeneous magnetic
    field, quenched state
    [see the panel (b) in Fig. \ref{Fig-Dolan-Clinton}] for different
    intensities of rf-irradiation,
    adapted from Clinton and Johnson \cite{Clinton-JAP-99}. The experiment was
    carried out at $T=5$ K, $T/T{c0}\simeq 0.76$, $H_{ext}=0$,
    frequency $f$ of the radio signal equal to 0.75 GHz. \newline
    (b) The $I-V$ dependence obtained for a superconducting Pb
    loop of the width 4.5 $\mu$m with rectangular hole 1.5$\times$7.0
    $\mu$m$^2$ covered by permalloy film [see the panel (c) in Fig. \ref{Fig-Dolan-Clinton}],
    adapted from Eom and Johnson \cite{Eom-APL-01}.
    This curve demonstrates the short period oscillations with the
    period determined by the area of the superconducting loop $S_{loop}$.}
    \label{Fig-Dolan-experiment}
    \end{figure}

An alternative method to obtain a field-induced weak link has been
more recently introduced by Clinton and Johnson
\cite{Clinton-APL-97,Clinton-JAP-99,Clinton-APL-00,Clinton-JAP-02}.
The basic device consist of a bilayer of a thin superconducting
strip and a ferromagnetic layer with in-plane magnetic moment
overlapping the width of the bridge [see panel (b) in Figure
\ref{Fig-Dolan-Clinton}]. When the magnetic moment is parallel to
the superconducting bridge the dipolar fringe is strong enough to
locally suppress the superconducting order parameter across the
bridge (quenched state) and thus create a weak link. This effect
can be turned off by simply magnetizing the ferromagnetic layer
perpendicular to the transport bridge with an external in-plane dc
field or by a current pulse in a separate transport line
\cite{Clinton-JAP-02}. Clearly, the proposed switchable Josephson
junction seems to be very attractive for potential technological
applications, since energy is required only to change the magnetic
states, which are thereafter maintained in thermodynamic
equilibrium. Later on based on the same idea Eom and Johnson
\cite{Eom-APL-01} proposed a switchable superconducting quantum
interferometer consisting of a ferromagnetic Py film partly
covering two parallel superconducting Pb bridges fabricated in a
loop geometry. The dependence of the voltage $V$, induced on this
superconducting loop at injection of stationary bias current, on
the perpendicular magnetic field $H_{ext}$ is shown in Fig.
\ref{Fig-Dolan-experiment}(b) and it reminds the standard
Fraunhofer diffraction pattern (Barone and Paterno
\cite{Barone-Paterno}).

\section{Hybrid structures: superconductor -- soft magnets}
\label{Soft-SF-systems}

Thus far, we have discussed the influence that a ferromagnet has
on the superconducting properties of S/F hybrids, assuming that
the magnetization of the ferromagnet ${\bf M}$ remains practically
unaltered. In this last section, we consider the possibility that
the magnetization ${\bf M}$ can be changed either due to the
external magnetic field or by the superconducting screening
currents induced by the magnetic subsystem, which are particularly
relevant at low temperatures. This situation could, in principle,
be achieved by using paramagnetic materials or soft ferromagnetic
materials with a low coercive field.

The equilibrium properties of ``superconductor -- soft magnet"
hybrid structures (so-called soft S/F hybrids) can be obtained
phenomenologically by the minimization of the Ginzburg-Landau
energy functional Eq. (\ref{GibbsEnergyExpansion1}) or the London
energy functional Eq. (\ref{GibbsEnergyLondon}), in which the term
$G_m$ responsible for the self-energy of the ferromagnet becomes
important
    \begin{eqnarray}
    G_m = \frac{1}{2M_s^2}\int\limits_{V_f} \Big(J_{\parallel}\,
    |\nabla M_x|^2+J_{\parallel}\,|\nabla M_y|^2+J_{\perp}\,|\nabla
    M_z|^2\Big)\,dV - \int\limits_{V_f} 2\pi Q M_z^2\,dV,
    \label{GibbsEnergyExpansionFerromagnet}
    \end{eqnarray}
where $J_{\parallel}$ and $J_{\perp}$ characterize the exchange
interaction between spins in a uniaxial ferromagnet with respect
to the in-plane and out-of-plane direction, $Q$ is a quality
factor taking into account the internal anisotropy of the
ferromagnet and determining the preferable orientation of the
magnetization (either in-plane or out-of-plane). Equation
(\ref{GibbsEnergyExpansionFerromagnet}) describes the energy cost
for having a slowly varying spatial distribution of the
magnetization\footnote[1]{The theory of
    superconductor -- soft ferromagnet systems near the
    ``ferromagnet--paramagnet" transition has been considered by Li {\it et
    al.} \cite{Li-PRB-06} within Ginzburg-Landau formalism.}
and, in particular, it describes the energy of a domain wall in a
ferromagnet. In some cases (for instance, for rapid ${\bf M}$
variations typical for ferromagnets with domain walls of rather
small width), in order to simplify the problem, the increase of
the free energy given by Eq.
(\ref{GibbsEnergyExpansionFerromagnet}) can be taken into account
phenomenologically by substituting $G_m$ by a fixed term $G_{dw}$
representing the energy of a domain wall.

\vspace*{0.3cm} \noindent {\it Modification of the domain
structure in a ferromagnetic film by the superconducting screening
currents}

\noindent The influence of superconducting environment (both
substrate or coating) on the equilibrium width of magnetic domains
in ferromagnetic films was considered theoretically by Sonin
\cite{Sonin-PRB-02b}, Genkin {\it et al.} \cite{Genkin-JMMM-94},
Sadreev \cite{Sadreev-PSS-93}, Bespyatykh {\it et al.}
\cite{Bespyatykh-PSS-94,Bespyatykh-PSS-98b}, Stankiewicz {\it et
al.} \cite{Stankiewicz-JAP-97,Stankiewicz-JPCM-97}, Bulaevskii and
Chudnovsky \cite{Bulaevskii-PRB-00,Bulaevskii-PRB-02}, Daumens and
Ezzahri \cite{Daumens-PLA-03}. In particular, one can expect a
prominent change in the equilibrium period of a one-dimensional
domain structure at $H_{ext}=0$ for rather thick ferromagnetic
films ($D_f\gg w$) with out-of-plane magnetization. Indeed, the
Meissner currents, induced by the ferromagnet, will decrease the
magnetic field inside the superconductor (usual flux expulsion
effects) and significantly increase the magnetic field inside the
ferromagnet. As a consequence, the density of the free energy of
the ferromagnet, given by \mbox{${\bf B}^2/8\pi-{\bf B}\cdot{\bf
M}$} or, equivalently, by \mbox{${\bf H}^2/8\pi-2\pi M_z^2$},
raises for a given $M_z$ distribution. However, the total energy
of the S/F system can be lowered by a decrease of the period of
the ferromagnetic domains: the smaller the period, the faster the
decay of ${\bf H}$ away from the surfaces of the ferromagnetic
film. Thus, it is expected that the equilibrium width of magnetic
domains in planar S/F bilayer becomes smaller below the critical
temperature of the superconducting transition as compared with the
state $T>T_{c0}$. In contrast to that, for thin ferromagnetic
films ($D_f\ll w$) the opposite behavior is predicted: the domain
width in the free ferromagnetic film should be smaller than that
for the same film on top of a superconducting substrate
\cite{Stankiewicz-JPCM-97}. This can be understood by taking into
account the change of the far-zone demagnetizing field
characteristics. In addition, Stankiewicz {\it et al.}
\cite{Stankiewicz-JPCM-97} argued that the effect of the
superconducting substrate on the period of domain structure in
ferromagnetic films with in-plane magnetization is rather small as
compared with that for the out-of-plane magnetized ferromagnets.
However, an increase of the magnetostatic energy of the S/F
hybrids at $T<T_{c0}$ due to the Meissner currents results also in
a shrinkage of the equilibrium width of an isolated 180$^{\circ}$
Bloch wall, separating two ferromagnetic domains with in-plane
magnetization, in the vicinity of the superconducting substrate,
as was predicted by Helseth {\it et al.} \cite{Helseth-PRB-02a}.

\begin{figure}[t!]
    \begin{center}
    \epsfxsize=80mm \epsfbox{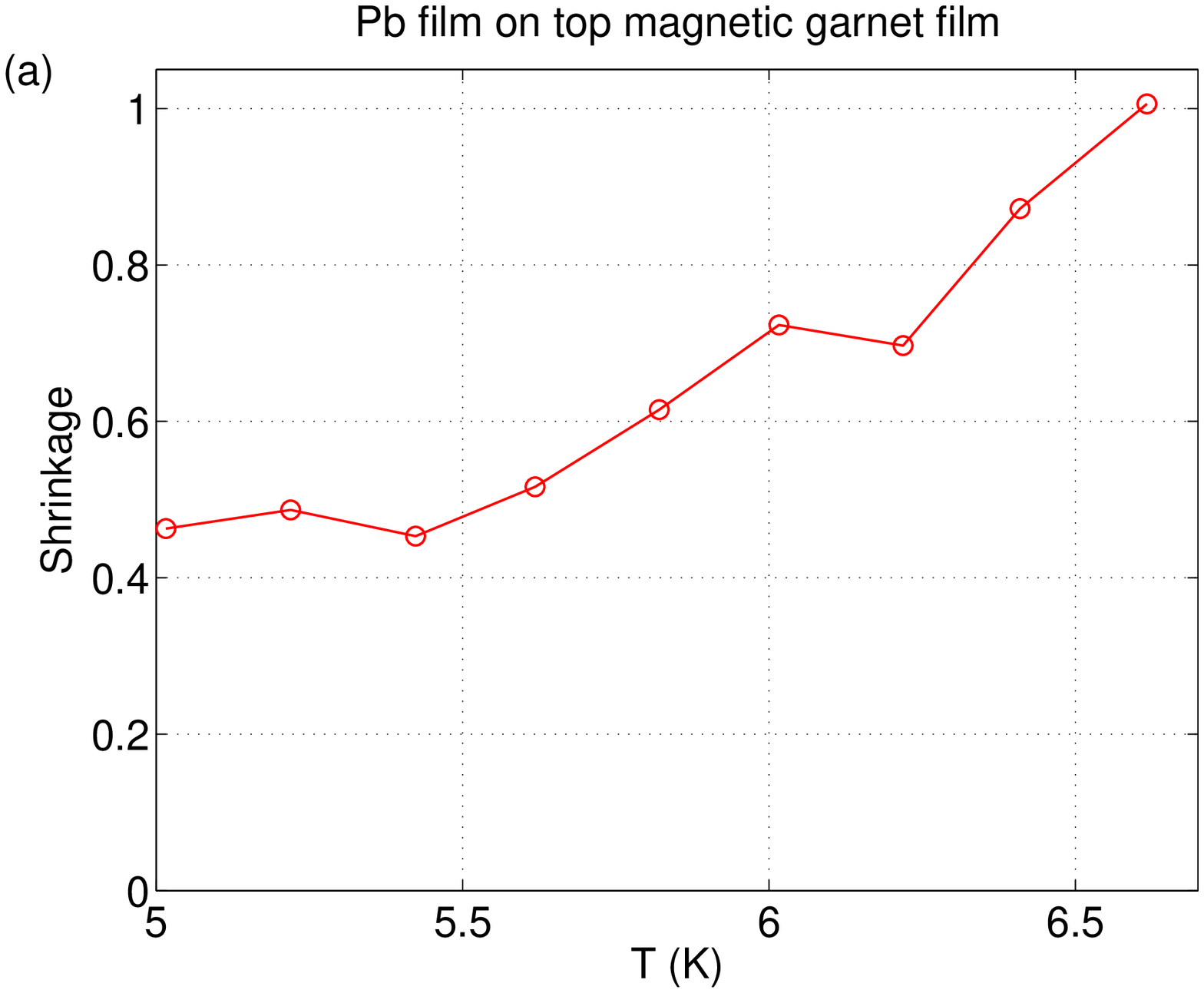}
    \epsfxsize=80mm \epsfbox{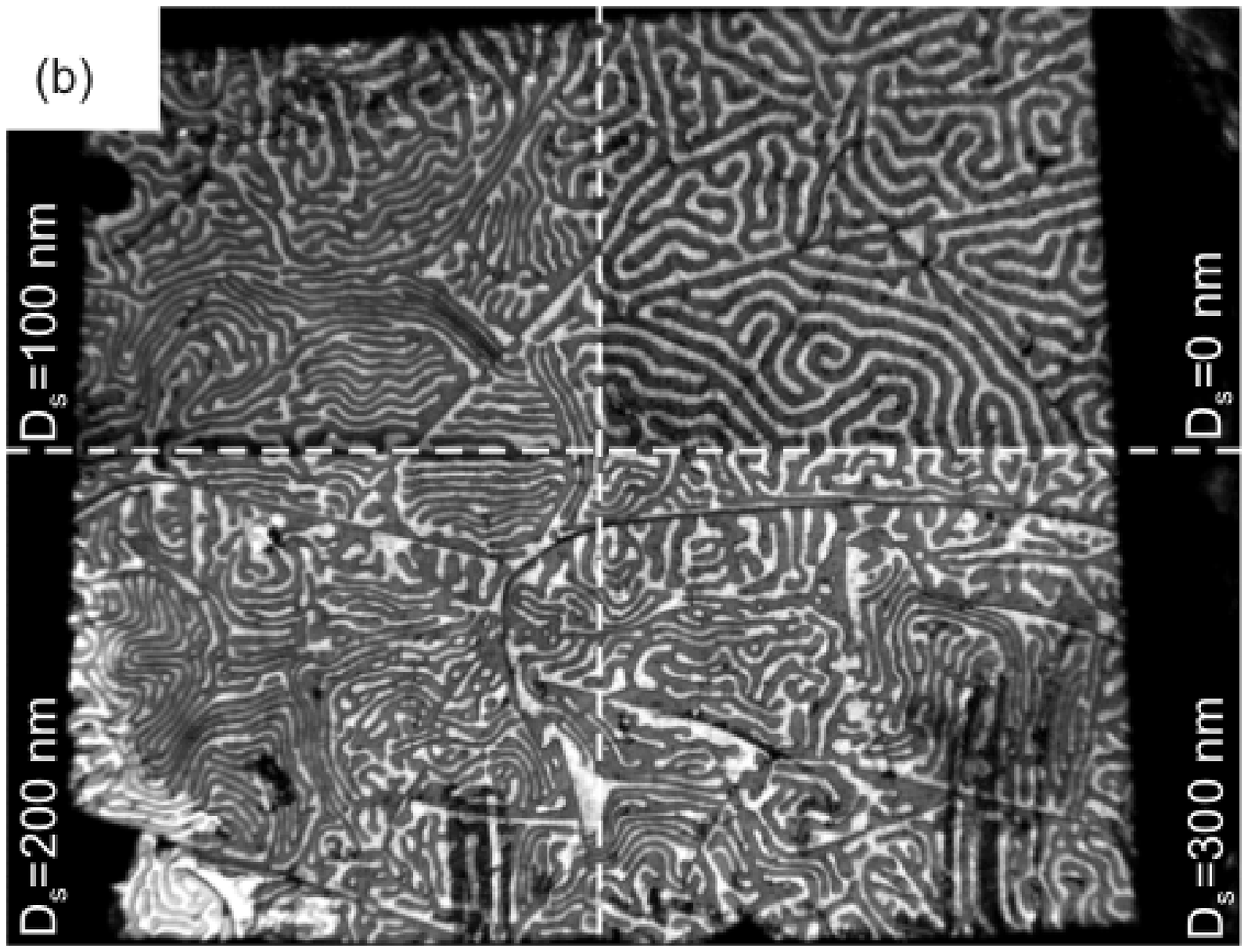}
    \end{center}
    \caption{(color online) (a) The temperature dependence of the
    shrinkage ratio of the width of magnetic domains
    in a hybrid structure consiting of a Pb film ($D_s=300$ nm) on top of a magnetic garnet
    film in comparison with the same ferromagnetic film without superconducting coating, adapted from
    Tamegai {\it et al.} \cite{Tamegai-JPCS-08}. \newline
    (b) The magneto-optical
    image of four segments of the superconducting Pb/garnet film
    structure, differing by the thickness of the Pb film ($D_{s}=0$ nm, 100
    nm, 200 nm, 300 nm in a counter-clock-wise direction), taken at $T=5.0$ K, adapted from
    Tamegai {\it et al.} \cite{Tamegai-JPCS-08}.}
    \label{Fig:Tamegai-08}
    \end{figure}

The foreseen decrease of the period of the domain structure in a
ferromagnetic garnet film in contact with a superconducting Pb
film was recently investigated by magneto-optical imaging (Tamegai
{\it et al.} \cite{Tamegai-JPCS-08}). It was demonstrated that the
shrinkage depends both on temperature and the thickness of
superconducting coating layer. The temperature dependence of the
shrinkage factor $s$ evaluated by comparing the average width
$\langle w\rangle$ of the magnetic domain width in regions with
and without the superconducting Pb film is shown in Fig.
\ref{Fig:Tamegai-08}(a). It points out that the lower the
temperature, the narrower the magnetic domains are ($s=0.47$ at
$T=5.0$ K).

\vspace*{0.3cm} \noindent {\it Alteration of magnetization of
ferromagnetic dots by the superconducting screening currents}

\begin{figure}[b!]
    \begin{center}
    \epsfxsize=70mm \epsfbox{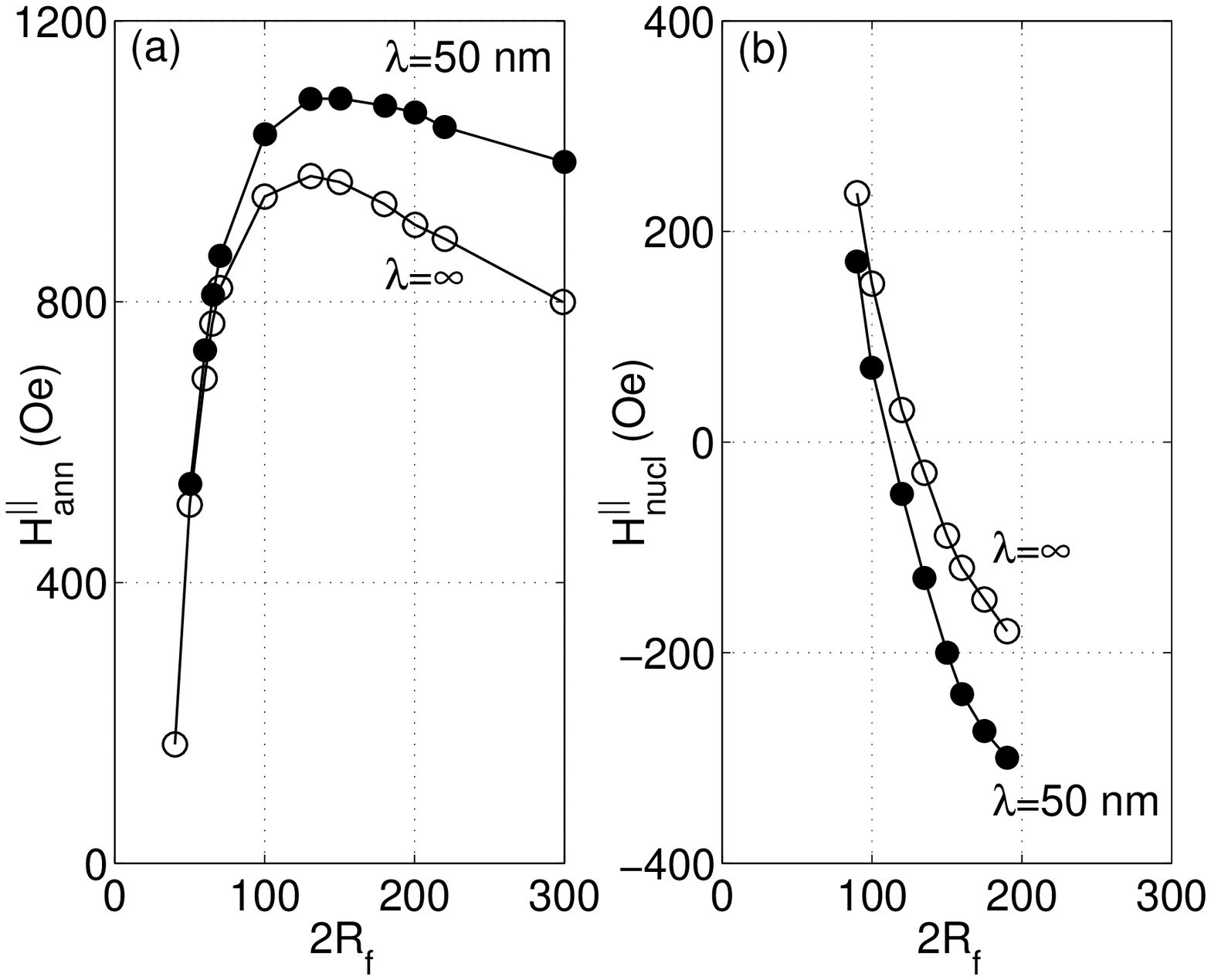}
    \epsfxsize=75mm \epsfbox{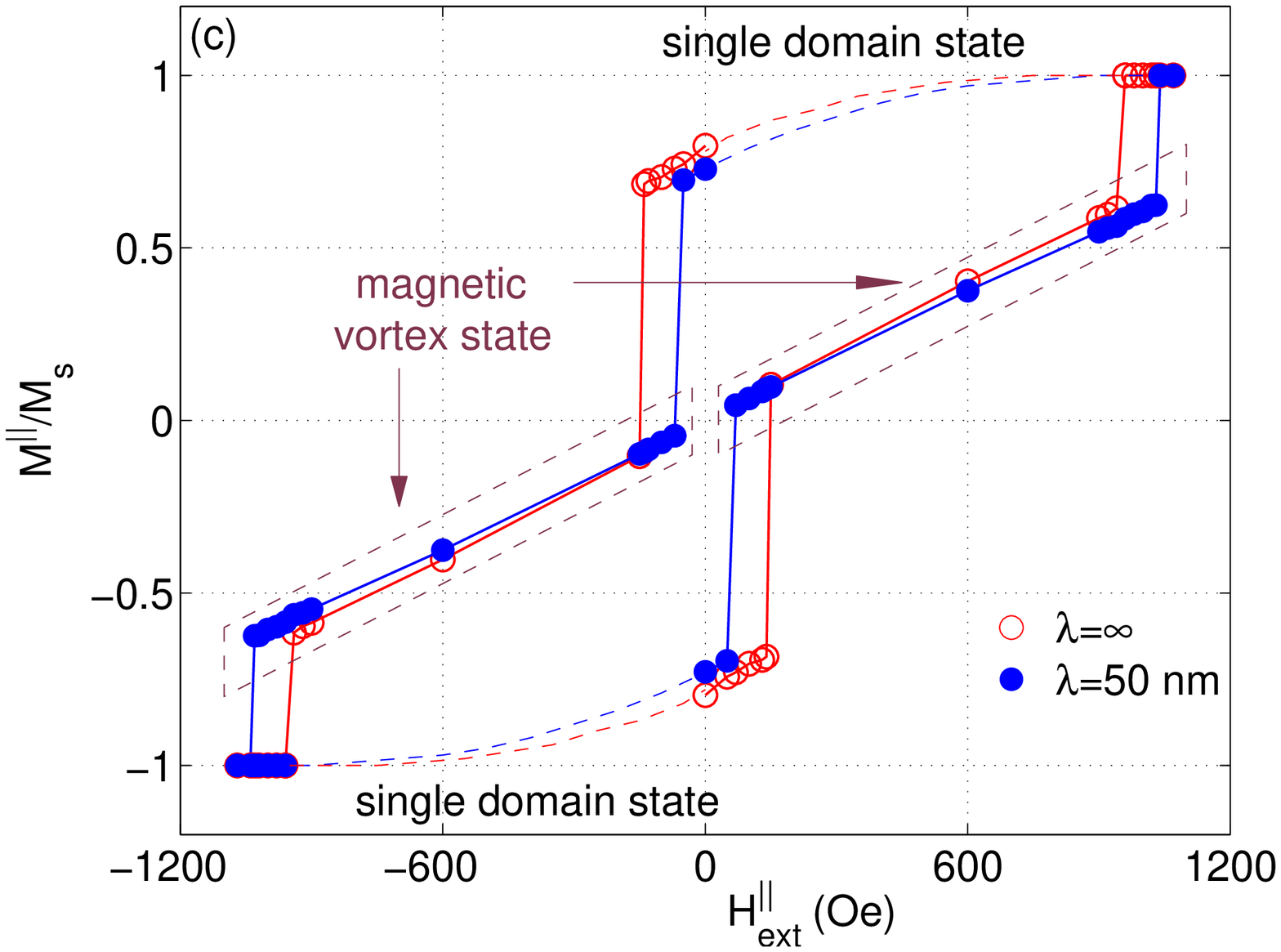}
    \end{center}
    \caption{(color online) (a) The dependence of the magnetic
    vortex annihilation field $H^{\parallel}_{ann}$, corresponding to the transition from the
    magnetic vortex state to a single-domain state,
    on the diameter of disk $2R_f$, calculated for an isolated ferromagnetic disk of 20 nm thickness
    (i.e, without superconductor,
    $\lambda=\infty$, open circles) and for the same disk placed above a bulk superconductor
    ($\lambda=50$ nm, filled circles), adapted from
    Fraerman {\it et al.} \cite{Fraerman-PRB-05}.  \newline
    (b) The dependence of the magnetic
    vortex nucleation field $H^{\parallel}_{nucl}$, corresponding to the transition from single-domain state to the
    magnetic vortex state for the same problem, adapted from
    Fraerman {\it et al.} \cite{Fraerman-PRB-05}.  \newline
    (c) The magnetization curve $M^{\parallel}/M_s$ vs. in-plane external field $H^{\parallel}_{ext}$
    demonstrating the process of the magnetization reversal for the magnetic disk (20 nm thickness and 100 nm
    diameter) for $\lambda=\infty$ (open circles) and $\lambda=50$ nm (filled circles), adapted from
    Fraerman {\it et al.} \cite{Fraerman-PRB-05}. Thus, the screening effect
    increases the width of the $H_{ext}$ interval at the ascending and descending branches
    of the magnetization curve where magnetic vortex state
    is energetically favorable.}
    \label{Fig:Fraerman-PRB-05}
    \end{figure}

\noindent The Meissner currents also influence the magnetic states
and the process of magnetization reversal in ferromagnetic disks
placed above a superconductor. It is well known that a uniformly
magnetized (single domain) state is energetically favorable for
radii $R_f$ smaller than some threshold value $R_f^*$ (for a given
thickness of the dot $D_f$), while the magnetic vortex state can
be realized for $R_f>R_f^*$. The typical $M(H_{ext})$ dependence
for ferromagnetic disks for $R_f>R_f^*$ was already shown in Fig.
\ref{Fig:Villegas-PRL-07}.  The dependence $R_f^*$ vs. $D_f$ (the
phase diagram in the ``diameter--height" plane) in the presence of
a bulk superconductor, characterized by the London penetration
depth $\lambda$, was investigated numerically by Fraerman {\it et
al.} \cite{Fraerman-PRB-05} and later analytically by Pokrovsky
{\it et al.} \cite{Pokrovsky-JMMM-06}. It was shown that the
smaller $\lambda$, the smaller the critical diameter $R_f^*$
becomes for a given dot's thickness. The transitions between the
two magnetic states can be induced also by increasing the external
magnetic field: the magnetic vortex, possessing an excess energy
at zero field, becomes energetically favorable for finite external
fields (the magnetic-vortex nucleation field). Although the energy
of the interaction between the superconductor and ferromagnet is
expected to be much smaller than the self-energy of the
ferromagnetic particle (for realistic $\lambda$ values), it could
lead to an experimentally observable decrease in the magnetic
vortex nucleation field $H^{\parallel}_{nucl}$ and increase in the
magnetic vortex annihilation field $H^{\parallel}_{ann}$ (Fig.
\ref{Fig:Fraerman-PRB-05}).

The appearance of a spontaneous magnetization of individual S/F
hybrids, consisting of an Al bridge and demagnetized Ni dots on
top, upon cooling through the superconducting transition
temperature at $H_{ext}=0$, was reported by Dubonos {\it et al.}
\cite{Dubonos-PRB-02}. Indeed, the reshuffling of magnetic domains
in the submicron ferromagnetic disk, caused by
temperature-dependent screening of the domain's stray fields by
the superconductor, can explain the observed appearance of nonzero
magnetization of the ferromagnet at low temperatures. More
recently, the modification of the magnetic state of Nb/Co and
Nb/Py superlattices induced by screening currents in the
superconducting Nb films was studied experimentally by Monton {\it
et al.} \cite{Monton-PRB-07,Monton-PRB-08,Monton-APL-08} and Wu
{\it et al.} \cite{Wu-PRB-07}.

Kruchinin {\it et al.} \cite{Kruchinin-SuST-06} demonstrated
theoretically that a superconducting environment modifies the
magnetostatic interaction between localized magnetic moments
(embedded small ferromagnetic particles), resulting either in
parallel or antiparallel alignment of neighbor dipolar moments at
$H_{ext}=0$. The crossover between these regimes depends on the
ratio of the interparticle spacing and the London penetration
depth, and thus preferable ``magnetic'' ordering (ferromagnetic
vs. antiferromagnetic arrangements) can be tuned by varying
temperature.

\vspace*{0.3cm} \noindent {\it Mixed state of soft S/F hybrid structures}

\begin{figure}[b!]
    \begin{center}
    \epsfxsize=65mm \epsfbox{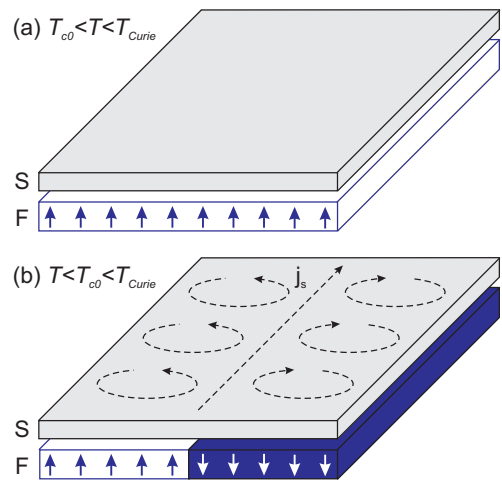}
    \end{center}
    \caption{(color online) (a) Uniform state of the S/F bilayer
    above the superconducting critical temperature, (b)
    Spontaneously formed magnetic domain structure and coupled chains of superconducting
    vortices with alternating vorticity at $T<T_{c0}$, adapted from
    Erdin {\it et al.} \cite{Erdin-PRL-01}.
    Solid arrows correspond to the magnetization of the ferromagnet,
    while black arrows schematically show the
    circulating superconducting currents.}
    \label{Fig:Erdin-PRL-02}
    \end{figure}

\noindent The magnetostatic interaction between a vortex-free
superconducting film and a uniformly magnetized ferromagnetic film
at $H_{ext}=0$ may cause the spontaneous formation of vortices in
the superconductor and magnetic domains in the ferromagnet in the
ground state of planar S/F bilayers with perpendicular
magnetization. Lyuksyutov and Pokrovsky \cite{Lyuksyutov-MPLB-00},
and Erdin {\it et al.} \cite{Erdin-PRL-01} argued that the ground
state of the S/F system could be unstable with respect to the
formation of superconducting vortices. Indeed, for a uniformly
magnetized S/F bilayer, characterized by a magnetization of the
ferromagnetic film per unit area $m=M_sD_f$, the magnetostatic
interaction between the superconductor and the ferromagnet changes
the total energy of an isolated vortex line to
$\varepsilon_v=\varepsilon_v^{(0)}-m\Phi_0$ \cite{Erdin-PRB-02} as
compared with the self-energy of the vortex in the superconducting
film $\varepsilon_v^{(0)}$ without a ferromagnetic layer. As a
consequence, the formation of vortices becomes energetically
favorable as soon as $\varepsilon_v<0$ (either for rather large
$m$ values or at temperatures close to $T_{c0}$ where
$\varepsilon_v^{(0)}$ vanishes). However, as the lateral size of
the S/F system increases, the averaged vortex density $n_v$ would
generate a constant magnetic field $B_z\simeq n_v\Phi_0$ along the
$z-$direction which can lead to an energy increase larger than the
gain in energy due to creation of vortices. Hence, in order for
the vortex phase to survive, the ferromagnetic film should split
in domains with alternating magnetization in a finite temperature
range at $T<T_{c0}$. As long as the magnetic domain width exceeds
the effective penetration depth, the energy of the stripe domain
structure seems to be minimal (Fig. \ref{Fig:Erdin-PRL-02}).
Interestingly, the interaction between a single vortex in a
superconducting film and the magnetization induced by this vortex
in the adjacent ferromagnetic film can cross over from attractive
to repulsive at short distances (Helseth \cite{Helseth-PLA-03}).

Carneiro studied the interaction between superconducting vortices
and a superparamagnetic particle with constant dipolar moment,
which is assumed to be able to freely rotate, in the London model
\cite{Carneiro-PRB-05,Carneiro-EPL-05,Carneiro-PhysC-06}. It was
found that, due to the rotational degree of freedom, the pinning
potential for superconducting vortices differs significantly from
that for a permanent dipole. In particular, the interaction
between the superconducting vortex and the magnetic dipole can be
tuned by applying an in-plane external field: the corresponding
depinning critical current was shown to be anisotropic and its
amplitude potentially varies by as much as one order of magnitude.
Later on, this approach was generalized by Carneiro
\cite{Carneiro-PRB-07} for hybrid systems consisting of thin
superconducting film and a soft ferromagnetic disks in the
magnetic vortex state (similar to that for Refs.
\cite{Fraerman-PRB-05,Pokrovsky-JMMM-06} but considering a vortex
line inside the superconducting sample).

A new method of pinning vortices in S/F epitaxial composite
hybrids consisting of randomly distributed Gd particles
incorporated in a Nb matrix was reported by Palau {\it et al.}
\cite{Palau-PRL-07,Palau-SuST-07}. Since the size of Gd particles
are much smaller than the coherence length and the interparticle
distance is much shorter than the penetration depth, this regime
of collective magnetic pinning differs both from conventional core
and magnetic pinning mechanisms. In this case, since a vortex
``feels" a homogeneous superconductor (for length scales on the
order of $\lambda$), pinning effects are expected to be small.
However, due to the local field of a vortex, the Gd particles can
be magnetized and a moving vortex would lead to hysteretic losses
in the magnetic particles, which in turn results in an increased
pinning (for decreasing magnetic fields).

\vspace*{0.3cm} \noindent {\it Superconductor -- paramagnet hybrid structures}

\noindent An alternative way of modifying the superconducting
properties of soft hybrid structures is by using paramagnetic
constituents, characterized by zero or very low remanent
magnetization. Such superconductor-paramagnet hybrids with a
magnetization ${\bf M}=(\mu-1)\,{\bf H}/4\pi$ depending on the
external field ($\mu$ is the magnetic permeability) in the
presence of transport current were considered theoretically by
Genenko \cite{Genenko-PSS-02}, Genenko and Snezhko
\cite{Genenko-JAP-02}, Genenko {\it et al.}
\cite{Genenko-PRL-99,Genenko-PRB-00,Genenko-PhysC-00} for $\mu\gg
1$. It was predicted that the paramagnetic material placed near
superconducting stripes and slabs can drastically modify the
current distribution in such hybrids, thus, suppressing the
current enhancement near the superconducting sample's edges
inherent for any thin-film superconductor in the flux-free
current-carrying state. As a consequence, the current
redistribution leads to an increase of the threshold value of the
total bias current corresponding to the destruction of the
Meissner state. In other words, the magnetically shielded
superconductors even in the Meissner state are able to carry
without dissipation rather high transport current comparable with
the typical current values for a regime of strong flux pinning
\cite{Genenko-PSS-02,Genenko-PRL-99,Genenko-PRB-00,Genenko-PhysC-00}.
A survival of the Meissner state for thin-film superconducting
rings carrying a current and placed between two coaxial
cylindrical soft magnets was studied by Genenko {\it et al.}
\cite{Genenko-SuST-01,Genenko-PhysC-02} and Yampolskii {\it et
al.} \cite{Yampolskii-PhysC-04}. The similar problem concerning
with the distribution of magnetic field inside and outside a
superconducting filament sheathed by a magnetic layer, as well as
the magnetization of such a structure in the region of reversible
magnetic behavior in the Meissner state was considered by Genenko
{\it et al.} \cite{Genenko-APL-04}. The formation of the mixed
state in various superconductor/paramagnet structures in the
presence of transport current, or an external magnetic field
$H_{ext}$ or the field of hard magnetic dipoles, were analyzed by
Genenko et al. \cite{Genenko-JPCM-05}, Genenko and Rauh
\cite{Genenko-PhysC-07}, Yampolskii and Genenko
\cite{Yampolskii-PRB-05}, and Yampolskii {\it et al.}
\cite{Yampolskii-EPL-06,Yampolskii-PhysC-07a,Yampolskii-PhysC-07b}
in the framework of the London model. The Bean-Livingston barrier
against the vortex entry in shielded superconducting filaments was
shown to strongly depend on the parameters of the paramagnetic
coating and, as a result, the critical field at which the first
vortex enters can be enhanced \cite{Genenko-JPCM-05}.

Since for the superconductor/paramegnet hybrids there are no
changes neither in the vortex structure in the Meissner state of
superconductor or in the magnetic state of paramagnetic elements,
characteristics of superconductor--paramagnet hybrids are presumed
to be reversible, and ac losses should be minimal for such
structures. It stimulated the implementation of paramagnetic and
ferromagnetic coatings in high-$T_c$ superconductors in order to
improve the critical current and reduce the ac-losses (Majoros
{\it et al.} \cite{Majoros-PhysC-00}, Glowacki {\it et al.}
\cite{Glowacki-PhysC-01}, Horvat {\it et al.}
\cite{Horvat-APL-02,Horvat-APL-05}, Duckworth
\cite{Duckworth-SuST-03}, Kovac {\it et al.} \cite{Kovac-SuST-03},
Touitou {\it et al.} \cite{Touitou-APL-04}, Pan {\it et al.}
\cite{Pan-SuST-04}, Jooss {\it et al.} \cite{Jooss-PRB-05}, Gu
{\it et al.} \cite{Gu-SuST-07}, \mbox{G\"{o}m\"{o}ry} {\it et al.}
\cite{Gomory-APL-07,Gomory-JPCS-08}).

Although, strictly speaking, permalloy is a ferromagnet with
rather low-coercive field, it can behaves qualitatively similar to
paramagnetic materials. Indeed, the magnetization vector for
thin-film structures deviates from in-plane orientation if a
perpendicular external field is applied. Such rotation of the
magnetization of the dot toward the out-of-plane direction while
sweeping the external field gives rise to a $z-$component of
magnetization depending on the external field. The effect of the
stray field generated by soft permalloy dots on the critical
current $I_c$ of the superconducting Al loops was experimentally
studied by Golubovi\'{c} and Moshchalkov \cite{Golubovic-APL-05}.
The monotonous decrease in the period of the oscillation on the
$I_c(H_{ext})$ with increasing the $H_{ext}$ value was interpreted
as a flux enhancement due to the increase of the out-of-plane
component of the dot's magnetic moment. In this sense soft
magnetic materials are promising candidates for the design of a
linear magnetic flux amplifier for applications in superconducting
quantum interference devices.

\section{Conclusion}
\label{conclusion}

We would like to conclude this work by formulating what, we
believe, are the most exciting unsolved issues and possible
interesting directions for further studies of S/F hybrid systems.

\vspace{0.1cm}

{\it Spontaneous formation of a vortex lattice and domain
patterns.} As we discussed in Section \ref{Soft-SF-systems}, the
magnetostatic interaction between a vortex-free superconducting
film and a uniformly and perpendicularly magnetized ferromagnetic
film at zero external field, may lead to a spontaneous formation
of vortex-antivortex pairs in the superconductor accompanied by
alternating magnetic domains in the ferromagnet in the ground
state (see Fig. \ref{Fig:Erdin-PRL-02} and Refs.
\cite{Lyuksyutov-AdvPhys-05,Erdin-PRL-01}). To the best of our
knowledge, thus far there are no experimental results confirming
this prediction. In part, this is likely due to the difficulties
of combining a very low coercive field ferromagnetic materials,
needed to guarantee the free accommodation of magnetic domains,
and a well defined out-of-plane magnetic moment.

\vspace{0.1cm}

{\it Magnetic pinning.} Based on London equations we have clearly
defined the magnetic pinning energy as ${\bf m}_0 \cdot {\bf B}_v$
(see section \ref{LondonSection}). Since London equations are
valid as long as core contributions are negligible, in principle
this simple relationship holds for materials with
$\kappa=\lambda/\xi\gg 1$ and low temperatures. Unfortunately in
the vast majority of the experimental reports so far, it seems
that these conditions are not satisfied. Any other contributions
such as local suppression of $T_c$, proximity effect, or local
changes in the mean free path, which are not accounted for within
London approximation, could lead to a deviation from the treatment
in the framework of London model. The problem that remains
unsolved so far is the identification of the most relevant
mechanisms of vortex pinning in S/F hybrid systems.

\vspace{0.1cm}

{\it Thermodynamic properties of S/F heterostructures.} Although
the electric transport in superconductor/ferromagnet hybrid
systems has been intensively studied during the past decades, very
little is known about their thermodynamic and thermal properties
such as their entropy, specific heat, thermal conductivity, etc.
From an academic point of view it would be very relevant to
investigate the nature of the phase transitions, or present
thermodynamic evidence of confinement of the superconducting order
parameter. On the other hand, estimating the heat released when
the system changes its state might also provide useful information
for devices based on S/F heterostructures.

\vspace{0.1cm}

{\it Electromagnet-superconductor hybrids.} Most of the research
performed so far has been focussed on the effects of an
inhomogeneous field generated by a ferromagnetic layer onto a
superconducting film. As was early demonstrated by Pannetier {\it
et al.} \cite{Pannetier-mohograph-95} in principle there is no
difference whether this inhomogeneous field is the stray field
emanating from a permanent magnet or the magnetic field generated
by micro (nano) patterned current carrying wire on top of the
surface of the film. The great advantage of the latter is the
degree of flexibility and control in the design of the magnetic
landscape and the external tuneability of its intensity. Such
electromagnet-superconductor hybrids represent a promising
alternative for further exploring the basic physics behind S/F
hybrid.


\vspace{0.1cm}

{\it Time resolved vortex creation and annihilation.} Josephson
$\pi$-junctions give rise to spontaneous formation of half-integer
flux quanta, so called semifluxons (Hilgenkamp {\it et al.}
\cite{Hilgenkamp-Nature-03}). It has been theoretically
demonstrated that for long Josephson junctions with zigzag
$\pi$-discontinuity corners the ground state corresponds to a flat
phase state for short separation between corners whereas an array
of semifluxons is expected for larger separations. Interestingly,
by applying an external bias current it is possible to force the
hopping of these semifluxons between neighboring discontinuities
(Goldobin {\it et al.} \cite{Goldobin-PRB-03}. This hopping of
semifluxons could be identified through time resolve ac
measurements with drive amplitudes above the depinning current.
There are clear similarities between these arrays of semifluxons
in zigzag Josephson systems and the vortex-antivortex arrays in
S/F systems. Indeed, recently Lima and de Souza Silva
\cite{Lima-arxiv-08} have shown theoretically that the dynamics of
the vortex-antivortex matter is characterized by a series of
creation and annihilation events which should be reflected in the
time dependence of the electrical field. Experimental work
corroborating these predictions are relevant for understanding the
dynamics of flux annihilation and creation in S/F systems.

\vspace{0.1cm}

All in all, the strive to comprehend the ultimate mechanisms
ruling the interaction between ferromagnets and superconductors
has made this particular topic an active theoretical and
experimental line of research. We believe that these vigorous
efforts will inspire further developments in this area of solid
state physics and perhaps motivate new applications of
technological relevance.

\appendix
\section*{Summary of experimental and theoretical researches}

\begin{table}[h!]
\centering
    \begin{tabular}{||l||c|c|c|c|c||}
    \hline
                                        & Co/Pt                                                                                                 &   BaFe$_{12}$O$_{19}$                                                 & Co, Fe, Ni                                                                           & Fe/Ni (Py)                                                & Other\\
                                        & Co/Pd                                                                                                 &   PbFe$_{12}$O$_{19}$                                                 &                                                                                      &                                                           & ferromagnets\\
                                        &                                                                                                       &                                                                       &                                                                                      &                                                           & \\ \hline

          \multicolumn{6}{||c||}{\textbf{ }} \\
          \multicolumn{6}{||c||}{\textbf{S/F hybrids consisting of large-area superconducting and plain non-patterned  }} \\
          \multicolumn{6}{||c||}{\textbf{ferromagnetic films (single crystals) with domain structure }} \\ \hline

          Pb films                      & \cite{Lange-JLTP-05,Lange-APL-02,Lange-MPLB-03,Lange-PhysC-04}                                        &  \cite{Yang-APL-06}                                                   &                                                                                      &  \cite{Vlasko-Vlasov-PRB-08b}                             & \cite{Tamegai-JPCS-08}\\
                                        & \cite{Lange-PRB-03}                                                                                   &                                                                       &                                                                                      &                                                           & \\ \hline

          Nb films                      & \cite{Gillijns-PRL-05,Gillijns-PhysC-06,Zhu-PRL-08,Cieplak-APP-04}                                    & \cite{Yang-Nature-04,Yang-PRB-06}                                     &  \cite{Monton-PRB-07,Monton-PRB-08,Monton-APL-08,Hinoue-JMMM-01}                     & \cite{Rusanov-PRL-04,Wu-PRB-07}                           & \cite{Haindl-PhysC-07,Matsuda-JAP-08,Feigenson-JAP-05} \\
                                        & \cite{Cieplak-JAP-05,Singh-APA-07}                                                                    & \cite{Fritzsche-PRL-06}                                               &  \cite{Lemberger-JAP-08,Joshi-JAP-07,Kobayashi-PRB-02,Kobayashi-PhysB-04}            &                                                           & \\ \hline

          Al films                      & \cite{Gillijns-PRB-07a,Aladyshkin-PhysC-08}                                                           &                                                                       &                                                                                      &                                                           & \\ \hline

          Other low$-T_c$               & \cite{Rakshit-PRB-08a,Rakshit-PRB-08b}                                                                &                                                                       & \cite{Kuroda-SuST-06,Horvat-APL-02,Horvat-APL-05,Pan-SuST-04}                        & \cite{Artley-APL-66,Vlasko-Vlasov-PRB-08a}                & \cite{Bell-PRB-06,Goa-APL-03}\\
           films                        &                                                                                                       &                                                                       &                                                                                      & \cite{Belkin-APL-08,Belkin-PRB-08}                        & \\ \hline

          High$-T_c$ films              & \cite{Jan-APL-03,Touitou-APL-04}                                                                      & \cite{Garsia-APL-00}                                                  & \cite{Majoros-PhysC-00,Glowacki-PhysC-01,Kovac-SuST-03,Gu-SuST-07}                   & \cite{Rubinstein-PRB-93}                                  & \cite{Zhang-EPL-01,Laviano-PRB-07,Duckworth-SuST-03}\\
                                        &                                                                                                       &                                                                       & \cite{Gomory-APL-07,Gomory-JPCS-08,Rubinstein-PRB-93}                                &                                                           & \cite{Jooss-PRB-05,Yuzhelevski-PhysC-99,Yuzhelevski-PRB-99}\\
                                        &                                                                                                       &                                                                       &                                                                                      &                                                           & \cite{Habermeier-SuST-04,Abd-Shukor-AIP-07}\\\hline


          \multicolumn{6}{||c||}{\textbf{ }} \\
          \multicolumn{6}{||c||}{\textbf{S/F hybrids consisting of large-area superconducting film and ferromagnetic elements: }} \\
          \multicolumn{6}{||c||}{\textbf{single particles, periodic arrays of magnetic dots (antidots) and stripes}} \\ \hline

          Pb films                      & \cite{Lange-PRL-03,vanBael-PhysC-00a,vanBael-PhysC-00b,vanBael-JC-01}                                 &                                                                       & \cite{vanBael-PRB-99,vanBael-PRL-01,vanBael-PhysC-00a,vanBael-PhysC-00b}            &                                                                & \\
                                        & \cite{vanBael-PhysC-02,vanBael-PhysC-01,vanBael-PRB-03,Lange-JLTP-05}                                 &                                                                       & \cite{vanBael-JC-01,vanBael-PhysC-02,vanLook-PhysC-00,Lange-JLTP-05}                &                                                                & \\
                                        & \cite{Lange-PRB-05,Neal-PRL-07,Lange-EPL-01,Lange-EPL-02}                                             &                                                                       & \cite{vanBael-JAP-02,Raedts-PhysC-02,Silhanek-APL-06,Xing-PRB-08}                   &                                                                & \\
                                        & \cite{Lange-JMMM-02,Gheorghe-PRB-08,Teniers-PhysC-02,Moshchalkov-CRP-06}                              &                                                                       & \cite{Xing-CondMat-08,Xing-CondMat-09,deSouzaSilva-PRL-07,Bending-PhysC-00}         &                                                                & \\ \hline

          Nb films                      & \cite{Stamopoulos-PRB-04,Stamopoulos-SuST-04,Stamopoulos-PRB-05,Stamopoulos-PhysC-06}                 &                                                                       & \cite{Martin-PRL-97,Martin-PRL-99,Martin-JMMM-98,Martin-PRB-00}                     & \cite{Pannetier-mohograph-95,Hoffmann-PRB-08}                  & \cite{Otani-JMMM-93,Geoffroy-JMMM-93,Koch-JAP-69} \\
                                        & \cite{Stamopoulos-SuST-05}                                                                            &                                                                       & \cite{Morgan-PRL-98,Morgan-JLTP-01,Hoffmann-PRB-00,Jaccard-PRB-98}                  & \cite{Otani-PhysC-94,Sun-PRL-04}                               & \cite{Palau-SuST-07,Palau-PRL-07,Haindl-PhysC-07} \\
                                        &                                                                                                       &                                                                       & \cite{Villegas-PRB-05b,Villegas-PRB-03,Villegas-PRB-05c,Stoll-PRB-02}               &                                                                & \cite{Haindl-SuST-08,Nozaki-JAP-96a,Nozaki-JAP-96b} \\
                                        &                                                                                                       &                                                                       & \cite{Velez-PRB-02b,Jaque-APL-02,Terentiev-PhysC-99,Terentiev-PhysC-00}             &                                                                & \\
                                        &                                                                                                       &                                                                       & \cite{Terentiev-PRB-00,Villegas-Science-03,Otani-PhysC-94,Silevitch-JAP-01}         &                                                                & \\
                                        &                                                                                                       &                                                                       & \cite{Velez-PRB-02a,Villegas-PRB-05a,Villegas-PRL-06,Montero-EPL-03}                &                                                                & \\
                                        &                                                                                                       &                                                                       & \cite{Dinis-PRB-2007,Dinis-NJP-2007}                                                &                                                                & \\ \hline

          Al films                      & \cite{Gillijns-PRB-07a,Aladyshkin-PhysC-08,Gillijns-PRB-06,Gillijns-PRB-07b}                          &                                                                       & \cite{Villegas-PRL-07,Villegas-PRB-08,deSouzaSilva-PRL-07}                          & \cite{Verellen-APL-08,Silhanek-APL-07,Silhanek-PhysC-08}      & \cite{Snezhko-PRB-05} \\
                                        & \cite{Gillijns-PhysC-08,Gillijns-PRL-07,Silhanek-PRB-07}                                              &                                                                       &                                                                                     &                                                               & \\ \hline

          Other low$-T_c$               & \cite{Rakshit-PRB-08a}                                                                                &                                                                       & \cite{Fasano-PRB-99,Strongin-RMP-64,Alden-APL-66,Alden-JAP-66}                      &                                                               & \\
          films                         &                                                                                                       &                                                                       & \cite{Togoulev-PhysC-06,Rizzo-APL-96}                                               &                                                               & \\\hline

          High$-T_c$ films              &                                                                                                       &                                                                       & \cite{Suleimanov-PhysC-04,Cheng-JSNM-08}                                            &                                                               & \cite{Melnikov-PRB-98,Aladyshkin-JETP-99,Nozdrin-IEEE-99}\\
                                        &                                                                                                       &                                                                       &                                                                                     &                                                               & \cite{Moser-JMMM-98,Gardner-APL-02,Auslaender-NatPhys-09}\\ \hline


        \multicolumn{6}{||c||}{\textbf{ }} \\
        \multicolumn{6}{||c||}{\textbf{Laterally confined and mesoscopic S/F hybrids}} \\ \hline

          Pb films                      &                                                                                                       &                                                                       &  \cite{Clinton-APL-00,Petrashov-JETPLett-94}                                         & \cite{Clinton-JAP-99,Clinton-APL-00,Clinton-JAP-02}            & \\
                                        &                                                                                                       &                                                                       &                                                                                      & \cite{Eom-APL-01}                                              & \\ \hline

          Nb films                      &                                                                                                       &                                                                       &  \cite{Vodolazov-PRB-05b,Vodolazov-APL-09}                                           & \cite{Clinton-JAP-02}                                          &   \\\hline

          Al films                      & \cite{Golubovic-APL-03,Golubovic-EPL-04,Golubovic-PRB-03,Golubovic-PRL-04}                            &                                                                       & \cite{Dubonos-PRB-02}                                                                & \cite{Schildermans-JAP-09,Golubovic-APL-05,Kerner-APL-04}      & \cite{Dolan-IEEE-1977} \\
                                        & \cite{Schildermans-PRB-08,Golubovic-PRB-05,Morelle-APL-06,Moshchalkov-CRP-06}                         &                                                                       &                                                                                      &                                                                & \\
                                        & \cite{Moshchalkov-EPJB-04,Golubovic-JAP-05}                                                           &                                                                       &                                                                                      &                                                                & \\ \hline

          Other low$-T_c$               &                                                                                                       &                                                                       & \cite{Togoulev-PhysC-06,Genenko-APL-04}                                              & \cite{Clinton-APL-97}                                                               & \\
          films                         &                                                                                                       &                                                                       &                                                                                      &                                                                & \\\hline

   \end{tabular}

\caption{Summary of experimental research on vortex matter in the
S/F hybrids with dominant orbital interaction (suppressed
proximity effect).} \label{Table-SF-experiment}
\end{table}

\begin{table}[h!]
\centering
    \begin{tabular}{||l||c|c|c||}
    \hline
                                        & \textbf{Ginzburg--Landau}                                                                             &   \textbf{London}                                                                                                                     & \textbf{Newton--like}      \\
                                        & \textbf{formalism}                                                                                    &   \textbf{formalism}                                                                                                                  & \textbf{description of}    \\
                                        &                                                                                                       &                                                                                                                                       & \textbf{vortex dynamics}   \\ \hline

          Bilayered and multilayered    & \cite{Aladyshkin-PRB-03,Buzdin-PRB-03,Samokhin-PRB-05,Aladyshkin-PRB-06}                              &  \cite{Sonin-JTP-88,Sonin-PRB-02b,Erdin-PRB-02,Erdin-PRB-04,Santos-PRB-01,Genkin-JMMM-94}                                             &   \\
          large-area S/F structures     & \cite{Gillijns-PRB-07a,Gillijns-PRL-05,Gillijns-PhysC-06,Li-PRB-06}                                   &  \cite{Bespyatykh-PSS-01a,Bespyatykh-PSS-01b,Helseth-PRB-02a,Laiho-PRB-03,Traito-PhysC-03,Erdin-PRB-06}                               &   \\
          (superconducting ferromagnets)& \cite{Radzihovsky-PRL-01}                                                                             &  \cite{Bulaevskii-PRB-00,Bulaevskii-PRB-02,Kayali-PRB-04b,Burmistrov-PRB-05,Bulaevskii-APL-00,Sadreev-PSS-93}                      &   \\
                                        &                                                                                                       &  \cite{Bespyatykh-PSS-94,Bespyatykh-PSS-98b,Stankiewicz-JAP-97,Stankiewicz-JPCM-97,Daumens-PLA-03,Lyuksyutov-MPLB-00}                      &   \\
                                        &                                                                                                       &  \cite{Erdin-PRL-01,Bespyatykh-JTP-97,Bespyatykh-PSS-97,Bespyatykh-PSS-00,Genkin-JMMM-95,Bespyatykh-PSS-98a}                      &   \\
                                        &                                                                                                       &  \cite{Pokrovsky-PRB-04,Sonin-PhysB-03,Faure-PRL-05a,Sonin-PRL-05,Faure-PRL-05b,Sonin-PRB-02a}                                          &   \\
                                        &                                                                                                       &  \cite{Sonin-PRB-98}                                                                                                                  &   \\ \hline

          Individual F elements         & \cite{Aladyshkin-JPCM-03,Aladyshkin-PRB-07,Matulis-PRL-94,Reijniers-PRB-99}                           &  \cite{Erdin-PRB-02,Wei-PRB-96,Carneiro-PRB-04,Xu-PRB-95,Coffey-PRB-95,Haley-PRB-96b}                                                 &   \\
          over large-area S films       & \cite{Milosevic-PRB-03a,Milosevic-JLTP-03a,Milosevic-PhysC-04b}                                       &  \cite{Wei-PhysC-97,Melnikov-PRB-98,Aladyshkin-JETP-99,Milosevic-PRB-02b,Milosevic-JLTP-03b,Carneiro-PhysC-04}                        &   \\
          (inside bulk superconductors) &                                                                                                       &  \cite{Carneiro-PRB-05,Carneiro-EPL-05,Carneiro-PhysC-06,Tokman-PLA-92,Milosevic-PRB-03b,Milosevic-PRB-04}                            &   \\
                                        &                                                                                                       &  \cite{Erdin-PRB-05,Kayali-PLA-02,Kayali-PRB-04a,Helseth-PLA-03,Kayali-PRB-05,Erdin-PRB-04}                                           &   \\
                                        &                                                                                                       &  \cite{Carneiro-PRB-07,Kruchinin-SuST-06,Snezhko-PRB-05,Carneiro-arxiv-08,Fraerman-PRB-05,Pokrovsky-JMMM-06}                          &   \\
                                        &                                                                                                       &  \cite{Pokrovsky-PRB-04}                                                                                                              &   \\ \hline

          Arrays of F dots elements     & \cite{Priour-PRL-04,Priour-PhysC-04,Milosevic-PRL-04,Milosevic-JLTP-05}                               &  \cite{Martin-PRB-00,Helseth-PRB-02b,Lyuksyutov-PRL-98,Sasik-CondMat-00,Erdin-PhysC-03,Wei-PRB-05}                                    & \cite{Chen-PRB-06a,Carneiro-PhysC-05,Carneiro-arxiv-08}                                 \\
          over large-area S films       & \cite{Milosevic-PhysC-06,Milosevic-PhysC-04a,Milosevic-PRL-05a,Milosevic-EPL-05}                      &  \cite{Wei-PhysC-06,Carneiro-PhysC-05,Cheng-JSNM-08,Teniers-PhysC-02,Pokrovsky-PRB-04}                                                & \cite{Lima-arxiv-08,deSouzaSilva-PRL-07,Teniers-PhysC-02} \\
          (inside bulk superconductors) & \cite{Gillijns-PRB-07b,Doria-PhysC-04a,Doria-PhysC-04b,Autler-JLTP-72}                                &                                                                                                                                       & \cite{Dinis-PRB-2007,Dinis-NJP-2007,Laguna-PRB-02}\\
                                        & \cite{Autler-HPA-72,Silhanek-PRB-07}                                                                                                      &                                                                                                   & \\ \hline

          Laterally confined and        & \cite{Chibotaru-JMP-05,Aladyshkin-PRB-07,Marmorkos-PRB-96,Cheng-PRB-99}                               & \cite{Coffey-PRB-02,Haley-PRB-96a,Lyuksyutov-MPLB-99,Lyuksyutov-IJMPB-03a,Lyuksyutov-IJMPB-03b,Ainbinder-SuST-07}                     & \\
          mesoscopic S/F structures     & \cite{Milosevic-PRB-02a,Milosevic-PRB-07,Golubovic-EPL-04,Golubovic-PRB-03}                           & \cite{Maksimova-PRB-06,Maksimova-PRB-08,Vodolazov-PRB-05b,Mawatari-PRB-08}                                                            & \\
                                        & \cite{Schildermans-PRB-08,Carballeira-PRL-05,Chen-PRB-06b,Golubovic-PRB-05}                           &                                                                                                                                       & \\
                                        & \cite{Doria-EPL-07,Doria-PhysC-08,Milosevic-PRL-05b,Moshchalkov-EPJB-04}                                                                  &                                                                                                   & \\ \hline

          Hybrid structures with        &                                                                                                       & \cite{Genenko-PSS-02,Genenko-JAP-02,Genenko-PRL-99,Genenko-PRB-00,Genenko-PhysC-00,Genenko-SuST-01}                                   & \\
          paramagnetic elements         &                                                                                                       & \cite{Genenko-PhysC-02,Yampolskii-PhysC-04,Genenko-APL-04,Genenko-JPCM-05,Genenko-PhysC-07,Yampolskii-PRB-05}                         & \\
                                        &                                                                                                       & \cite{Yampolskii-EPL-06,Yampolskii-PhysC-07a,Yampolskii-PhysC-07b}                                                                    & \\ \hline

   \end{tabular}

\caption{Summary of theoretical research on vortex matter in the
S/F hybrids with dominant orbital interaction (suppressed
proximity effect)} \label{Table-SF-theory}
\end{table}

\section*{Acknowledgements}

The authors are grateful to G. Carneiro, M.M. Doria, A.A.
Fraerman, A.S. Mel'nikov, M.V. Milo\v{s}evi\'{c}, A.V. D.A.
Ryzhov, Samokhvalov, M.A. Silaev, T. Tamegai, J.E. Villegas, V.K.
Vlasko-Vlasov, D.Yu. Vodolazov, J. van de Vondel for the valuable
comments and remarks which certainly improved the quality of this
review.

This work was supported by the K.U. Leuven Research Fund
GOA/2004/02 program, NES--ESF program, the Belgian IAP, the Fund
for Scientific Research -- Flanders (F.W.O.--Vlaanderen), the
Russian fund for Basic Research, by Russian Academy of Sciences
under the program ``Quantum physics of condensed matter" and the
Presidential grant MK-4880.2008.2 (A.Yu.A.). A.V.S. and W.G. are
grateful for the support from the F.W.O.--Vlaanderen.

\vspace*{2 cm}

\bibliography{List_of_references_revised}
\bibliographystyle{unsrt}

\end{document}